\documentstyle[twoside,ijmpa1,epsf]{article}

\catcode`\@=11
\long\def\@makefntext#1{
\protect\noindent \hbox to 3.2pt {\hskip-.9pt  
$^{{\eightrm\@thefnmark}}$\hfil}#1\hfill}		

\def\@makefnmark{\hbox to 0pt{$^{\@thefnmark}$\hss}}	
	
\def\ps@myheadings{\let\@mkboth\@gobbletwo
\def\@oddhead{\hbox{}
\rightmark\hfil\eightrm\thepage}   
\def\@oddfoot{}\def\@evenhead{\eightrm\thepage\hfil
\leftmark\hbox{}}\def\@evenfoot{}
\def\sectionmark##1{}\def\subsectionmark##1{}}

\def\qed{\hbox{${\vcenter{\vbox{			
   \hrule height 0.4pt\hbox{\vrule width 0.4pt height 6pt
   \kern5pt\vrule width 0.4pt}\hrule height 0.4pt}}}$}}
\def\Journal#1#2#3#4{{#1} {\bf #2}, #3 (#4)}
 
\def\NPB{{\em Nucl. Phys.} B}
\def\NPBP{{\em Nucl. Phys.} B (Proc.\,Suppl.)}
\def\PLB{{\em Phys. Lett.}  B}
\def\PRL{\em Phys. Rev. Lett.}
\def\PRD{{\em Phys. Rev.} D}
\def\ZPC{{\em Z. Phys.} C}

\def\be{\begin{equation}}
\def\ee{\end{equation}}
\def\bea{\begin{eqnarray}}
\def\eea{\end{eqnarray}}

\begin{document}
\renewcommand{\textfraction}{0.0}
\renewcommand{\floatpagefraction}{1.0}
 
\newcommand{\mev}{{\rm MeV}}
\newcommand{\gev}{{\rm GeV}}
\newcommand{\csw}{c_{\rm sw}}
\newcommand{\kcrit}{\kappa_{\rm crit}}
\newcommand{\msbar}{\overline{{\rm MS}}}
\newcommand{\stg}{\sqrt{\sigma}}
\newcommand{\bbar}{\mbox{$B^0$--$\overline{B^0}${}}}
\newcommand{\bbard}{\mbox{$B_d^0$--$\overline{B_d^0}${}}}
\newcommand{\bbars}{\mbox{$B_s^0$--$\overline{B_s^0}${}}}
\newcommand{\fbrootb}[1]{f_{B_{{#1}}}\,\sqrt{\widehat{B}_{B_{{#1}}}}}
\newcommand{\amb}{{\alpha_s(m_b)}}
\newcommand{\amu}{{\alpha_s(\mu)}}
\newcommand{\w}{{\omega}}
\newcommand{\fd}{f_D}
\newcommand{\fds}{f_{D_s}}
\newcommand{\fb}{f_B}
\newcommand{\fbd}{f_{B_d}}
\newcommand{\fbs}{f_{B_s}}
\newcommand{\fbstat}{f_B^{\rm stat}}
\newcommand{\bb}{B_B}
\newcommand{\bbd}{B_{B_d}}
\newcommand{\bbs}{B_{B_s}}
\newcommand{\rgbb}{\widehat{B}_B}
\newcommand{\rgbbd}{\widehat{B}_{B_d}}
\newcommand{\rgbbs}{\widehat{B}_{B_s}}
\newcommand{\rgbk}{\widehat{B}_K}
\newcommand{\xisd}{\xi_{sd}}
\newcommand{\zastat}{Z_A^{\rm stat}}
\newcommand{\Vckm}{V_{{\rm CKM}}}
\newcommand{\etal}{{\em et~al.}}
\newcommand{\hepph}[1]{{\tt hep-ph/{#1}}}
\newcommand{\heplat}[1]{{\tt hep-lat/{#1}}}
\newcommand{\gtaeq}{\raisebox{-.6ex}{$\stackrel{\textstyle{>}}{\sim}$}}
\newcommand{\lesssim}{\raisebox{-.6ex}{$\stackrel{\textstyle{<}}{\sim}$}}
\newcommand{\SP}{\Delta_{\rm 1P-1S}}
\newcommand{\dleft}{\stackrel{\leftarrow}{D}}
\newcommand{\dright}{\stackrel{\rightarrow}{D}}
\newcommand{\plus}{\makebox[13pt][c]{$+$}}
\newcommand{\minus}{\makebox[13pt][c]{$-$}}
 
\newcommand{\err}[2]{\raisebox{0.08em}{\scriptsize
{$\,\begin{array}{@{}l@{}}
                          \plus\makebox[0.55em][r]{#1} \\[-0.12em]
                          \minus\makebox[0.55em][r]{#2}
                        \end{array}$}}}
\newcommand{\er}[2]{\raisebox{0.08em}{\scriptsize
{$\,\begin{array}{@{}l@{}}
                          \plus\makebox[0.15em][r]{#1} \\[-0.12em]
                          \minus\makebox[0.15em][r]{#2}
                        \end{array}$}}}
\newcommand{\ewxy}[2]{\setlength{\epsfxsize}{#2}\epsfbox[-50 0 420 420]{#1}}
\newcommand{\fwxy}[2]{\setlength{\epsfxsize}{#2}\epsfbox[-50 0 420 320]{#1}}
\begin{titlepage}

\begin{flushright}
{\large Oxford Preprint: OUTP--97--20P}\\
{\large May~1997}
\end{flushright}

\vspace*{15mm}

\begin{center}
{\huge Leptonic Decays of Heavy Quarks on the Lattice}\\[12mm]

{\large\bf Hartmut~Wittig}\\ [0.3cm]
Theoretical Physics, University of Oxford,\\
1~Keble Road, Oxford OX1~3NP, UK

\end{center}
\vspace{15mm}
\begin{abstract}
{\normalsize The status of lattice calculations of heavy-light decay
constants and of the $B$ parameter $B_B$ is reviewed. After describing
the lattice approach to heavy quark systems, the main results are
discussed, with special emphasis on the systematic errors in present
lattice calculations. A detailed analysis of the continuum limit for
decay constants is performed. The implications of lattice results on
studies of CP violation in the Standard Model are discussed.}
\end{abstract}

\vspace{2cm}

\begin{center}
{\large (Submitted to International Journal of Modern Physics A)}
\end{center}

\clearpage

\end{titlepage}

\clearpage
\begin{titlepage}
\mbox{}
\vfill
\clearpage
\end{titlepage}

\runninghead{Leptonic decays of heavy quarks on the lattice}
{Leptonic decays of heavy quarks on the lattice}

\normalsize\textlineskip
\thispagestyle{empty}
\setcounter{page}{1}

\copyrightheading{}			

\vspace*{0.88truein}

\fpage{1}
\centerline{\bf LEPTONIC DECAYS OF HEAVY QUARKS ON THE LATTICE}
\vspace*{0.37truein}
\centerline{\footnotesize HARTMUT WITTIG}
\vspace*{0.015truein}
\centerline{\footnotesize\it Theoretical Physics, University of
Oxford, 1~Keble Road}
\baselineskip=10pt
\centerline{\footnotesize\it Oxford OX1 3NP, UK}
\vspace*{0.225truein}
\publisher{(received date)}{(revised date)}

\vspace*{0.21truein}
\abstracts{
The status of lattice calculations of heavy-light decay
constants and of the $B$ parameter $B_B$ is reviewed. After describing
the lattice approach to heavy quark systems, the main results are
discussed, with special emphasis on the systematic errors in present
lattice calculations. A detailed analysis of the continuum limit for
decay constants is performed. The implications of lattice results on
studies of CP violation in the Standard Model are discussed.
}{}{}


\vspace*{1pt}\textlineskip	

\section{Introduction}

Heavy quark systems have attracted a lot of interest in the past
decade. Theoretical and experimental investigations of weak decays of
heavy quark systems offer the main sources of information on the
phenomenon of CP violation in the Standard Model. Our understanding of
CP violation is directly related to the knowledge of the elements of
the Cabibbo-Kobayashi-Maskawa (CKM) matrix, which describes the
pattern of quark mixing in flavour-changing charged current
interactions. Within the framework of the Standard Model the CKM
matrix is unitary, and, provided that one can determine its elements
with high enough precision, any deviation from unitarity would be
interpreted as a hint of ``new physics''. The theoretical treatment of
weak decay amplitudes involving heavy quarks is, however, hampered by
large uncertainties due to strong interactions. Since quarks are
confined within hadrons, the exchange of soft gluons between them
makes weak decay matrix elements intractable in perturbation
theory. Therefore, other theoretical tools have to be applied in order
to deal with the intrinsic non-perturbative nature of weak decays of
hadronic systems, e.g. QCD sum rules\,\cite{QCD_SR} or lattice gauge
theories\,\cite{wilson74,wilson75}. Another formalism which can be
applied specifically to heavy quark systems is provided by the Heavy
Quark Symmetry and the Heavy Quark Effective Theory
(HQET)\,\cite{Shur_82}${}^{-}$\cite{HQET_reviews}. 

Lattice gauge theories offer the possibility of a systematic
non-perturbative treatment of strong interaction effects in weak decay
amplitudes. Indeed, Monte Carlo simulations of lattice QCD have made
important contributions to our understanding of hadronic physics by
providing non-perturbative, model-independent estimates for a number
of quantities, which are relevant for the spectrum and weak decays of
hadronic systems in general. An important ingredient in
non-perturbative studies of weak decays is the operator product
expansion\,\cite{wilson_ope}. In general, a weak amplitude, which
usually involves the product of two hadronic currents, can be written
as a sum over local operators, viz.
\be
   {\cal A}({\rm f}\leftarrow {\rm i}) = \sum_n\,C_n(\mu)\,
  \langle{\rm f}|\widehat{O}_n(\mu)|{\rm i}\rangle.
\ee
Here, $\mu$ is the renormalisation scale, and the Wilson coefficients
$C_n(\mu)$, which contain the short-distance physics, are independent
of the initial and final states and can be computed in perturbation
theory for $\mu\gg\Lambda_{\rm QCD}$. The matrix element $\langle{\rm
f}|\widehat{O}_n(\mu)|{\rm i}\rangle$, on the other hand, describes
the long-distance physics, contains strong interaction contributions
and must be calculated non-perturbatively, e.g. in a lattice
simulation.

The r\^ole of lattice simulations in the analysis of weak matrix
elements is two-fold. First, they serve to test QCD, as lattice data
can be compared to experiment or to results obtained using HQET or QCD
sum rules. Second, they can make predictions for yet unmeasured
quantities like the decay constant $\fb$. In fact, $\fb$ (or rather
the combination $\fb\sqrt{\bb}$, where $\bb$ is the $B$ parameter
relevant for $\bbar$ mixing) is the principal unknown quantity in the
analysis of CP violation in the heavy quark sector. If lattice QCD is
to have a serious impact on testing the consistency of the Standard
Model, then it is clear that the control over systematic errors in the
lattice approach is of paramount importance.

There exists by now a vast amount of literature on lattice results for
heavy-light decay constants, which will be discussed in detail in this
review, pointing out the differences between various calculations and
assessing their inherent systematic errors. Another aim of this study
is to combine as many lattice results as possible in a systematic
fashion, and to provide the currently best estimate for
$\fb\sqrt{\bb}$ computed on the lattice. The present status of
lattice calculations of $B$ meson decay constants and $B$ parameters
can be summarised as follows
\bea
    \fb & = & 172\err{27}{31}\,\mev,\qquad \fbs/\fbd=1.14\pm0.08,\\
    \fb\sqrt{\rgbb} & = & 195\err{30}{40}\,\mev,\qquad\rgbb=1.3\er{2}{3},
\eea
where $\rgbb$ is the renormalisation group invariant $B$ parameter. In
the following sections we will review lattice data from several groups
and describe the derivation of the global results above.

An overview of the development of this area of lattice gauge theories
can be gained by consulting various review talks in the proceedings of
recent lattice conferences\,\cite{cts_lat_92}${}^{-}$\cite{flynn_lat96}.
Other reviews and pedagogical introductions can be found in
refs.\,\cite{bernard_TASI89}${}^{-}$\cite{flynn_ICHEP_96}. 

We begin this review in section\,2 with a brief introduction to
lattice QCD for the non-specialist and describe the lattice approach
to heavy quark systems. In section\,3 we discuss lattice results for
heavy-light decay constants. A detailed, if somewhat technical
analysis of the continuum limit is presented in subsection\,3.4, which
may be skipped if the reader is not so interested in the technical
details. Section\,4 contains the discussion of lattice results for the
$B$ parameter $\bb$. The implications of our findings on the study of
CP violation is presented in section\,5. Finally, section\,6 contains
some concluding remarks. Two appendices are added, which present
details of the matching between lattice matrix elements and their
continuum counterparts, as well as a summary of lattice data in the
light hadron sector.

\section{Lattice Approach to Heavy Quark Systems}

In this section we present the basic concepts of lattice QCD before
various formulations of heavy quarks on the lattice are discussed in
some detail. Furthermore, the relation between weak matrix elements on
the lattice and their continuum counterparts is discussed, and at the
end of this section we summarise the main systematic errors which
affect lattice results for heavy quark physics.

\subsection{Lattice QCD and numerical simulations} 
\label{SEKLQCD}

Since its original foundation by Wilson
in~1974\,\cite{wilson74,wilson75}, lattice QCD has developed into a
mature area of research in elementary particle physics. In general,
the lattice approach now forms an important part in the study of a
large class of quantum field theories, ranging from scalar field
theory to quantum gravity.

One important aspect in the study of lattice field theories is that
they are suitable for numerical methods such as Monte Carlo
simulations. Indeed, the biggest impact lattice QCD has had on
phenomenology in the past decade has originated from large-scale
numerical simulations. The strength of this approach is based on the
fact that the lattice is currently the only known non-perturbative
regulator of a quantum field theory which is systematically
improvable.

In this subsection we will briefly discuss the basic features of the
formulation and the implementation of lattice QCD in Monte Carlo
simulations. For more detailed information, the reader is referred to
several good textbooks\,\cite{creutz,rothe,mont_muen}.

In order to be easily accessible to computer simulations, lattice QCD
is usually formulated in euclidean space-time. For a finite system,
the lattice size is $L^3\cdot T$, where $L^3$ is the spatial volume,
and~$T$ is the extension of the lattice in the time direction. The
lattice sites are separated by the lattice spacing~$a$. It is easy to
show that the lattice spacing acts as an UV cut-off. Present computers
can handle lattices of typically $16^4$ to $48^4$ lattice sites. Thus,
to be able to accommodate a hadron, the physical length of the lattice
should not be smaller than about 1.5\,fm. This implies that values of
the lattice spacing in physical units must lie in the range of
$a=0.1$--0.05\,fm.

On the lattice, non-abelian gauge fields are represented by so-called
{\it link variables}, which connect neighbouring sites.  These link
variables, which carry a Lorentz index~$\mu$, are elements of the
gauge group (here: SU(3)). Given a link variable $U_\mu(x)$ emanating
from site~$x$ in direction~$\mu=1,\ldots,4$, one can define a vector
field $A_\mu(x)$ as an element of the Lie algebra of the gauge group
SU(3) via 
\be  \label{GLlink}
   e^{iaA_\mu(x)} \equiv U_\mu(x) \in {\rm SU(3)}.  
\ee
The simplest lattice version of the action of a Yang-Mills theory is
the Wilson plaquette action\,\cite{wilson74} 
\be
S_G[U] = \beta\sum_{x}\sum_{\mu<\nu}\Big( 1-\frac{1}{3}{\rm
       Re\,Tr}\,P_{\mu\nu}(x)\Big), \qquad \beta=6/g_0^2 
\ee
where $P_{\mu\nu}(x)\equiv U_\mu(x)U_\nu(x+\hat{\mu})
U_\mu^\dagger(x+\hat{\nu})U_\nu^\dagger(x)$ is the product of links
around an elementary square of the lattice, the ``plaquette", and
$g_0$ is the bare gauge coupling. Here, $\hat{\mu}$ denotes a vector
in direction~$\mu$ of length~$a$, such that $x+\hat{\mu}$ is the
neighbouring site of~$x$ in direction~$\mu$. Using the definition of
the field $A_\mu(x)$ in eq.\,(\ref{GLlink}), one can show that for
small lattice spacings, the term in $S_G[U]$ reduces to the familiar
expression $-\frac{1}{g_0^2}{\rm Tr}\,F_{\mu\nu}^2$.

Quark and antiquark fields, $\psi(x),\,\overline{\psi}(x)$ are
associated with the lattice sites. Using the link variable $U_\mu(x)$,
a possible definition of the covariant derivative is given by
\bea   \
   D_\mu\psi(x) & \equiv & \frac{1}{a}
   \big\{U_\mu(x)\psi(x+\hat{\mu})-\psi(x)\big\} \label{GLcov_dev}\\
   \overline{\psi}(x)\dleft_\mu & \equiv & \frac{1}{a}
      \big\{\overline{\psi}(x+\hat\mu)U_\mu^\dagger(x) -
            \overline{\psi}(x)\big\}.
\eea
In euclidean space-time, the Dirac matrices can be defined to satisfy
\be
  \left\{\gamma_\mu,\gamma_\nu\right\} = 2\delta_{\mu\nu}.
\ee
Now we are in a position to write down a latticised version of the QCD
Lagrangian. It turns out, however, that the na\"\i{}ve lattice
transcription of the fermionic part suffers from the notorious fermion
doubling problem: in the case of a free Dirac particle it simply means
that, in four dimensions of space-time, the propagator contains
16~poles, resulting in a 16-fold degeneracy of the spectrum. Following
Wilson's proposal\,\cite{wilson75}, the degeneracy can be lifted by
adding a counterterm (i.e. the Wilson term) to the action which has
the effect of pushing the masses of the unwanted doublers to the
cut-off scale.  However, the Wilson term breaks chiral symmetry
explicitly, which has wide-ranging consequences as we shall see below.

Another method for the removal of doubler states is the use of
``staggered" or Kogut-Susskind fermions\,\cite{staggered}. In this
approach, the individual spin components of a Dirac spinor are spread
over several lattice sites within a hypercube. Thereby the number of
degenerate states is reduced from~16 to~4, which are then interpreted
as different flavours. It should be emphasised, however, that the
spin/flavour assignment is only valid in the continuum limit. This is
different for Wilson fermions, where spin/flavour assignments on the
lattice are exactly as in the continuum. This makes the identification
of local operators much easier, and therefore the Wilson formulation
appears to be better suited for the study of weak matrix elements of
local composite fields. However, staggered fermions leave a $\rm
U(1)\times U(1)$ subgroup of chiral symmetry invariant, and therefore
the use of staggered fermions may, after all, be advantageous for
studying quantities such as the $B$~parameter $B_K$, whose chiral
behaviour is a central issue. As most results for heavy quark physics
have been obtained using Wilson quarks, we shall concentrate on this
approach in the remainder of this review.

The Wilson action for lattice QCD reads
\be   \label{GLwilsonqcd}
   S[U,\overline{\psi},\psi] = S_G[U]+S_F^W[U,\overline{\psi},\psi],
\ee
where
\bea
   S_F^W[U,\overline{\psi},\psi] &=& a^4\sum_x\bigg\{
  -\kappa\sum_{\mu=1}^4 \frac{1}{a}\Big[ 
     \overline{\psi}(x)(r-\gamma_\mu)U_\mu(x)\psi(x+\hat{\mu}) \nonumber\\
    & &+\overline{\psi}(x+\hat{\mu})(r+\gamma_\mu)U_\mu^\dagger(x)\psi(x)
    \Big] + \overline{\psi}(x)\psi(x)
  \bigg\}, \label{GLwilsonterm} \\
  & \equiv & a^4\sum_{x,y}\overline{\psi}(x)\,{\cal M}(x,y)\,\psi(y),
\eea
and in the last line we have introduced the Wilson-Dirac operator
${\cal M}$. The Wilson term is the piece proportional to~$r$ in the above
expression. The hopping parameter $\kappa$ is related to the bare mass
$m_0$ via
\be \kappa=\frac{1}{2am_0+8r},
\ee
and the Wilson parameter $r$ is usually set to one. The Wilson action
is thus conveniently parametrised in terms of bare parameters
$(\beta,\kappa)$ instead of the bare gauge coupling and quark mass
$(g_0,m_0)$. The parametrisation of the fermionic part of the action
in eq.\,(\ref{GLwilsonterm}) implies that the quark and antiquark
fields have undergone a rescaling according to
\be  \label{GL2kappa}
   \psi(x)\rightarrow\sqrt{2\kappa}\,\psi(x),\qquad
   \overline{\psi}(x)\rightarrow\overline{\psi}(x)\,\sqrt{2\kappa}.
\ee
This normalisation is referred to as the relativistic norm of quark
fields. 

The addition of the Wilson term proportional to~$r$ to the na\"\i{}ve
lattice action leads to an additive renormalisation of the quark
mass. This implies that there exists a critical value of the hopping
parameter, $\kcrit$, at which the quark mass vanishes and chiral
symmetry is restored. The subtracted quark mass is defined by
\be \label{GLam}
   am = \frac{1}{2}\bigg(\frac{1}{\kappa}-\frac{1}{\kcrit}\bigg),
\ee
and in the free theory the critical value of $\kappa$ occurs at 
\be \label{GLkcrit_free}
   \kcrit =\frac{1}{8},\qquad g_0=0.  
\ee 

The r\^ole of $a^{-1}$ as an UV cut-off can also be seen from the fact
that at small coupling \be g_0^2 \sim 1/\ln\,a,\qquad \beta=6/g_0^2,
\ee and therefore the continuum limit, $a\rightarrow0$, is formally
reached as $\beta\rightarrow\infty$.  Note that typical values of the
lattice spacing in physical units are $a^{-1}\simeq1.5$--$4.0\,\gev$ in
current simulations, which corresponds to $\beta\simeq5.7$--6.4.

By means of the lattice discretisation procedure, one has suitably
altered the UV (short-distance) behaviour in order to obtain a
non-perturbatively regularised theory, whilst preserving the
long-distance physics. The continuum result for, say, hadron masses
computed on the lattice is then obtained in the limit $a\rightarrow0$.
At this point it is worth noting that the Wilson fermion action
differs from the classical continuum action by terms of order~$a$. In
contrast to this, the leading discretisation errors (lattice
artefacts) for the pure gauge action are only $O(a^2)$, and there are
arguments that this is also true for staggered
fermions\,\cite{staggered_oa2}. The predictive power of lattice QCD
depends crucially on the degree to which lattice artefacts can be
controlled. This is of particular importance for heavy quarks, where
the effects due to finite lattice spacing are large, as we shall see
below.

The lattice action $S[U,\overline\psi,\psi]$  in
eq.\,(\ref{GLwilsonqcd}) can now be used to define a partition
function
\be
   Z=\int D[U]D[\overline\psi]D[\psi]\,e^{-S[U,\overline\psi,\psi]},
\ee
where, on a finite lattice
\be
   \int D[U] = \prod_{x,\mu}\int\,dU_\mu(x),
\ee
and $dU_\mu(x)$ is the invariant group measure. The vacuum expectation
value (VEV) of an observable $\cal O$ is defined as
\bea
   \langle{\cal O}\rangle & \equiv & 
   \frac{1}{Z}\int D[U]D[\overline\psi]D[\psi]\,{\cal O}\,
    e^{-S_G[U]-S_F^W[U,\overline\psi,\psi]} \nonumber\\
    & = & \frac{1}{Z}\int D[U]\,{\cal O}\,\det{\cal M}[U]\,e^{-S_G[U]},
\eea
where in the last equation we have introduced the determinant of the
Wilson-Dirac operator by integrating out the fermion fields. The aim
of any Monte Carlo simulation is to evaluate $\langle\cal O\rangle$
stochastically through an average $\overline{{\cal O}}$ of individual
measurements ${\cal O}\{U_i\}$, computed on a set of gauge configurations
$\{U_i\},\,i=1,\ldots,N_c$
\be
   \langle{\cal O}\rangle \simeq \overline{{\cal O}}
      = \frac{1}{N_c}\sum_i^{N_c}\,{\cal O}\{U_i\},
\ee
where a sequence of $N_c$ gauge configuration has been generated in a
Markov process using $\det{\cal M}[U]\,e^{-S_G[U]}$ as probability
measure.  This implies that Monte Carlo results have a statistical
error that decreases proportionally to $1/\sqrt{N_c}$. In the limit of
infinite statistics, the relation between $\langle\cal O\rangle$ and
$\overline{{\cal O}}$ becomes exact.

If the determinant $\det{\cal M}[U]$ is to be included in the
probability measure, then, roughly speaking, it needs to be evaluated
in every update step in the Markov chain. Since $\det{\cal M}[U]$ is
highly non-local this is prohibitively costly on present computers.
Various algorithms have therefore been constructed which should allow
for a more efficient evaluation\,\cite{HMC,kramers,lusch_bos}, but
still the computational effort is very large.

It has therefore been proposed to set $\det{\cal M}[U]\equiv1$ in the
generation of gauge configurations, which corresponds to neglecting
effects due to quark loops. This procedure defines the so-called
{\it quenched approximation\/}\,\cite{parisi81,weing82}, which is still
most widely used in current simulations. Consequently, the
overwhelming part of the material covered in this review was obtained
using the quenched approximation. As far as hadronic quantities are
concerned, the quenched approximation amounts to computing observables
in QCD in the presence of a Yang-Mills background field.

We end this subsection with a few more technical remarks. Hadronic
observables are extracted from mesonic or baryonic two- or three-point
functions, which are constructed from quark propagators $S_q(x,y)$.
The quark propagators themselves are obtained as the inverse of the
Wilson-Dirac operator~${\cal M}$ by solving
\be  \label{GLquark_prop}
   {\cal M}(x,y)\,S_q(y,z) = \delta_{xz}.
\ee
Hence, on every gauge configuration $\{U_i\},\,i=1,\ldots,N_c$, one
needs to invert the matrix that couples the quark fields in the
lattice action.  Therefore, the use of an efficient inversion
algorithm is of great importance. The most widely used inversion
algorithms\,\cite{inv_alg} include the Conjugate Gradient, Minimal
Residual and, more recently, the Stabilised Biconjugate Gradient
method\,\cite{bicgstab}.

As an example of a mesonic two-point function constructed from quark
propagators, we now discuss the euclidean correlation function of the
axial current $A_\mu=\overline{\psi}\gamma_\mu\gamma_5\psi$, which is
used to extract properties of pseudoscalar mesons. For euclidean times
$t\equiv x_4>0$, the correlation function is defined by
\be  \label{GLCtp}
   C(t;\vec{p}) \equiv \sum_{\vec{x}}\,e^{-i\vec{p}\cdot\vec{x}}
   \langle0|A_4(\vec{x},t)A_4^\dagger(0)|0\rangle,
\ee
which, after performing the Wick contractions, can be written in terms
of quark propagators $S_q(y,x)$ as
\be
   C(t;\vec{p}) \equiv -\sum_{\vec{x}}\,e^{-i\vec{p}\cdot\vec{x}}
   \big\langle{\rm Tr}\left\{\gamma_4 S_q(0,x)\gamma_4 S_q^\dagger(0,x)
   \right\}\big\rangle.
\ee
Here we have used that $S_q(x,0)=\gamma_5 S_q^\dagger(0,x)\gamma_5$.
Inserting a complete set of intermediate states
in eq.\,(\ref{GLCtp}), one obtains the spectral decomposition of the
correlation function, viz.
\be
   C(t;\vec{p}) = \sum_{n}\,
   \frac{|\langle0|A_4(0)|n;\vec{p}\,\rangle|^2}{2E_n(\vec{p})}\,
   e^{-E_n(\vec{p})t}.
\ee
For large $t$, the lightest state in the spectral decomposition
dominates. For $\vec{p}=0$, the asymptotic form of $C(t;\vec{p})$ is
given by
\be
   C(t;\vec{0}) \stackrel{t\gg0}{\sim} 
   \frac{|\langle0|A_4(0)|P\rangle|^2}{2M_P}\,
   e^{-M_P t},
\ee
where $M_P$ is the mass of the pseudoscalar meson in the ground state.

In order to obtain reliable estimates of the mass $M_P$ and the matrix
element $\langle0|A_4(0)|P\rangle$ in a lattice calculation, one has
to be able to follow the signal for $C(t;\vec{0})$ up to large~$t$,
which requires good statistics. Furthermore, the lattice has to be
long enough for the lightest state to be observable, and therefore one
normally chooses $T=2L$. For heavy quark systems, the issue of
reaching the asymptotic behaviour of $C(t;\vec{0})$ is a particular
problem; here, the correlation function falls off rapidly and may have
dived into the statistical noise before the ground state can be
observed. Therefore, one has to choose operators in the evaluation of
$C(t;\vec{0})$ which have a good overlap onto the ground state. Using
point-like, local operators such as
$\overline{\psi}(x)\gamma_\mu\gamma_5\psi(x)$ above, results in a very
poor signal.  This can be intuitively understood from the fact that
the wave function of the particle one wants to study is an extended
object and not point-like. Therefore, when solving for the quark
propagators, eq.\,(\ref{GLquark_prop}), one applies so-called {\it
smearing techniques\/}, which are designed to generate more extended
sources, having a better overlap onto the desired
state\,\cite{CUB}${}^{-}$\cite{FNAL,INV,PCW_stat_92,PCW_91,FNAL_94}.
Since the singal-to-noise ratio for hadron masses and decay constants
can be enormously improved using ``smeared" sources and/or sinks in
the propagator calculation, smearing techniques are now an
indispensable tool in lattice simulations of QCD.

In a simulation, dimensionful quantities like hadron masses or
pseudoscalar decay constants are obtained in lattice units. In order
to convert these lattice estimates into physical units, one needs to
set the lattice scale. This is usually done by comparing a low-energy
hadronic quantity computed on the lattice to its physical value. For
instance, using the mass of the $\rho$ meson one obtains
$a^{-1}\,[\gev]$ via
\be
   a^{-1}\,[\gev] = \frac{M_\rho\,[\gev]}{(aM_\rho)},
\ee
where $aM_\rho$ denotes the lattice result for the $\rho$ mass. Other
quantities used to set the scale include the pion decay constant
$f_\pi$, the nucleon mass, the string tension $\sqrt{\sigma}$ or the
hadronic scale $r_0$ discussed by Sommer\,\cite{sommer_r0_93}.

One complication that arises in the computation of $aM_\rho$ (and,
indeed, of all quantities that involve $u$ and $d$ quarks), is that
one cannot simulate directly at the $u$ and $d$ quark masses. This
comes from the fact that hadronic quantities involving very light
quarks have large correlation lengths, which would lead to severe
finite-size effects on currently accessible lattice sizes. Therefore,
one typically uses quark masses in the region of that of the strange
quark and extrapolates the results to $m_{u,d}$, or to the chiral
limit. Finite-size effects are discussed in more detail in
subsection\,2.4.

\subsection{The $b$ quark on the lattice} \label{SEKbquark}
As already mentioned in subsection\,2.1, current
simulations are carried out at typical values of the inverse lattice
spacing in physical units of $a^{-1}\simeq1.5$--4$\,\gev$. For heavy
quarks, this implies that one needs to worry about the effects of
finite cut-off already when one wants to study charm physics, since
$m_{\rm charm}$ is only about a factor of two smaller than $a^{-1}$.
Furthermore, it is evident that the $b$ quark cannot be studied
directly, since its Compton wavelength is smaller than the lattice
spacing. Hence, one concludes that lattice artefacts are large for
heavy quarks, and that special care is required to control them.

In the following we describe several methods which are employed to
circumvent the problem that the $b$ quark cannot be studied directly
or are designed to alleviate the problem of large discretisation
errors. 
%

\subsubsection{$O(a)$ improvement}
In this approach one seeks to reduce lattice artefacts by using
so-called {\it improved\/} actions and operators in order to cancel
the leading discretisation error in on-shell quantities. The concept
of perturbative improvement was first outlined by
Symanzik\,\cite{symanzik_impr} and developed further in
refs.\,\cite{LueWei_85}${}^{-}$\cite{heatlie_91}. For the Wilson
action in lattice QCD, Sheikholeslami and Wohlert have shown that the
$O(a)$ discretisation error in the action can be cancelled by adding a
local counterterm\,\cite{SW_85}
\be  \label{GLSW}
   S_{SW}[U,\overline\psi,\psi]=S_G[U]+S_F^W[U,\overline{\psi},\psi]
    + \csw\frac{i}{4}a^5\sum_{x,\mu,\nu}\overline\psi(x)\sigma_{\mu\nu}
      F_{\mu\nu}(x)\psi(x),
\ee
where the improvement coefficient $\csw$ depends on the gauge
coupling. In order to obtain $O(a)$ improved matrix elements of
composite operators, these composite fields have to be improved,
too\,\cite{heatlie_91,alphaI}. For instance, the improved axial
current reads
\be  \label{GLAimp}
   {A}^I_\mu(x) = A_\mu(x) + c_A a\partial_\mu P(x),
\ee
where $A_\mu(x)=\overline\psi(x)\gamma_\mu\gamma_5\psi(x)$,
$P(x)=\overline\psi(x)\gamma_5\psi(x)$, and $c_A$ denotes another
improvement coefficient. Recently, the ALPHA
Collaboration\,\cite{alphaI,alphaII,alphaIII} has proposed an $O(a)$
improved action for which the improvement coefficients $\csw$ and
$c_A$ were determined non-perturbatively for $\beta\ge6.0$. This
action has so far not been used in simulations of heavy quark systems. 

\subsubsection{Non-relativistic normalisation}
In a second approach, it has been
suggested\,\cite{ask_lat92}${}^{-}$\cite{KKM_96} to suitably
adapt the Wilson action, such that the Wilson propagator does not
deviate from the continuum behaviour even for quark masses
$am\,\gtaeq\,1$ in lattice units (i.e. for quark masses above the
cut-off). It has been argued that this can be achieved by a modified
rescaling of lattice quark fields (c.f. eq.\,(\ref{GL2kappa}))
according to
\be  \label{GLKLM}
   \psi(x)\rightarrow \sqrt{2\kappa}\,e^{am_{\rm P}/2}\psi(x),
\ee
where the ``pole mass'' $am_{\rm P}$ of the Wilson propagator is given by
\be
   am_{\rm P} = \ln(1+am),
\label{GLamP}
\ee
and $am$ is defined in eq.\,(\ref{GLam}). The factor
$\sqrt{2\kappa}\,e^{am_{\rm P}/2}$ is designed to interpolate smoothly
between the relativistic and non-relativistic regimes.

Consequently, in order to cancel the effects of large quark masses in
studies of matrix elements of composite operators, the normalisation
of quark fields is changed according to the above scale factor. This
non-relativistic normalisation of quark fields according to
eq.\,(\ref{GLKLM}) is referred to as the
Kronfeld-Lepage-Mackenzie~(KLM) norm. 

Higher-order corrections for large quark masses can also be
considered. As is argued in\,\cite{lepage_lat91,ask_lat92,mack_lat92},
this can be motivated by the observation that for quark masses $am>1$,
the mass $am_{\rm P}$ and the ``kinetic mass" $am_2$ appearing in the
non-relativistic dispersion relation, are no longer equal
\be
   aE(\vec{p}) = am_{\rm P} + \frac{{\vec{p}}^{\,2}}{2am_2} + \ldots,\qquad 
   m_2\not=m_{\rm P},
\ee
and the relevant quark mass is the kinetic mass $m_2$.  In the free
theory, the kinetic mass $am_2$ is obtained from the free Wilson
propagator in the non-relativistic limit
\be
    am_2=\frac{e^{am_{\rm P}}\sinh(am_{\rm P})}{\sinh(am_{\rm P})+1}.
\label{GLm2}
\ee

Both $O(a)$ improvement and the KLM norm are used for quark masses in
the region of that of the charm quark. Clearly, residual lattice
artefacts remain and must ultimately be extrapolated away (which,
however, can be performed much more reliably if they are small). The
results obtained around $m_{\rm charm}$ must also be extrapolated to
the mass of the $b$ quark, and clearly one needs to control this
extrapolation in order to obtain meaningful results. In what follows,
we shall refer to this approach, where relativistic heavy quarks are
used in conjunction with a prescription to reduce lattice artefacts
(either $O(a)$ improvement or the KLM-norm), as the ``conventional
method".
%

\subsubsection{Static approximation}
Here the $b$ quark is treated as infinitely
heavy\,\cite{eichten_81}. From an expansion of the heavy quark
propagagtor in the inverse heavy quark mass, $1/m_Q$, one obtains at
leading order\,\cite{eichten_lat87}
\be                \label{GLsb0}
S_Q(\vec x,t;\vec 0,
0)=\left\{\Theta(t)\,e^{-m_Qt}\frac{1+\gamma_4}{2} +
\Theta(-t)\,e^{m_Qt}\frac{1-\gamma_4}{2}\right\}\,
\delta(\vec x){\cal P}_{\vec 0}(t,0),
\ee
where ${\cal P}_{\vec 0}(t,0)$ is the product of links from $(\vec 0,t)$
to the origin, for example for $t>0$,
\be                \label{GLcalp}
{\cal P}_{\vec 0}(t,0) = U_4^\dagger(\vec 0,t-1)
U_4^\dagger(\vec 0, t - 2) \cdots U_4^\dagger(\vec 0, 0).
\ee
This form of the propagator is also obtained using a Lagrangian for heavy
quarks $Q(x)$ which simply reads\,\cite{eichten_hill_90_1}
\be  \label{GLstatic}
   {\cal L}^{\rm static}={Q}^\dagger(x){D}_4Q(x),
\ee
where the covariant derivative is defined in eq.\,(\ref{GLcov_dev}),
and $Q(x)$ is a two-component spinor describing the heavy quark.

One expects corrections of order $\Lambda_{\rm QCD}/m_Q$ to the
results in the static approximation, which can potentially be large.
With a few exceptions, such as the $B^*$--$B$ mass splitting, the
computation of higher order corrections in $1/m_Q$ to the static limit
is complicated due to the presence of power
divergencies\,\cite{power}.  Nevertheless, the static approximation is
a valuable tool in lattice studies of heavy quark systems. It plays
the crucial r\^ole of guiding the extrapolation of results obtained
with the conventional method to the mass of the $b$~quark by providing
direct information at infinite quark mass.
%

\subsubsection{Non-relativistic QCD (NRQCD)}
A non-relativistic formulation of heavy quark
systems\,\cite{lepthack_91} can be obtained by excluding relativistic
momenta through the introduction of a finite cut-off
$\Lambda_{\rm UV}\,\lesssim\, m_Q$ such that
\be
   p\sim m_Q v \ll m_Q,
\ee
where $v$ is the 4-velocity of the heavy quark. The loss of
relativistic states through this requirement can be compensated for by
adding new local interactions order by order in $p/\Lambda_{\rm UV}$.
In this way one obtains a cut-off theory as an expansion of the
original action in $p/\Lambda_{\rm UV}\sim v$. The non-relativistic
QCD Lagrangian is then obtained by applying a Foldy-Wouthuysen
transformation, which separates quark and antiquark fields, thereby
generating an expansion of the QCD Lagrangian in $1/m_Q$. For
instance, at order $1/m_Q$ the NRQCD Lagrangian reads
\be
   {\cal L}^{\rm NRQCD} = Q^\dagger
       \bigg(D_4-\frac{\vec{D}^2}{2m_Q}\bigg)Q
      -Q^\dagger\frac{\vec{\sigma}\cdot\vec{B}}{2m_Q}Q.
\ee
This approach is thus based on the {\it a priori\/} discretisation of
the non-relativistic formulation of the Wilson action.  Although the
above expression contains the term that defines ${\cal L}^{\rm
static}$ in eq.\,(\ref{GLstatic}), one should realise that there are
important differences between NRQCD and the static approximation: the
theory defined by the NRQCD Lagrangian is non-renormalisable, since a
finite cut-off has to be kept
\be
   \Lambda_{\rm UV} \sim a^{-1} \,\lesssim\, m_Q.
\ee
In the language of the lattice this means that the formal continuum
limit, $a\rightarrow0$, does not exist, in contrast to the case of the
static approximation. Hence, for NRQCD to work in a lattice
simulation, one needs to calculate at fairly large values of~$a$ (i.e.
at small $\beta$). Lattice artefacts (i.e. cut-off effects) have to be
reduced by including higher orders in $1/m_Q$.  For the study of
matrix elements of composite fields using NRQCD to $n$th order in
$1/m_Q$, it is important to take into account corrections of
$O(1/m_Q^n)$ in the operators as well.

From the above discussion of various methods to formulate heavy quarks
on the lattice, it is obvious that none of them is entirely
satisfactory. They are subject to rather different systematic effects
and thus provide complementary information on heavy quark systems. The
full picture will therefore only emerge when results from all methods
are compared. 

\subsection{Weak matrix elements on the lattice and their continuum
  counterparts} \label{SEKlatt_cont}

It has already been noted in subsection\,2.1 that the Wilson
action in eq.\,(\ref{GLwilsonterm}) breaks chiral symmetry
explicitly. Therefore, even in the massless theory, vector and axial
vector currents are not conserved, since the UV behaviour of the
theory has been changed as a consequence of the regularisation
procedure. In other words, the regularisation procedure conflicts with
the na\"\i{}ve conservation of the currents. The resulting
short-distance corrections between the vector and axial currents in
lattice and continuum theories can be absorbed into normalisation
factors $Z_V$ and $Z_A$, respectively.

The chiral Ward identities for the axial current normally ensure that
the axial current does not get renormalised. However, since chiral
symmetry is broken explicitly by the Wilson term, the Ward identity no
longer applies, as it is violated by terms of order~$a$.
Nevertheless, the condition that the correctly normalised lattice
axial current satisfies the Ward identities can be used to derive its
normalisation\,\cite{boch_etal_85,maimar_86}. Since the normalisation
factors incorporate short-distance effects, they can in principle be
calculated in lattice perturbation theory.

Using the unimproved Wilson action the one-loop perturbative results
for the vector and axial current normalisation
constants read\,\cite{MeySmi_83,MarChe_83,GHS_84}
\be \label{GLZwilson}
  Z_A=1-0.133\,g_0^2 + O(g_0^4),\qquad Z_V=1-0.174\,g_0^2 + O(g_0^4),
\ee
whereas for the $O(a)$ improved Sheikholeslami-Wohlert
action\,\footnote{Note that here we quote the expressions for the
  non-local currents which have been ``rotated" according to the
  prescription given in\,\cite{heatlie_91,BPFG_93}.}, eq.\,(\ref{GLSW})
one obtains\,\cite{GMPHS_91,BPFG_93}
\be
  Z_A=1-0.018\,g_0^2 + O(g_0^4),\qquad Z_V=1-0.100\,g_0^2 + O(g_0^4).
\ee
These expressions are written in terms of the bare gauge coupling
$g_0^2$, which is given by $g_0^2=6/\beta$. However, it has been
argued in\,\cite{lepenzie_92} that the bare gauge coupling is a bad
expansion parameter due to the appearance of large gluonic tadpole
contributions in the relation between the link variable $U_\mu(x)$ and
the continuum gauge field $A_\mu(x)$ (see eq.\,(\ref{GLlink})). The
poor convergence property of lattice perturbation theory can be
improved by absorbing these tadpole contributions into suitable
redefinitions of the bare parameters of the theory. The simplest
definition of an improved expansion parameter is the so-called
``boosted" coupling $\widetilde{g}^2$ introduced by
Parisi\,\cite{parisi80}
\be
   \widetilde{g}^2\equiv\frac{g_0^2}{u_0^4},
\label{GLParisi}
\ee
where $u_0^4$ is taken to be the measured average value of the
plaquette $\langle\frac{1}{3}{\rm Re\,Tr}P\rangle$
($u_0=\langle\frac{1}{3}{\rm Re\,Tr}P\rangle^{1/4}$ is a gauge
invariant estimate of the average link). Replacing the bare
coupling $g_0^2$ by $\widetilde{g}^2$ in the one-loop expressions for
$Z_A$ and $Z_V$ defines the so-called ``boosted'' perturbation
theory. 

In the fermionic sector, one can absorb tadpole contributions into a
redefinition of the hopping parameter $\kappa$ in a similar fashion
according to
\be  \label{GLkappatilde}
   \widetilde{\kappa}\equiv\kappa u_0.
\ee
It is then expected that the critical value of $\widetilde{\kappa}$ is
much closer to its tree-level value of 1/8, such that
$u_0\simeq1/8\kcrit$, c.f. eq.\,(\ref{GLkcrit_free}).  This is the basic
concept of {\it tadpole\/} or {\it mean field\/}
improvement\,\cite{lepenzie_92}. By using non-perturbative input for
$u_0$ such as the measured link or, alternatively, $\kcrit$, one can
improve the UV behaviour of the lattice theory, which should manifest
itself in a better convergence of lattice perturbation theory.

Tadpole improved expressions for the normalisation factors are
obtained by combining their perturbation expansions with those for
$u_0$ as well as the measured value of the
latter\,\cite{lepenzie_92}. For instance, the tadpole improved one-loop
expression for the axial current normalisation reads
\be
   \widetilde{Z}_A = u_0\Big(1+[Z_A^{(1)}-u_0^{(1)}]\widetilde{g}^2\Big),
\label{GLZAtad}
\ee
where $Z_A^{(1)}$ and $u_0^{(1)}$ are the one-loop coefficients in the
expansions of $Z_A$ and $u_0$, respectively. The numerical value of
$u_0^{(1)}$ depends on whether one identifies $u_0$ with the average
link or the critical hopping parameter, thus considering the perturbative
expansions of either $\langle\frac{1}{3}{\rm Re\,Tr}P\rangle^{1/4}$ or
$8\kcrit$. 

In conjunction with the KLM-norm defined in eq.\,(\ref{GLKLM}),
tadpole improvement also plays an important r\^ole in reducing
lattice artefacts due to large quark masses. As an example we consider
the case of the lattice axial current
$\overline{\psi}_1\gamma_\mu\gamma_5\psi_2$ of quark fields
$\psi_1,\,\psi_2$ with hopping parameters $\kappa_1$ and $\kappa_2$,
respectively. The normalisation factor of the axial current is then
supplemented by an extra factor of
\be
\exp\{a(m_{\rm P,1}+m_{\rm P,2})/2\}, 
\ee
which arises from the modified rescaling of quark fields in
eq.\,(\ref{GLKLM}). Inserting the definition of the pole mass and
applying tadpole improvement, the normalisation factor becomes
\begin{eqnarray}
e^{a(\widetilde{m}_{\rm P,1}+\widetilde{m}_{\rm P,2})/2} \widetilde{Z}_A
 & = &
 \sqrt{\Big(1+\frac{1}{2\widetilde{\kappa}_1}
             -\frac{1}{2\widetilde{\kappa}_{\rm crit}}\Big)
       \Big(1+\frac{1}{2\widetilde{\kappa}_2}
             -\frac{1}{2\widetilde{\kappa}_{\rm crit}}\Big)}
	\widetilde{Z}_A  \nonumber\\
 & = &  \sqrt{\Big(\frac{1}{2\widetilde{\kappa}_1}-3\Big)
              \Big(\frac{1}{2\widetilde{\kappa}_2}-3\Big)}
	\widetilde{Z}_A,
\label{GLZatilde}
\end{eqnarray}
where in the last step we have used $\widetilde{\kappa}_{\rm
crit}=\kcrit u_0\simeq1/8$, a relation that becomes exact if $u_0$ is
defined by $1/8\kcrit$ rather than the average link.

Apart from the coupling constant defined in eq.\,(\ref{GLParisi})
other definitions of a mean field improved coupling can be
used. Lepage and Mackenzie\,\cite{lepenzie_92} proposed a coupling
$\alpha_V$ defined through
\be
\alpha_V(3.41a^{-1})\Big(1-(1.19+0.017n_f)\alpha_V\Big)
= -\frac{3}{4\pi}\ln\langle\frac{1}{3}{\rm Re\,Tr}P\rangle.
\label{GLalphaV}
\ee
If $\alpha_V$ is to be used for the evaluation of a tadpole improved
normalisation factor, then, according to ref.\,\cite{lepenzie_92}, it
has to be computed at a scale $q^*$, which denotes the mean momentum
flow relevant for a given matrix element. Typical values of $q^*$ lie
in the range $1\leq aq^*\leq\pi$.

Given the potential uncertainty in the perturbative expressions from
higher orders, it is desirable to perform non-perturbative
determinations of the normalisation factors. For $Z_A$ and $Z_V$ this
can be done by imposing the chiral Ward identities as a normalisation
condition\,\cite{maimar_86,MPSV_93,PPTV_94,ZA_62}.  A more general
approach, which can be applied to a large class of
operators\,\cite{MPSTV_95}, imposes the normalisation condition
between quark states. Numerical values have so far been obtained for
the axial current for $\csw=0$\,\cite{QCDSF_lat95}, in the static
approximation for $\csw=1$\,\cite{mauro_lat95}, and also for the
$\Delta S=2$ four-fermion operator relevant for the Kaon
$B$\,parameter $B_K$\,\cite{mauro_lat95,Z_BK}. Systematic errors in
non-perturbative determinations of normalisation factors are, of
course, present and need to be controlled. Recently, the ALPHA
Collaboration\,\cite{alphaIV} has determined $Z_A$ and $Z_V$
non-perturbatively in the whole range $0\leq\ g_0^2\leq1$, with total
errors at the 1\% level.

Apart from the problem of the normalisation of lattice operators,
another consequence of explicit chiral symmetry breaking is the
possibility of mixing of operators with a definite chirality. For
instance, the $\Delta F=2$ four-fermion operator defined by
\be
   \widehat{O}_L \equiv 
	\big(\overline{\psi}\gamma_\mu(1-\gamma_5)\psi\big)\;
        \big(\overline{\psi}\gamma_\mu(1-\gamma_5)\psi\big)
\ee
can mix with its right-handed counterpart
\be
   \widehat{O}_R \equiv 
	\big(\overline{\psi}\gamma_\mu(1+\gamma_5)\psi\big)\;
        \big(\overline{\psi}\gamma_\mu(1+\gamma_5)\psi\big).
\ee
Furthermore, operators can mix with higher dimension operators, which
is also a consequence of the breaking of Lorentz invariance by
formulating QCD on a hypercubic lattice.

Hence, in general the relation between the matrix element of an
operator $\widehat{O}$ in the continuum and in the Wilson formulation
is given by
\be
   \langle f|\widehat{O}|i\rangle^{\rm cont} = \sum_\alpha
    Z_\alpha\langle f|\widehat{O}_\alpha^{\rm latt}|i\rangle
   +O(a),
\ee
where $\alpha$ labels the operators of the same dimension that can mix
with each other, and $Z_\alpha$ are the short-distance normalisation
factors. 

\subsection{Summary of systematic errors} \label{SEKsyst_err}
The main systematic effects that affect lattice results for heavy
quark physics can be broadly divided into lattice artefacts,
finite-size effects, quenching errors and normalisation of lattice
operators.

In any simulation in physical volumes, the following inequalities must
be satisfied
\be 
   a\ll M^{-1}\ll L,
\ee
where $M$ is a typical hadronic mass, and $L$ denotes the spatial
length of the lattice. The left part of the inequality constrains the
possible values of heavy quark masses that can be studied safely,
i.e. without suffering from large discretisation errors. On the other
hand, very light quark masses will give rise to finite-size effects,
as is evident from the right part of the above inequality. Hence, a
major limitation of current lattice simulations is that very different
scales need to be incorporated on a single lattice.

We now list and discuss the main systematic errors which affect
lattice results for heavy quark systems:
%

\subsubsection{Discretisation errors}
  Discretisation errors, i.e. the effects of the finiteness of the
  lattice spacing, are present in all simulations of lattice QCD, but
  are especially important for heavy quark systems as we have seen in
  subsection\,2.2. We have already discussed how $O(a)$
  improvement can be used to reduce these effects, and there are
  attempts to extend the improvement programme beyond
  $O(a)$\,\cite{D234,AKLeP96} or to even construct ``perfect" lattice actions
  that ideally are completely free of discretisation
  errors\,\cite{perf_sigma,perf_YM,perf_ferm}. Once these cut-off
  effects are reduced, either by employing the Symanzik improvement
  programme or by absorbing effects of large quark masses into the
  normalisation of quark fields (KLM-norm), residual discretisation
  errors have to be extrapolated away by taking the limit
  $a\rightarrow0$.

  It should be emphasised that this procedure does {\it not\/} solve
  the problem that the $b$ quark cannot be studied directly in current
  simulations using relativistic quarks.

\subsubsection{Finite-size effects}
  Lattice estimates of hadron masses and matrix elements are distorted
  due to the finiteness of the lattice volume. An analytic calculation
  by L\"uscher\,\cite{lusch_86} showed that the mass shift in finite
  volume decreases exponentially with $L$
\be  \label{GLlusch_finvol}
   \delta M(L) \equiv M(L)-M(\infty) \sim e^{-L/L_0}.
\ee
Numerical studies, however, revealed that for spatial lengths of
$L\,\lesssim\,2.0$\,fm the observed mass shift deviates from the
exponential decrease and is very well described by a power law
behaviour. Using a simple model, the effect can be ascribed to the
``sqeezing" of hadronic wave functions for small
volumes\,\cite{fuku_92,aoki_lat93}, and for values
of~$L\,\gtaeq\,2$\,fm, one recovers the exponential behaviour of
eq.\,(\ref{GLlusch_finvol}).

In practice one obtains estimates of hadronic observables in infinite
volume by repeating the simulations at different values of $L$ and
extrapolating to the infinite volume limit.

Finite-size effects also prevent light quark propagators from being
calculated at realistic values of the masses of~$u$ and~$d$ quarks,
whose bound states have large correlation lengths. Therefore, light
quark masses are usually taken around the mass of the strange quark,
and hadronic observables are then extrapolated to the chiral limit,
using the quark mass dependence deduced from chiral perturbation
theory as a guide for the extrapolation.

\subsubsection{Quenching}
  As most lattice results for heavy quark systems are still obtained
  in the quenched approximation, attempts must be made to quantify the
  effects due to neglecting quark loops, in order to provide
  meaningful information for phenomenology. In the past few years,
  several collaborations have reported results for heavy quark
  physics using dynamical quarks, but much more precise simulations
  have yet to be performed.

  An indirect manifestation of quenching effects is the uncertainty in
  the lattice scale $a^{-1}\,[\gev]$. This uncertainty arises because
  different quantities used to set the scale give different results,
  which is understood from the fact that quark loops make different
  contributions to the various quantities used to compute
  $a^{-1}\,[\gev]$.

\subsubsection{Normalisation and mixing of lattice operators}
  Uncertainties in the normalisation factors relating lattice matrix
  elements to their continuum counterparts come from the fact that
  these factors are, in most cases, only known perturbatively to
  leading order in $g_0^2$. In a few cases, e.g. for the heavy-light
  axial current in the static approximation, this uncertainty may be
  of the order of 10\%. Non-perturbative estimates of these constants
  are therefore highly desirable. For vector and axial vector currents
  the chiral Ward identities can easily be employed as normalisation
  conditions in a non-perurbative determination. For more complicated
  operators or scale-dependent renormalisations, a more general
  condition needs to be applied\,\cite{MPSTV_95}, which, however, may
  be less accurate numerically.

  In order to check the consistency of different methods, one needs to
  make contact between perturbative and non-perturbative estimates of
  normalisation constants. If $g_0^2$ is used as an expansion
  parameter in the evaluation of the perturbative expressions, this is
  only possible at large values of $\beta$, where non-perturbative
  methods normally fail. Using a ``boosted'' or tadpole-improved
  coupling $\widetilde{g}^2$ instead, one may try to make a comparison
  in a region where current simulations are being performed.  
  In some cases it was found that ``boosted'' perturbation theory yields
  values close to the non-perturbative determinations, but the issue
  remains inconclusive as long as small values of the couplings cannot
  be investigated numerically.
  A systematic non-perturbative analysis for $0\leq g_0^2\leq1$ has
  been performed in the framework of the Schr\"odinger
  functional\,\cite{alphaIII,alphaIV}.

\section{Leptonic Decays of Heavy Mesons}

In this section we review lattice results for heavy-light decay
constants such as $\fd$, $\fds$, $\fb$ and $\fbs$. After introducing
the basic definitions, we present and discuss the results obtained
using different formulations of heavy quarks on the
lattice. Subsection\,3.4 contains a detailed discussion of the
approach of the continuum limit for unimproved Wilson
fermions. Readers who are not interested in the details of the
continuum extrapolation will find the main results listed in
subsection\,3.5, where also a comparison to other theoretical methods
is presented. Throughout this review we use a convention for which
$f_\pi=131\,\mev$.

\subsection{Basic definitions}

In the continuum, the pseudoscalar decay constant $f_P$ is defined by
\be
   \langle0|A_\mu(0)|P(\vec{p}\,)\rangle = ip_\mu f_P,
\ee
whereas the vector decay constant $1/f_V$ is usually defined by
\be
   \langle0|V_\mu(0)|V\rangle = \epsilon_\mu\frac{M_V^2}{f_V}.
\ee
Hence, in order to extract the decay constants of heavy-light mesons,
one needs to compute the matrix elements of the axial and vector
currents between a mesonic state and the vacuum.

On a lattice, this is achieved by analysing various mesonic two-point
functions of the form
\be
   C^{QR}_{J_1 J_2}(t) \equiv \sum_{\vec{x}}\,
   \langle0|J_1^Q(x)J_2^{\dagger R}(0)|0\rangle,
\ee
where $J_1$ and $J_2$ are interpolating operators which can annihilate
or create the heavy-light pseudoscalar or vector meson under study.
The superscripts $Q,\,R$ denote whether a local~($L$) or smeared~($S$)
interpolating operator is used.

One possibility to extract the heavy-light pseudoscalar decay constant
is to consider the ratio
\bea
    \frac{C_{AP}^{LS}(t)}{C_{PP}^{SS}(t)} & \equiv &
\frac{\sum_{\vec{x}}\,\langle0|A_4^L(\vec{x},t)P^{\dagger S}(0)|0\rangle}
     {\sum_{\vec{x}}\,\langle0|P^S(\vec{x},t)P^{\dagger S}(0)|0\rangle}
\nonumber\\
& \stackrel{t\gg0}{\sim} & 
\frac{\langle0|A_4^L(0)|P\rangle}{\langle0|P^S(0)|P\rangle}\,
  \tanh\big(M_P(T/2-t)\big),
\eea
where we have used the expressions for the asymptotic behaviour for
both $C_{AP}^{LS}(t)$ and $C_{PP}^{SS}(t)$ on a finite lattice with
time extension~$T$ and periodic boundary conditions. Using the smeared
pseudoscalar density $P^S(x)$ to create the pseudoscalar meson in
$C_{AP}$ and $ C_{PP}$ results in a much better signal for these
correlation functions. This is of particular importance in the static
approximation discussed below, where it is notoriously difficult to
obtain a reliable signal, and where calculations using purely local
operators are known to
fail\,\cite{FNAL,Wupp,boucaud_89,Hashi_saeki_92}.

In order to determine $\langle0|A_4^L(0)|P\rangle$ which contains the
decay constant, one also needs to know the pseudoscalar mass $M_P$ and
the matrix element $\langle0|P^S(0)|P\rangle$, which are extracted
from separate fits to
\be
   C_{PP}^{SS}(t) \stackrel{t\gg0}{\sim} 
   \frac{|\langle0|P^S(0)|P\rangle|^2}{2\,M_P}\,
   e^{-M_P T/2}\cosh\big(M_P(T/2-t)\big).
\ee
From $\langle0|A_4^L(0)|P\rangle$ one can then extract $f_P$ via
\be
   \langle0|A_4^L(0)|P\rangle = M_P\,f_P/Z_A,
\ee
where $Z_A$ is the normalisation constant of the axial current
discussed in subsection\,2.3, which is usually
evaluated in (``boosted'' or tadpole-improved) perturbation
theory. The actual value of $Z_A$ depends on whether an improved
action has been used (i.e. on the choice of $\csw$), on the details of
the implementation of tadpole improvement, and also on whether the
relativistic, non-relativistic or static formulation of heavy quarks
has been used.  

In the two-spinor formalism of the static approximation (see
eq.\,(\ref{GLstatic})), there is no distinction between $A_4(x)$ and
$P(x)$, such that the pseudoscalar decay constant is extracted from
combinations of the following correlation functions
\begin{eqnarray}
  C^{SS}(t) & = &
\sum_{\vec{x}}\langle0|A_4^S(\vec{x},t)A_4^{\dagger S}(0)|0\rangle 
\stackrel{t\gg0}{\sim} (A^S)^2\,e^{-{\cal E}t}, \label{GLstatcorrSS}
\\
  C^{SL}(t) & = &
\sum_{\vec{x}}\langle0|A_4^S(\vec{x},t)A_4^{\dagger L}(0)|0\rangle 
\stackrel{t\gg0}{\sim} A^S\,A^L\,e^{-{\cal E}t},
\label{GLstatcorrSL}
\end{eqnarray}
where ${\cal E}$ is the unphysical difference between the meson mass
and the mass of the heavy quark. The decay constant $\fbstat$ is
related to the matrix element $A^L$ via
\be	\label{GLstatAL}
  \fbstat=A^L\sqrt{2/M_B}\,\zastat.
\ee
The one-loop expressions for $\zastat$ were calculated in
refs.\,\cite{BouLinPene_89,eichten_hill_90_1,eichten_hill_90_2}, and
later extended to the $O(a)$ improved
case\,\cite{BorPit92,HerHil93}. Here we do not list the expressions
but refer the reader to appendix\,A, where we have listed the
normalisation factors for some operators of interest.

A similar procedure as the one described above can be applied to
determine the vector decay constant $1/f_V$. 

\subsection{Decay constants and heavy quark symmetry}
\label{SEKHQS}
\begin{figure}[tb]
\vspace{2.0cm}
\hspace{-1.4cm}
\ewxy{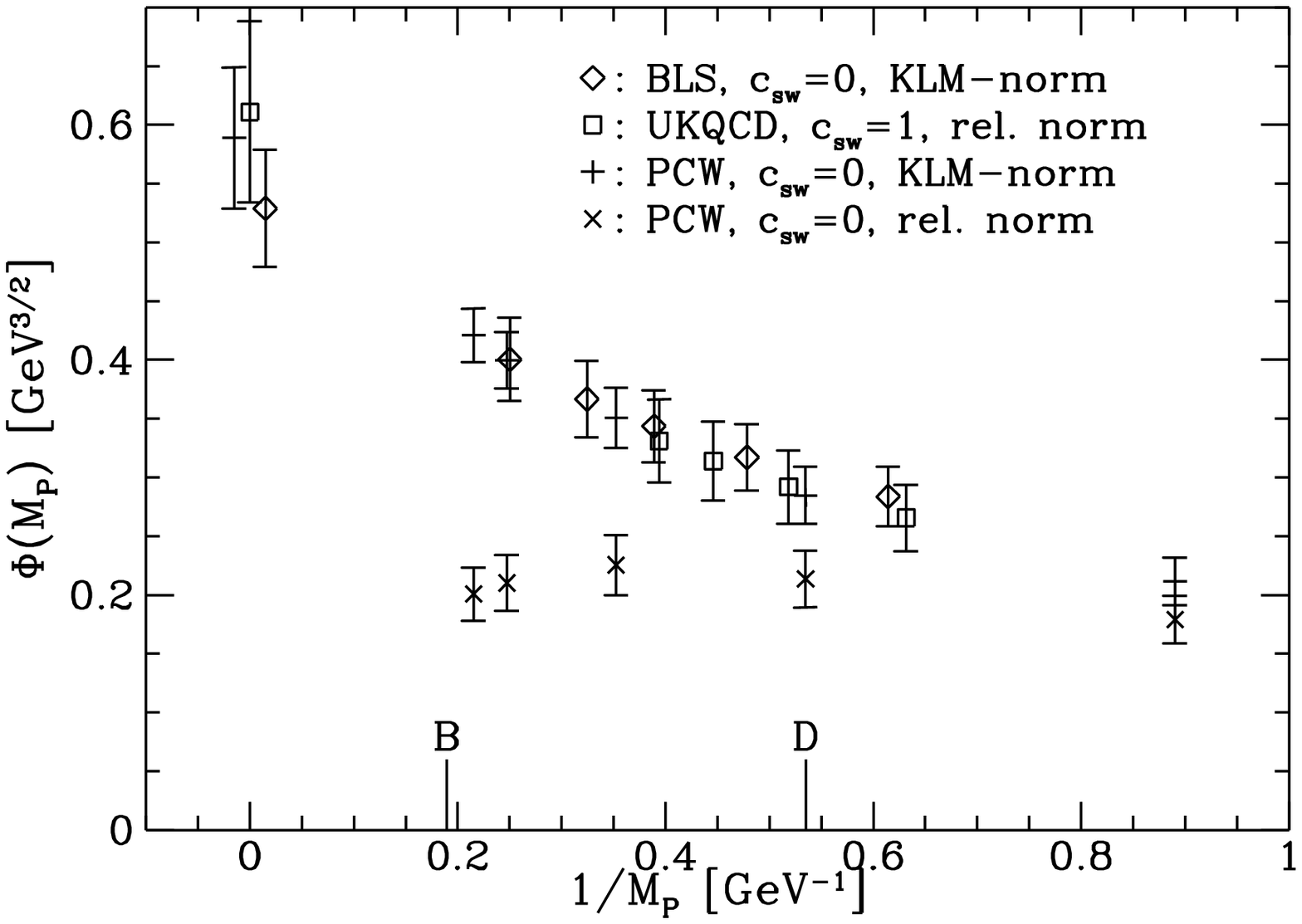}{95mm} 
%
%
%
\vspace{-2.6cm}
\fcaption{The quantity $\Phi(M_P)$ plotted against the inverse
  pseudoscalar mass. Diamonds represent the data from
  ref.\,\protect\cite{BLS_93}, obtained at $\beta=6.3$, squares denote
  the results from UKQCD\,\protect\cite{quenched,bbar}, using the SW
  action at $\beta=6.2$. The data from
  refs.\,\protect\cite{PCW_stat_92,PCW_prop_93} at $\beta=6.26$ are
  represented by plus signs (KLM-norm) and crosses (relativistic
  norm). The data obtained in the static approximation are slightly
  shifted for clarity. Also the positions of the $B$ and $D$ meson
  masses are indicated.} 
\label{frootm} 
\end{figure}

Lattice results for decay constants using the conventional or
non-relativistic formulations can be used to test predictions of the
Heavy Quark Effective Theory. It is well known that HQET predicts a
scaling law for heavy-light decay constants in the limit of infinite
quark mass, $m_Q\rightarrow\infty$
\be
   \frac{M}{f_V} \sim f_P \sim 
   \frac{{\rm const}}{\sqrt{M}}\alpha_s^{-2/\beta_0},
\ee
where $M_P\sim\ M_V\sim\ M\sim\ m_Q$. Hence, the heavy quark flavour
symmetry  implies that $f_P\sqrt{M_P}$ behaves like a constant (up to
logarithmic corrections)
\be
   f_P \sqrt{M_P} \sim {\rm const}\,\alpha_s^{-2/\beta_0},
\label{GLHQ_scaling}
\ee
where $\beta_0=11-\frac{2}{3}n_f$, and $n_f$ is the number of active
quark flavours. The heavy quark spin symmetry predicts the vector and
pseudoscalar decay constants to be degenerate.  Defining $U(M)$ as
the ratio of matrix elements of the axial and vector currents one
expects
\be
   U(M)\equiv\frac{f_V\,f_P}{M} \sim 1,\qquad m_Q\rightarrow\infty.
\ee
In order to study the mass dependence of $f_P\sqrt{M_P}$, one divides
out the scaling factor $\alpha_s^{-2/\beta_0}$ and defines the
quantity
\be  \label{GLfrootm}
   \Phi(M_P)\equiv f_P\sqrt{M_P}
   \,\left(\frac{\alpha_s(M_P)}{\alpha_s(M_B)}\right)^{2/\beta_0}.
\ee
In figure\,\ref{frootm} we compare some results for $\Phi(M_P)$
obtained using the static
approximation\,\cite{PCW_stat_92,BLS_93,bbar} with those from the
conventional formulation\,\cite{BLS_93,quenched,PCW_prop_93}. One
observes rather high values in the static approximation, whereas the
results using relativistic heavy quarks are much lower. The figure
also illustrates that improvement is a crucial ingredient for
detecting deviations from the scaling law: using either the
Sheikholeslami-Wohlert (SW) action in eq.\,(\ref{GLSW}) or employing
the KLM-norm, it seems possible to interpolate between the static
approximation and the conventional approach. Using the standard
relativistic norm in eq.\,(\ref{GL2kappa}) instead leads first to a
flattening and then to a decrease of $\Phi(M_P)$ as the heavy quark
mass is increased, as is shown in figure\,\ref{frootm} using the
results of the PSI-CERN-Wuppertal (PCW)
group\,\cite{PCW_prop_93}. Hence, the infinite mass limit of
$\Phi(M_P)$ would be in complete disagreement with the results
obtained directly at infinite $b$ quark mass.  Figure\,\ref{frootm}
also illustrates our earlier remark that different formulations of
heavy quarks have to be used simultaneously in order to reveal the
full picture. Another remarkable observation is that results from
simulations with different systematics agree very well within errors
(at similar values of the lattice spacing), which underlines the
consistency of the different methods.

From the slope of $\Phi(M_P)$ one can infer the size of $1/M_P$
corrections to the scaling law, eq.(\ref{GLHQ_scaling}). They amount
to about 15\% at the mass of the $B$ meson and about 40\% at the mass
of the~$D$. Thus it appears that there are large corrections to the
predictions of the heavy quark flavour symmetry for decay constants.
Recent studies using
NRQCD\,\cite{arifa_lat95}${}^{-}$\cite{SGO_NRQCD_97} have reported
even larger corrections at $M_B$.

A test of the heavy quark spin symmetry can be performed by computing
the quantity $U(M)$ and studying its mass dependence. In the heavy
quark limit, $U(M)$ differs from one by short-distance
corrections\,\cite{neub_92_2}${}^{-}$\cite{BroadGroz_92}
\be  \label{GLum_corr}
   U(M) \equiv \frac{f_V\,f_P}{M} = \left(1+
  \frac{8}{3}\frac{\alpha_s(M)}{4\pi} + O(1/M) \right).
\ee
The UKQCD collaboration performed this analysis using the $O(a)$
improved SW action; the short-distance corrections were divided out,
and the resulting data were fitted to either a linear or quadratic
function of the inverse, spin-averaged mass $1/M$.

\begin{figure}[tb]
\vspace{2cm}
\ewxy{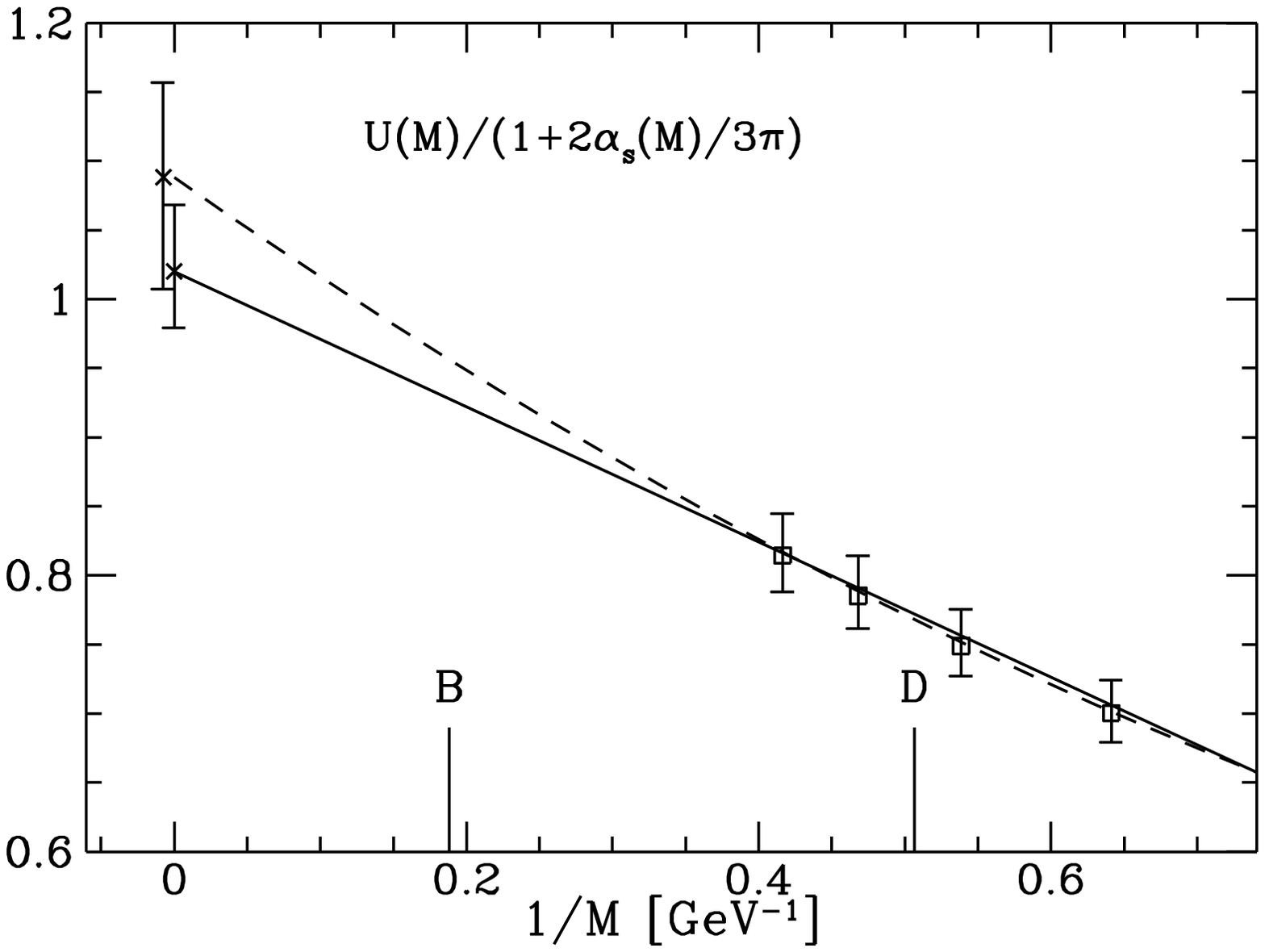}{95mm} 
%
%
%
\vspace{-2.6cm}
\fcaption{Data for the quantity $U(M)$ from the 
  UKQCD collaboration\,\protect\cite{quenched}. The solid and
  dashed lines represent the linear and quadratic fits to the
  data. Radiative corrections to $U(M)$ have been divided out.}
\label{um} 
\end{figure}

Figure\,\ref{um} shows that $U(M)$ indeed is consistent with one in
the heavy-quark limit. This result demonstrates not only the
manifestation of the heavy-quark spin symmetry in the infinite mass
limit, it also provides support for the parametrisation of the
non-scaling behaviour of $f_P\sqrt{M_P}$ as a power series in $1/M_P$.

Despite these encouraging results, heavy quark scaling laws for decay
constant require further investigation. In particular, the size of
$1/M$ corrections has to be studied as one approaches the continuum
limit, in order to detect to what extent the slope in $f_P\sqrt{M_P}$
is influenced by lattice artefacts. Also, it is {\em a priori} not
clear whether the interpolation of $f_P\sqrt{M_P}$ between the results
obtained in the static limit and the conventional approach is feasible
at non-zero values of the lattice spacing. Since discretisation errors
are potentially very different in the two approaches, a meaningful
interpolation may only be possible after extrapolation to the
continuum limit. We will return to this point in subsect.\,3.4.

\subsection{Results for pseudoscalar decay constants}

We now compile and discuss the results for the pseudoscalar decay
constants $\fd,\,\fb,\,\fds$ and $\fbs$ from various authors and using
different formulations of heavy quarks on the lattice.

\subsubsection{The conventional approach}
We start with the decay constants $\fd$ and $\fds$, which have only
been calculated using the conventional (relativistic) formulation. The
determination of $\fd$ proceeds by interpolating the lattice data for
$\Phi(M_P)$ (c.f. eq.\,(\ref{GLfrootm})) to the mass of the $D$~meson,
as either a linear or a quadratic function of $1/M_P$. In
table\,\ref{TABfD} we list the results, also indicating the value of
$\beta$, the normalisation of the quark fields and whether an improved
action was used (i.e. $\csw>0$). The first error on the decay constant
is the statistical error, whereas the others are an estimate of
systematic errors. The MILC
collaboration\,\cite{MILC_umh,MILC_lat9495,MILC_lat96} quote a third
error which is an estimate of quenching effects, and which is obtained
by comparing the results in the quenched approximation at
$a^{-1}\simeq2\,\gev$ to a simulation with $n_f=2$ flavours of
dynamical quarks at approximately the same value of~$a^{-1}$.

\begin{table}[tb]
\tcaption{Results for $\fd$ and $\fds$ using the conventional
  formulation of heavy quarks. All values were obtained in the
  quenched approximation, except for those from the HEMCGC
  collaboration which were computed for $n_f=2$. Infinite values of
  $\beta$ symbolise results obtained after extrapolation to the
  continuum limit. Results by MILC and JLQCD are preliminary.}
\begin{center}
\begin{tabular}{lcllcc}
\hline
\hline
Collab. & $\beta$ & $\fd\,[\mev]$ & $\fds\,[\mev]$ 
        & norm & $\csw$ \\
\hline
APE\,\cite{APE_prop_97} & 6.2 & 221(17) & 237(16) & rel. & 1 \\
MILC\,\cite{MILC_lat96} & ``$\infty$" & 
  196(9)(14)(8) & 211(7)(25)(11) & KLM & 0 \\
JLQCD\,\cite{JLQCD_lat95,JLQCD_lat96} & ``$\infty$" &
  202(8)\err{24}{11} & 216(6)\err{22}{15} & KLM & 0 \\[0.8ex]
LANL\,\cite{BG_95,BG_lat95} & 6.0 &
  229(7)\err{20}{16} & 260(4)\err{27}{22} & KLM & 0 \\
  & ``$\infty$" & 186(29) & 218(15) & KLM & 0 \\[0.8ex]
PCW\,\cite{PCW_prop_93} & ``$\infty$" &
  170(30) &  & KLM & 0 \\
PCW\,\cite{PCW_91} & 6.0 &
  198(17) & & rel. & 0 \\[0.8ex]
HEMCGC & 5.6 &
  200--287 & 220--320 &  & 0 \\
\,\cite{HEMCGC_stag93,HEMCGC_wil93} & 5.3 &
  215(5)(40)(35) & 287(5)(45)(40) & KLM & 0 \\[0.8ex]
UKQCD\,\cite{quenched} & 6.2 &
  185\er{4}{3}\err{42}{7} & 212\er{4}{4}\err{46}{7} & rel. & 1 \\
  & 6.0 & 199\err{14}{15}\err{27}{19} 
  & 225\err{15}{15}\err{30}{22} & rel. & 1 \\[0.8ex]
BLS\,\cite{BLS_93} & 6.3 &
  208(9)(35)(12) & 230(7)(30)(18) & KLM & 0 \\
ELC\,\cite{abada_92} & 6.4 & 210(40) & 230(50) & rel. & 0 \\[0.8ex]
ELC\,\cite{gavela_88} & 6.2 & 181(27) & & rel. & 0 \\
                      & 6.0 & 197(14) & & rel. & 0 \\
LDeG\,\cite{DeGrand_88} & 6.0 & 134(23) & 157(11) & rel. & 0 \\[0.8ex]
BDHS\,\cite{BDHS88} & 6.1 & 174(26)(46) & 234(46)(55) & rel. & 0 \\
\hline
\hline
\end{tabular}
\end{center}
\label{TABfD}
\end{table}

Table\,\ref{TABfB} shows the corresponding results for $\fb$ and
$\fbs$. Furthermore, the SU(3)-flavour breaking ratios $\fds/\fd$ and
$\fbs/\fb$ are listed in table\,\ref{TABfPsoverfP}. The extraction of
$\fb$ using the conventional method involves the extrapolation of
$\Phi(M_P)$ in $1/M_P$ to the $B$ system, which is an additional
source of uncertainty. In order to have better control over this
extrapolation, some authors prefer to interpolate between the values
obtained in the static approximation and those from the conventional
method\,\cite{BLS_93,PCW_prop_93,MILC_umh,MILC_lat9495,MILC_lat96}.
As discussed in subsection\,3.2, at a fixed value of $\beta$
this appears only possible if either an improved action or the KLM
norm is employed. Using instead the relativistic norm with $\csw=0$
results in a flattening of $\Phi(M_P)$ as $M_P$ is increased, which is
interpreted as a lattice artefact. This is consistent with the
observation that early calculations using the conventional method
obtained rather low values for $\fd$ and
$\fb$\,\cite{BDHS88,DeGrand_88}, and underlines the importance of
addressing the effects of large quark masses in the conventional
approach.

In addition to using the KLM norm in conjunction with the otherwise
unimproved Wilson action ($\csw=0$), some
collaborations\,\cite{BLS_93,MILC_umh,MILC_lat9495,MILC_lat96,JLQCD_lat96}
also include higher-order corrections for large quark mass, as
explained in subsect.\,2.2.  The pseudoscalar meson mass
$M_P$ has been shifted accordingly, i.e.
\be	\label{GLMkinetic}
M_P\to M_P^\prime \equiv M_P + (\widetilde{m}_2-\widetilde{m}_{\rm P}),
\ee
where $m_{\rm P}$ and $m_2$ are defined in eqs.\,(\ref{GLamP}) and
(\ref{GLm2}), respectively, and also tadpole improvement has been
applied to\,$m_{\rm P}$ and\,$m_2$. The mass dependence of $\Phi(M_P)$
is then studied in terms of the shifted mass $M_P^\prime$, which is
the tree-level estimate of the kinetic mass appearing in the
non-relativistic dispersion relation.

\begin{table}[tb]
\tcaption{Results for $\fb$ and $\fbs$ using the conventional
  formulation of heavy quarks. All values were obtained in the
  quenched approximation, except for those from the HEMCGC
  collaboration. Results by MILC and JLQCD are preliminary.}
\begin{center}
\begin{tabular}{lcllcc}
\hline
\hline
Collab. & $\beta$ & $\fb\,[\mev]$ & $\fbs\,[\mev]$ 
        & norm & $\csw$ \\
\hline
APE\,\cite{APE_prop_97} & 6.2 & 180(32) & 205(35) & rel. & 1 \\
MILC\,\cite{MILC_lat96} & ``$\infty$" & 
  166(11)(28)(18) & 181(10)(36)(18) & KLM & 0 \\
JLQCD\,\cite{JLQCD_lat95,JLQCD_lat96} & ``$\infty$" &
  179(11)\err{2}{31} & 197(7)\err{0}{35} & KLM & 0 \\
PCW\,\cite{PCW_prop_93} & ``$\infty$" &
  180(30) &  & KLM & 0 \\[0.8ex]
HEMCGC & 5.6 &
  152--235 & & & 0 \\
\,\cite{HEMCGC_stag93,HEMCGC_wil93} & 5.3 &
  150(10)(40)(40) &  & KLM & 0 \\[0.8ex]
UKQCD\,\cite{quenched} & 6.2 &
  160\er{6}{6}\err{59}{19} & 194\er{6}{5}\err{62}{9} & rel. & 1 \\
  & 6.0 & 176\err{25}{24}\err{33}{15}   &  & rel. & 1 \\[0.8ex]
BLS\,\cite{BLS_93} & 6.3 &
  187(10)(34)(15) & 207(9)(34)(22) & KLM & 0 \\
ELC\,\cite{abada_92} & 6.4 & 205(40) & & rel. & 0 \\
BDHS\,\cite{BDHS88} & 6.1 & 105(17)(30)  &  & rel. & 0 \\
\hline
\hline
\end{tabular}
\end{center}
\label{TABfB}
\end{table}

\begin{table}[tb]
\tcaption{Results for the SU(3)-flavour breaking ratios $\fds/\fd$ and
  $\fbs/\fb$ using the conventional formulation of heavy quarks.
  All values were obtained in the quenched approximation.
  Results by MILC are preliminary.} 
\begin{center}
\begin{tabular}{lcll}
\hline
\hline
Collab. & $\beta$ & $\fds/\fd$ & $\fbs/\fb$ \\
\hline
APE\,\cite{APE_prop_97} & 6.2 & 1.07(4) & 1.14(8) \\
MILC\,\cite{MILC_lat96} & ``$\infty$" & 1.09(2)(5)(5) & 1.10(2)(5)(8) \\
LANL\,\cite{BG_95,BG_lat95} & 6.0 & 1.135(21)\err{23}{6} & \\
PCW\,\cite{PCW_prop_93} & ``$\infty$" & 1.09(2)(5) & 1.09(2)(5) \\[0.8ex]
UKQCD\,\cite{quenched} & 6.2 & 1.18\er{2}{2} & 1.22\er{4}{3} \\
  & 6.0 & 1.13\er{6}{7} & 1.17(12) \\[0.8ex]
BLS\,\cite{BLS_93} & 6.3 & 1.11(6) & 1.11(6) \\
\hline
\hline
\end{tabular}
\end{center}
\label{TABfPsoverfP}
\end{table}

Tables\,\ref{TABfD} and\,\ref{TABfB} show that the results among
different collaborations are broadly consistent, provided that the
effects of large quark masses have been treated, either by employing
an improved action, or by using the KLM norm. The error analyses by
the different groups reveal that the statistical errors amount to
5--10\%, so that the uncertainty in pseudoscalar decay constants is
now dominated by systematic effects. In view of the large overall
errors and the different treatment of the systematics, it would be
premature to conclude on the basis of tables\,\ref{TABfD}
and\,\ref{TABfB} that discretisation errors are under control. This
requires a detailed analysis of the continuum limit, which we shall
attempt in the next subsection.

The ratios $\fds/\fd$ and $\fbs/\fb$ listed in
table\,\ref{TABfPsoverfP} vary by about 10\% among different groups,
which is slightly larger than the typical statistical error.  This is
inspite of the fact that some of the systematic errors in these ratios
(e.g. the renormalisation factor $Z_A$) are expected to cancel, such
that they are determined much more reliably.

Here we wish to make a few comments about the estimation of systematic
errors. Most collaborations quote a systematic error coming from the
uncertainty in the lattice scale. Apart from the statistical error in
the quantity that is used to set the scale, this may also include the
effects of choosing different quantities to estimate
$a^{-1}\,[\gev]$. The MILC Collaboration\,\cite{MILC_lat96,CB_priv}
include this uncertainty in their estimation of quenching
errors. Furthermore, MILC perform a detailed analysis of systematic
errors after the extrapolation to the continuum limit. Their estimate
yields a total uncertainty in $\fb$ due to using large quark masses of
about 10\% of the result. Further sizeable systematic effects are
ascribed to the extrapolations in the lattice spacing and the light
quark mass. By comparing results for $\fb$ for $n_f=0$ and $n_f=2$ at
similar values of $a^{-1}\,[\gev]$, MILC conclude that $\fb$ may be
larger by around 10\% in the unquenched theory.

Bhattacharya and Gupta\,\cite{BG_95} have analysed different
prescriptions in the definition of the tadpole improved normalisation
factor $\widetilde{Z}_A$. Using either $u_0=\langle\frac{1}{3}{\rm
Re\,Tr}\,P\rangle$ or $u_0=1/8\kcrit$ in eq.\,(\ref{GLZatilde}) leads
to an uncertainty of 14\,\mev\ in $\fd$, which is considered a
conservative estimate by the authors. They also ascribe an error of
around 5\% to the uncertainty in fixing the charm mass, using either
the kinetic mass $M_P^\prime$ or the conventional ``pole'' mass $M_P$
for the heavy-light pseudoscalar meson. However, since only a single
mass value for the heavy quark is used in\,\cite{BG_95}, the authors
are not able to perform a detailed analysis of the mass behaviour, and
thus the problem of using different definitions of heavy masses cannot
be fully addressed in their study.

In fact, Allton\,\etal\,\cite{APE_prop_97} have found that the mass
dependence of $\Phi(M_P)$ is only marginally changed by considering
the ``kinetic'' mass instead of the ``pole'' mass. They quote their
best estimates from their run at $\beta=6.2$, $\csw=1$, using results
at $\beta=6.0$ as well as data with $\csw=0$ to estimate systematic
errors. Furthermore, Allton\,\etal\ emphasise that the data for
$\Phi(M_P)$ using $\csw=1$ are indistinguishable from those for
$\csw=0$ computed with the KLM prescription, in good agreement with
ref.\,\cite{quenched} and this work (c.f. figure\,\ref{frootm}).

JLQCD investigate mass effects over a range of quark masses and
lattice spacings using the unimproved Wilson
action\,\cite{JLQCD_lat95,JLQCD_lat96}. At fixed values of $\beta$
they analyse the effects of using different field normalisations
(relativistic or KLM) and definitions of the pseudoscalar mass
(``pole'' or ``kinetic''). Whilst different prescriptions produce a
fair amount of variation in the value of the decay constant at
non-zero lattice spacing, JLQCD's preliminary
findings\,\cite{JLQCD_lat96} indicate that these deviations are
decreasing as $a\rightarrow0$.  Similar observations were made earlier
by the PCW collaboration\,\cite{PCW_prop_93}.

\subsubsection{The static approximation}
Now we turn to the discussion of results for $\fb$ obtained in the
static approximation. In table\,\ref{TABfBstatic} we list the results
from several simulations. In general, $\fbstat$ turns out much
larger than the results from the conventional approach. Especially
some of the earlier simulations produced very large
values\,\cite{PCW_91,APE_stat60_93}${}^{-}$\cite{ELC_stat_91}, which
may partly be due to the fact that the axial current normalisation
factor in the static approximation, $\zastat$, was evaluated in bare
perturbation theory, thus producing a much larger value.

In order to avoid discussing many different systematic effects in
$\fbstat$ when comparing different simulations, we have also listed
the matrix element $A^L$ in lattice units in table\,\ref{TABfBstatic}
(see eq.\,(\ref{GLstatAL})). By comparing results at fixed $\beta$,
one observes consistency in $A^L$ among many different collaborations
over a wide range of $\beta$ values. This is remarkable as even the
use of an improved action for the light quark does not seem to have a
discernible effect on $A^L$ (see also the discussion
in\,\cite{bbar}). Thus, whilst the numerical values for $A^L$ obtained
using either $\csw=0$ or $\csw=1$ are totally consistent at fixed
$\beta$, the one-loop estimates of the renormalisation factor
$\zastat$ are larger by 10--15\% for $\csw=1$ as compared to
$\csw=0$. Therefore, as long as there is no non-perturbative
determination of $\zastat$, which may clarify the issue, lattice
determinations of $\fbstat$ will always differ in the unimproved and
improved theories at non-zero values of~$a$. It is therefore of great
importance to study the approach to the continuum limit of both cases.

As the signal-to-noise ratio in simulations of the static
approximation is notoriously bad\,\cite{ELC_stat_91,Hashi_saeki_92},
it is important to use an efficient smearing technique in order to
extract a reliable signal. The most advanced calculations in the
static approximation\,\cite{Ken_lat93,FNAL_94,bbar} use a variational
approach in which a matrix correlator is constructed by computing
cross correlations from different smearing functions. One can then
diagonalise the matrix correlator and project onto the ground state,
thereby eliminating approximately the contamination of the signal from
higher excited states\,\cite{matrix_corrs}.

\begin{table}[h]
\tcaption{Results for $\fb$, $\fbs/\fb$ and the matrix element in
  lattice units, $a^{3/2}A^L$, obtained in the static
  approximation. All data are from quenched simulations. The results
  in\,\protect\cite{APE_lat94} are preliminary.}
\begin{center}
\begin{tabular}{lclllc}
\hline
\hline
Collab. & $\beta$ & $\fbstat\,[\mev]$ & $\fbs/\fb$ 
        & $a^{3/2}A^L$ & $\csw$ \\
\hline
UKQCD\,\cite{bbar} & 6.2 & 266\err{18}{20}\err{28}{27} & 1.16\er{4}{3}
  & 0.112\er{8}{8} & 1 \\
APE\,\cite{APE_lat94} & 6.4 & 235(9)  & 1.13(2) & 0.075(3) & 1 \\
                      & 6.4 & 209(14) & 1.12(7) & 0.076(5) & 0 \\
                      & 6.2 & 221(12) & 1.12(7) & 0.109(6) & 1 \\
                      & 6.1 & 190(10) & 1.13(5) & 0.135(7) & 0 \\
                      & 6.0 & 258(9)  & 1.19(3) & 0.201(7) & 1 \\
FNAL\,\cite{FNAL_94} & ``$\infty$" & 188(23)(15)\err{26}{0}(14) 
  & 1.22(4)(2) & & 0 \\
               & 6.3 & 225(17)(14) & 1.17(3)(1) & 0.099(8)  & 0 \\
               & 6.1 & 215(21)(14) & 1.23(3)(2) & 0.135(13) & 0 \\
               & 5.9 & 241(13)(13) & 1.21(2)(1) & 0.250(14) & 0 \\
               & 5.7 & 271(13)(20) & 1.18(3)(1) & 0.564(28) & 0 \\
APE\,\cite{APE_stat62_94} & 6.2 & 290(15)(45) & 1.11(3) 
  & 0.111(6) & 1 \\ 
UKQCD\,\cite{quenched}  & 6.0 & 286\err{8}{10}\err{67}{42}
  & 1.13\er{4}{3} & 0.211\er{6}{7} & 1 \\
BLS\,\cite{BLS_93} & 6.3 & 235(20)(21) & 1.11(2)(2) & 0.092(6) & 0 \\
Ken\,\cite{Ken_lat93,Ken_lat94}
       & 6.0 & 224\er{9}{7} & 1.22(1) & 0.184(7) & 0 \\
SH\,\cite{Hashi_prd94} & 6.0 & 297(36)\err{15}{30} & & 0.206(25)& 0 \\
APE\,\cite{APE_stat60_93} & 6.0 & 370(40) & 1.19(5) & 0.22(2) & 1 \\
                          & 6.0 & 350(40)(30) & 1.14(4) & 0.23(2) & 0
                          \\
PCW\,\cite{PCW_stat_92} & ``$\infty$" & 230(22)(26) & 1.16(5) & & 0 \\
PCW\,\cite{PCW_91} & 6.0 & 366(22)(55) & 1.10(9) & & 0 \\
HS\,\cite{Hashi_saeki_92} & 6.0 & 386(15) & & & 0 \\
ELC\,\cite{ELC_stat_91} & 6.0 & 310(25)(50) &  & 0.22(2) & 0 \\
\hline
\hline
\end{tabular}
\end{center}
\label{TABfBstatic}
\end{table}
\begin{table}[tb]
\tcaption{Results for $\fb$ using NRQCD. The results by the SGO
Collaboration\,\protect\cite{collins_lat95,SGO_decay_96} were obtained on
dynamical configuration with $n_f=2$.}
\begin{center}
\begin{tabular}{lclllc}
\hline
\hline
 Collab.  & $\beta$ & $a^{3/2}A^L$ & $\fb\,[\mev]$ & $\fbs/\fb$ 
& $n_f$ \\
\hline
SGO\,\cite{SGO_NRQCD_97} & 6.0 & & 183(32)(28)(16) & 1.17(7) & 0 \\
SGO\,\cite{SGO_decay_96} & 5.6 & & 126--166 & 1.24(4)(4) & 2 \\
Ken\,\cite{Ken_lat95} & 6.0 & 0.110\er{4}{5} & & 1.15\er{2}{1} & 0 \\
SH\,\cite{Hashi_prd94} & 6.0 & 0.156(8) & 171(22)\err{19}{45} & & 0 \\
\hline
\hline
\end{tabular}
\end{center}
\label{fB_NRQCD}
\end{table}
A possible interpretation of the very low value for $A^L$ observed by
the Kentucky group\,\cite{Ken_lat93}, who use a large basis of
different smearing functions, is that there may be some residual
contamination from higher excited states in the results of
refs.\,\cite{ELC_stat_91,APE_stat60_93,PCW_stat_92}. This is not
implausible, since typical values for the binding energy extracted
from the exponential fall-off of the correlation function in
eq.\,(\ref{GLstatcorrSS}) are around ${\cal E}\simeq0.7$ in lattice
units at $\beta=6.0$. This is still rather large and thus great care
is required in isolating the ground state. Therefore, despite using
smeared sources, and in view of large overall statistical
fluctuations, the results at the lower end of the range in $\beta$
could still be affected by higher excited states if no variational
approach is applied.  The situation improves above $\beta\geq6.0$,
where the binding energies are lower such that not so much can be
gained by employing variational smearing. At $\beta=6.2$, UKQCD have
explicitly verified that the contamination from higher states was
negligible, and that their results with and without variational
smearing were consistent\,\cite{bbar}.

The dependence of $\fbstat$ on the number of dynamical quark flavours
$n_f$, has been studied in refs.\,\cite{ROME2_96,MILC_stat_lat96}. The
results by the Rome\,2 group\,\cite{ROME2_96} indicate that at
$a^{-1}=1.1\,\gev$ the value of $\fbstat$ increases by 15--20\,\% for
$n_f=3$. Further calculations are clearly needed to confirm this
sizeable flavour dependence.

\subsubsection{Non-relativistic QCD}
So far there are relatively few estimates of $\fb$ using NRQCD. For
all but one study\,\cite{SGO_NRQCD_97}, an estimate for the axial
current normalisation constant in NRQCD was not available, such that
most previous studies had resorted to using $\zastat$. The major
results are listed in table\,\ref{fB_NRQCD}. Quantities in which the
current normalisation cancels, such as $\fbs/\fb$ are consistent with
the results from the static approximation and the conventional
approach.

A major advantage of NRQCD in studies of the scaling law for
$f_P\sqrt{M_P}$ is that it is possible to study the corrections in the
inverse meson mass $1/M$ starting from the static approximation as the
limiting case. Indeed, very large $1/M$ and $1/M^2$ corrections to the
scaling law eq.\,(\ref{GLHQ_scaling}) have been observed in
NRQCD\,\cite{arifa_lat95,collins_lat95,SGO_decay_96}. However, as the
NRQCD hamiltonian in those studies was obtained after truncation at
order $1/m_Q$ in the heavy quark mass, one may question the
observation of large $1/M^2$ corrections. More recent studies have
therefore included all $1/m_Q^2$ terms in the NRQCD action and
operators. Preliminary results by Ali\,Khan and
Bhattacharya\,\cite{arifa_bhatt_lat96} indicate that the individual
corrections of $1/M^2$ to the NRQCD hamiltonian amount to about 20\%
of the $1/M$ corrections, although their effect on the slope of
$f_P\sqrt{M_P}$ is marginal, as the various $1/m_Q^2$ corrections to
the hamiltonian and operators come in with opposite signs.

All of these findings must, however, be re-evaluated in the light of a
recent study in which $Z_A$ computed in NRQCD at one-loop order, has
for the first time been applied. This has a dramatic effect on the
$1/M$ corrections to the scaling law, which turn out to be much
smaller when $Z_A$ is taken into account. Furthermore, the scale $q^*$
at which the mean field improved coupling $\alpha(q^*)$ in the
one-loop expression for $Z_A$ is evaluated (c.f. subsection\,2.3) has
a big influence on the mass behaviour of $f_P\sqrt{M_P}$. Choosing
$q^*=a^{-1}$, $f_P\sqrt{M_P}$ is almost constant in $1/M$, whereas for
$q^*=\pi/a$ the slope of $f_P\sqrt{M_P}$ in $1/M$ is consistent with
the findings using the conventional
approach\,\cite{BLS_93,quenched}. Performing a careful estimation of
systematic errors, the authors of ref.\,\cite{SGO_NRQCD_97} quote
\bea
    f_B & = & 174(28)(26)(16)\,\mev,\qquad q^*=1/a,  \\
    f_B & = & 183(32)(28)(16)\,\mev,\qquad q^*=\pi/a,
\eea
where the first error combines the statistical and the fitting error,
the second comes from the uncertainty in the lattice scale, and the
third is an estimate of uncertainties due to higher-order
contributions to $Z_A$ and neglected higher orders in $1/M$.

There are indications that consistency in the mass behaviour of
$f_P\sqrt{M_P}$ from the region of charm to the static limit is
observed using different formalisms of treating heavy quarks on the
lattice. It should, however, be emphasised that it is crucial to
investigate the continuum limit of the mass dependence of
$f_P\sqrt{M_P}$, which we will attempt in the next subsection.

\subsection{The continuum limit} \label{SEKcontlim}

We are now in a position to study the continuum limit of heavy-light
decay constants using results discussed in the previous
subsection. Although some of the more recent
simulations
\cite{PCW_stat_92,PCW_prop_93,FNAL_94,MILC_lat9495,MILC_lat96,JLQCD_lat96}
have already performed an extrapolation to the continuum limit, we
want to treat as many different results as possible on an equal
footing and perform a thorough analysis of systematic errors in the
final results.

As is apparent from tables\,\ref{TABfD}\,--\,\ref{TABfBstatic}, most
results have been obtained for Wilson fermions with $\csw=0$. At
present it is therefore not possible to perform the extrapolation
$a\rightarrow0$ of the data with $\csw=1$, which would be an important
consistency check. Furthermore, since the data of the
MILC\,\cite{MILC_lat9495,MILC_lat96} and JLQCD\,\cite{JLQCD_lat96}
collaborations are still preliminary and are likely to be updated, we
leave them out in the following analysis, despite their high
statistics and large range in $\beta$.

Since the systematic effects in the results listed in
tables\,\ref{TABfD}\,--\,\ref{TABfBstatic} are quite different,
e.g. the choice of lattice scale, the details of the quark field
normalisation and in the evaluation of $Z_A$, a straightforward
extrapolation of all listed results would be unreliable. Therefore, in
order to study the mass dependence of $f_P\sqrt{M_P}$ and obtain
estimates for decay constants in the continuum, we adopt a similar
strategy as described in\,\cite{sommer_review,PCW_prop_93}. The
various steps of the procedure are as follows:
\begin{enumerate}
\item[(a)] At a fixed value of $\beta$ one forms the dimensionless
ratio $\Phi(M_P)/\stg^{3/2}$ using lattice data for
$\Phi(M_P)$ defined in eq.\,(\ref{GLfrootm}) and the string tension
$\sqrt{\sigma}$.
\item[(b)] The ratio $\Phi(M_P)/\stg^{3/2}$ is then interpolated to
common values of $M_P/\stg$. 
\item[(c)] At every fixed value of $M_P/\stg$ one extrapolates
$\Phi(M_P)/\stg^{3/2}$ to the continuum limit as a function of
$a\stg$.
\item[(d)] Now one sets the scale using extrapolated lattice data for,
e.g. the pion decay constant, $f_\pi/\stg|_{a=0}$ of the $\rho$
meson mass $M_\rho/\stg|_{a=0}$. The continuum value of
$\Phi(M_P)/\stg^{3/2}$ is interpolated to the desired value of the
heavy-light meson mass, and finally one converts back to the decay
constant in physical units using the definition of $\Phi(M_P)$.
%
%
\end{enumerate}
This procedure implies that in step\,(d) $f_B$ is obtained through an
interpolation between the result in the static approximation and those
using relativistic heavy quarks. Therefore, a separate extrapolation
of the data for $\fbstat$ is required (i.e step\,(c) for
$\stg/M_P=0$). 

The difference between this procedure and the extrapolations performed
in refs.\,\cite{FNAL_94,MILC_lat96,JLQCD_lat96} is that in the latter
the conversion of heavy-light decay constants into physical units
precedes the extrapolation to $a=0$. This is feasible if the scale is
measured as well at every value of $\beta$ considered. The universal
treatment of a number of different simulations, however, forces one
to use the extrapolated lattice scales, since the values of
$\beta$ at which the quantity to set the scale is measured do not
necessarily coincide with the $\beta$ values used in the computation
of heavy-light decay constants.

Lattice data for the string tension $a\stg$ used in the following
extrapolations are listed in appendix\,B. In order to
perform a controlled extrapolation to the continuum limit of a number
of hadronic quantities, it has been suggested\,\cite{sommer_r0_93}
that the hadronic scale $r_0$ is to be preferred over the string
tension $\stg$. However, despite the merits of choosing $r_0$ to
monitor the approach of the continuum limit, it has so far only been
determined in a fairly limited range of $\beta$ (see
refs.\,\cite{UKQCD_pot65}${}^{-}$\cite{HW_lat94,SESAM_pot_96}).
Therefore we stick to the string tension for our purposes.

Appendix\,B also describes the details of the continuum
extrapolation of quantities in the light quark sector, such as
$f_\pi$, $M_\rho$ and the 1P--1S splitting in charmonium, $\SP$. The
pion decay constant $f_\pi$ is widely used to set the scale, because
both $f_\pi$ and the heavy-light decay constant $f_P$ are subjected to
similar systematic errors -- at least in the conventional approach --
such as quenching errors and uncertainties in $Z_A$. It can therefore
be expected that some of these effects cancel if the conversion is
performed using $f_\pi$. In ref.\,\cite{FNAL_alpha_92} it was argued
that the 1P--1S splitting in charmonium is largely insensitive to
lattice artefacts, which also makes it a reliable quantity to set the
scale. The mass of the $\rho$ meson, on the other hand, shows a strong
dependence on the lattice spacing for unimproved Wilson fermions (see
e.g. refs.\,\cite{sommer_review}, such that it is not
clear {\it a priori\/} whether or not one controls its extrapolation
to the continuum limit well enough. We include the mass of the $\rho$
meson nevertheless in this study.

In this review we choose to quote our best estimates using $f_\pi$ to
set the scale. Systematic errors are estimated by varying
\begin{itemize}
\item the details in the evaluations of $Z_A$ and $\zastat$
\item the field normalisation of propagating heavy quarks (KLM or
relativistic) 
\item the quantity that sets the scale
\item the minimum value $\beta_{\rm min}$ which is used in the
extrapolation in step\,(c) above (i.e. the maximum value of $a$\,[fm])
\end{itemize}

\subsubsection{Extrapolation of the data obtained in the static approximation}
We start our analysis by considering extrapolations of $\fbstat$ for
$\csw=0$, using the tadpole improved value for $\zastat$ (see
appendix~A). In figure\,\ref{extfBstatic} we have plotted the ratio
$f_B\sqrt{M_B}|^{\rm stat}/\stg^{3/2}$ as a function of the string
tension in lattice units. Taking the data from
refs.\,\cite{PCW_stat_92,BLS_93,APE_lat94,FNAL_94,Ken_lat93,Hashi_prd94,APE_stat60_93,ELC_stat_91},
it can be seen from the figure that the simultaneous extrapolation of
all data is not feasible and results in a large value of $\chi^2/\rm
dof$. We have therefore chosen to divide the data sample into two sets
that each can be extrapolated with acceptable $\chi^2/\rm
dof$. Combining the data from
refs.\,\cite{BLS_93,APE_lat94,FNAL_94,Ken_lat93,Hashi_prd94} into
``set\,1'' and those from
refs.\,\cite{PCW_stat_92,Hashi_prd94,APE_stat60_93,ELC_stat_91} into
``set\,2'', we note that the values in set\,2 are in general higher
than those in set\,1. As discussed in the previous subsection, a
possible explanation for this may be the presence of residual
contamination from excited states, because no variational smearing was
used in set\,2.
\begin{figure}[t]
%
\vspace{2.0cm}
\hspace{-1.4cm}
\ewxy{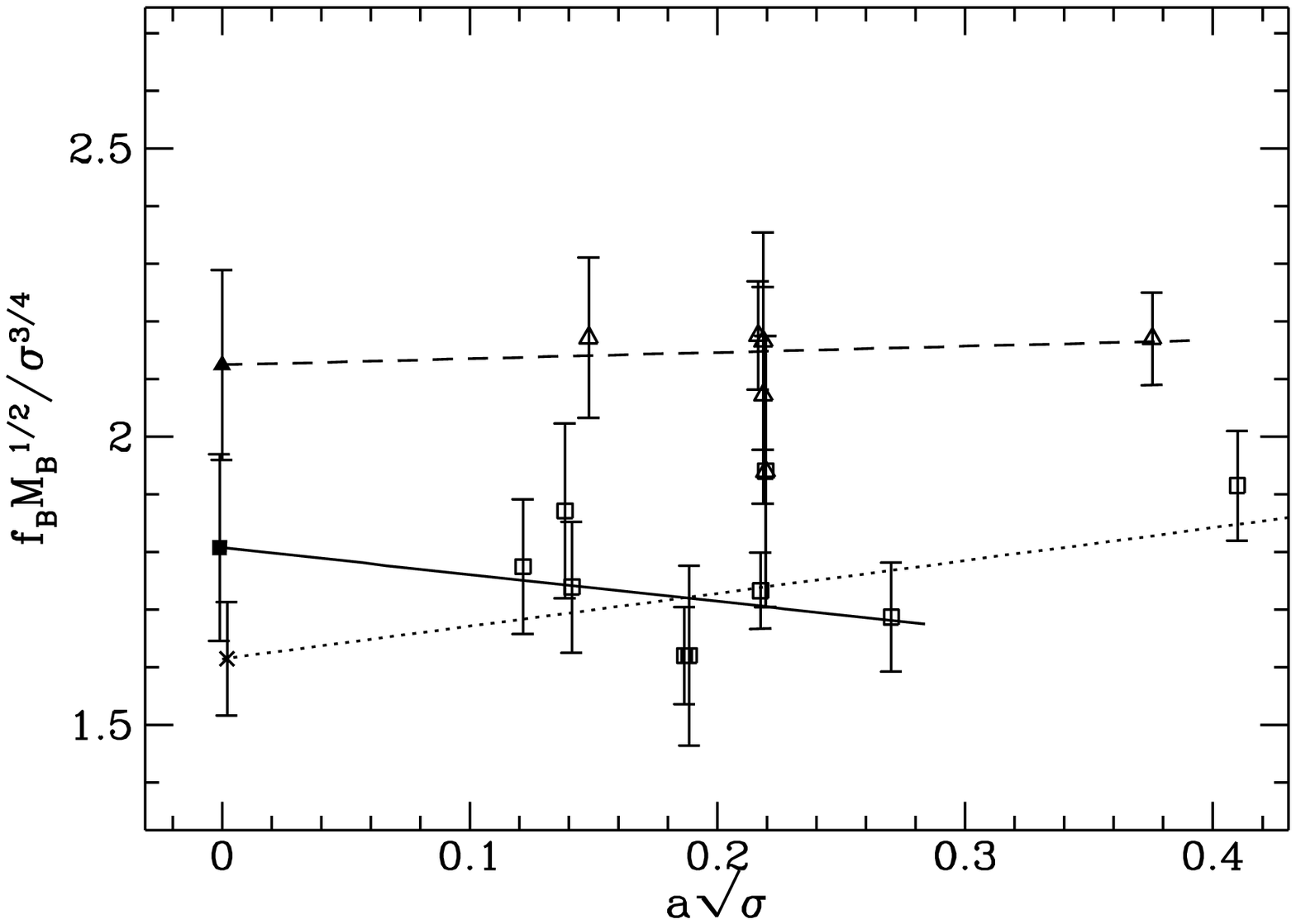}{95mm}
%
%
%
\vspace{-2.6cm}
\fcaption{The ratio $f_B\protect\sqrt{M_B}|^{\rm
static}/\protect\sqrt{\sigma}^{3/2}$ 
for $\csw=0$ plotted versus the string tension. Open squares denote
the results from refs.
\protect\cite{FNAL_94,APE_lat94,BLS_93,Ken_lat93,Hashi_prd94,%
APE_stat60_93,PCW_stat_92,ELC_stat_91}
(``set\,1''), whereas open triangles represent
refs. \protect\cite{PCW_stat_92,APE_stat60_93,ELC_stat_91,Hashi_prd94}
(``set\,2''). The solid line represents the linear extrapolation of the data
in set\,1 for $\beta_{\rm min}=5.9$, the dotted line is the
extrapolation of set\,1 for $\beta_{\rm min}=5.7$, and the dashed
curve denotes the extrapolation of the data in set\,2. The tadpole
improved normalisation factor of the axial current, $\zastat$ was used.}
\label{extfBstatic}
\end{figure}
However, since it is impossible at this stage to further investigate
the apparent discrepancies between sets\,1 and\,2, we will take set\,1
for our best estimate for $\fbstat$, whilst quoting the extrapolated
value from set\,2 as a systematic error.

We now discuss the dependence on the minimal $\beta$ value,
$\beta_{\rm min}$, used in the extrapolation.
Figure\,\ref{extfBstatic} shows that the extrapolated result for
$f_B\sqrt{M_B}|^{\rm stat}/\sqrt{\sigma}^{3/2}$ from set\,1 changes if
the point at $\beta=5.7$ is included or not. This may signal that
higher corrections in the lattice spacing may be important if one
includes data for which $a$ can be as large as 0.17\,fm. This
observation is consistent with the findings of
ref.\,\cite{cra_fB_cont}, where it was concluded that the expected
scaling behaviour of the quantity $A^L$ is observed down to values of
$\beta\simeq5.9 - 6.0$. However, since one obtains a perfectly good
fit of set\,1 if the result at $\beta=5.7$ is included, we again quote
the difference between results obtained for $\beta_{\rm min}=5.7$ and
$\beta_{\rm min}=5.9$ as a systematic error.

Our best estimate is computed for $\beta_{\rm min}=5.9$ and we obtain
\be
   \frac{f_B\sqrt{M_B}|^{\rm stat}}{\sqrt{\sigma}^{3/2}}
  = 1.81\pm0.16\,{\rm (stat)}\,{}^{+0.31}_{-0.20}\,{\rm (extr)}\pm0.07{\rm
(norm)}, 
\ee
where the first error is statistical, the second arises from the
difference between sets\,1 and\,2 as well as using $\beta_{\rm
min}=5.7$. The third error is due to the variations from using
different prescriptions in the numerical evaluation of $\zastat$.

Using $f_\pi/\sqrt{\sigma}|_{a=0}=0.285(13)$ from the extrapolation of
the data of the GF11 collaboration\,\cite{GF11_fpi} (see
appendix\,B) and $f_\pi=131\,\mev$ to convert into
physical units, we find
\be \label{GLresfBstatic}
   \fbstat = 245\pm27\,{\rm (stat)}\err{42}{39}\,{\rm (syst)}\,\mev,
\ee
where we have combined different systematic errors in quadrature,
including those from using $M_\rho$ and $\SP$ to set
the scale. Given the large errors, this result is not incompatible
with the fairly low result quoted by the FNAL group\,\cite{FNAL_94}
and that of the PCW collaboration\,\cite{PCW_stat_92} (see
table\,\ref{TABfBstatic}). It should be noted, however, that the
result in eq.\,(\ref{GLresfBstatic}) is a consequence of omitting the
results at low $\beta$ (i.e. $\beta=5.7$) in the extrapolation. If
those data are included one obtains the lower value of
$\fbstat=218\pm20\,{\rm (stat)}\,\mev$.

\subsubsection{Extrapolation of data using relativistic heavy quarks}
Now we proceed to discuss the extrapolation of heavy-light decay
constants using the conventional approach. This will finally enable us
to study the scaling law for $f_P\sqrt{M_P}$,
eq.\,(\ref{GLHQ_scaling}) in the continuum limit.

In order to be able to interpolate $f_P\sqrt{M_P}$ to common values of
$M_P$ whilst comparing the use of the kinetic versus the pole mass, we
need to resort to data whose mass dependence has been published (as a
function of the hopping parameter of the heavy quark). Therefore we
only use
refs.\,\cite{BLS_93,PCW_prop_93,abada_92,BG_95,APE_cra_lat93}, leaving
the data by MILC\,\cite{MILC_lat9495,MILC_lat96} and
JLQCD\,\cite{JLQCD_lat95,JLQCD_lat96} to be included in the
future. Furthermore we chose not to use the data of
ref.\,\cite{abada_92}. As already noted in\,\cite{sommer_review}, it
is questionable whether the asymptotic behaviour of the relevant
correlation functions has been reached in\,\cite{abada_92}, since no
smearing has been applied in the computation of the relevant
correlation functions. Failure to isolate the ground state leads to an
overestimate of the matrix element which is used to extract $f_P$. In
fact, the quantity $\Phi(M_P)$ from ref.\,\cite{abada_92} is very
different compared to
refs.\,\cite{BLS_93,PCW_prop_93,BG_95,APE_cra_lat93} when all data are
scaled appropriately and treated equally. Also some of the very early
calculations\,\cite{gavela_88,DeGrand_88,BDHS88} did not apply
smearing techniques to improve the isolation of the ground state, and
consequently we do not include them in our analysis.

The data for $f_P\sqrt{M_P}$ are used to compute
$\Phi(M_P)/\stg^{3/2}$, which is then interpolated to fixed values of
$\stg/M_P$ between\,0.12 and\,0.29, corresponding to heavy-light meson
masses of 1.4--3.5\,\gev. In ref.\,\cite{BG_95} only a single value of
the heavy quark mass was considered, which corresponds to an inverse
pole mass of $\stg/M_P\simeq0.283$ or to $\stg/M_P^\prime\simeq0.260$
if the definition of the kinetic mass, eq.\,(\ref{GLMkinetic}), is
used. Only at these values of the inverse pole and kinetic masses,
respectively could the results of ref.\,\cite{BG_95} be included in
the extrapolation to the continuum limit.

\begin{figure}[t]
%
\vspace{2.0cm}
\hspace{-1.4cm}
\ewxy{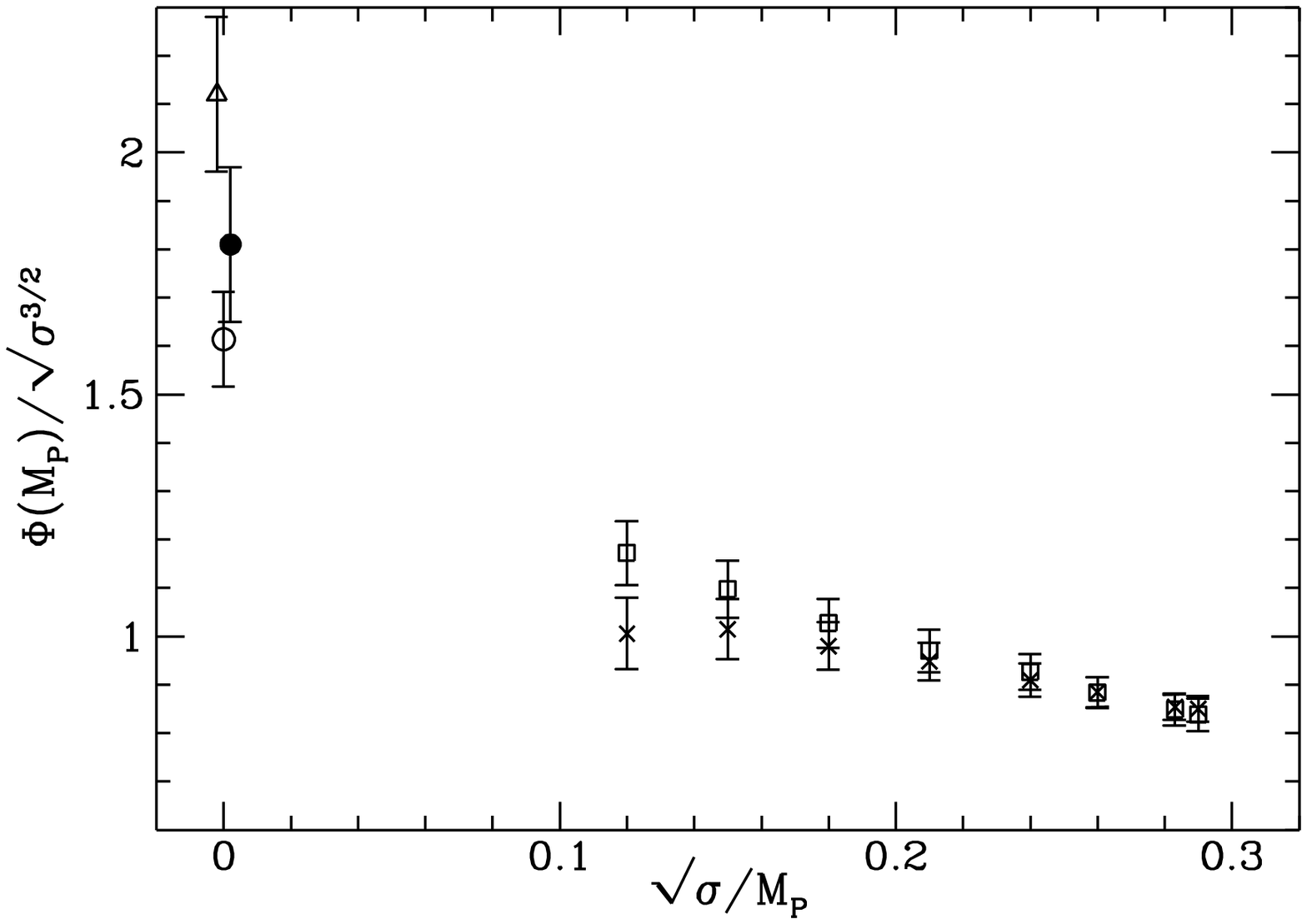}{95mm}
%
%
%
\vspace{-2.6cm}
\fcaption{The quantity $\Phi(M_P)/\protect\stg^{3/2}$ in the continuum
limit as a function of $\protect\stg/M_P$. Open squares denote the
data computed using procedure~(1), whereas the crosses represent the
points obtained using procedure~(3). Both sets were obtained using
$\beta_{\rm min}=5.7$ in the extrapolation. For
$\Phi(M_P)/\protect\stg^{3/2}$ in the static approximation, the
circles denote the extrapolated values for set\,1 using $\beta_{\rm
min}=5.7$ (open symbol) and $\beta_{\rm min}=5.9$ (full symbol). The
value obtained using set\,2 is denoted by the triangle.}
\label{frootm_cont}
\end{figure}

In order to test the stability of the extrapolations at fixed
$\stg/M_P$, we compare four procedures for which different evaluations
of $Z_A$ and different concepts to deal with heavy mass effects are
compared: 
\begin{itemize}
\item[(1)] KLM norm, kinetic mass, tadpole improved $Z_A$
\item[(2)] KLM norm, pole mass, tadpole improved $Z_A$
\item[(3)] relativistic norm, pole mass, tadpole improved $Z_A$
\item[(4)] relativistic norm, pole mass, $Z_A$ in boosted perturbation theory
\end{itemize}
Here, the kinetic mass is always evaluated using
eq.\,(\ref{GLMkinetic}), in contrast to some results in
refs.\,\cite{JLQCD_lat96,SGO_decay_96} where the kinetic mass was
extracted from the measured dispersion relation. It should be
mentioned at this point that using the KLM norm together with the pole
mass may, after all, be an ill-defined concept as large mass
corrections are only addressed at leading order, thus resulting in an
incomplete treatment of these effects.
\begin{figure}[t]
%
\vspace{2.0cm}
\hspace{-1.4cm}
\ewxy{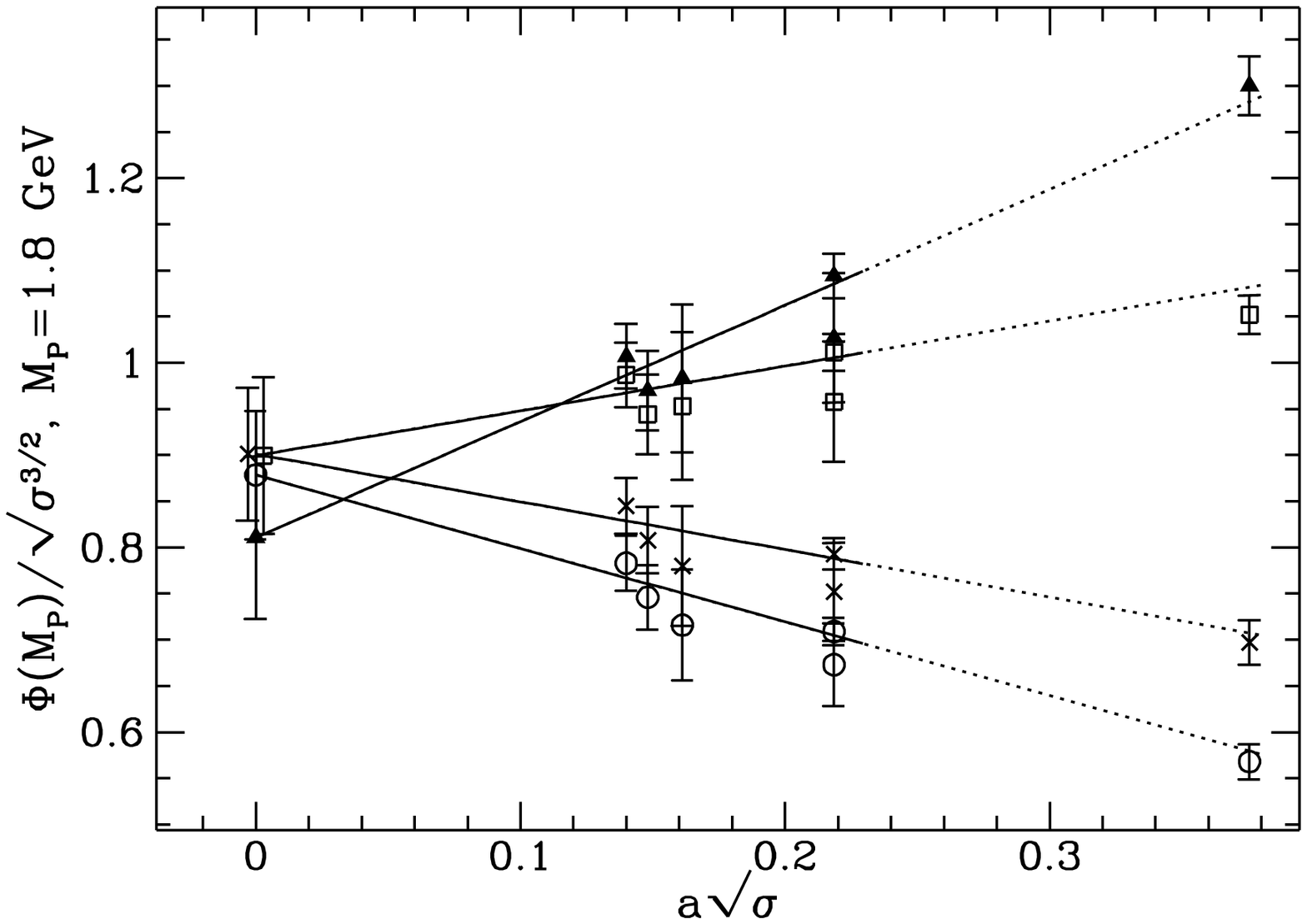}{95mm}
%
%
%
\vspace{-2.6cm}
\fcaption{The quantity $\Phi(M_P)/\protect\stg^{3/2}$ plotted as a
function of the string tension in lattice units.
Open squares represent the data computed procedure~(1) as described in
the text. Full triangles correspond to procedure~(2), and the crosses
and open circles represent the data using procedures~(3) and~(4),
respectively. The solid lines denote the linear extrapolations of
$\Phi(M_P)/\protect\stg^{3/2}$ to the continuum limit, using only the
data for which $\beta\ge6.0$. The dotted lines are the continuation of
the fits to the point at $\beta=5.74$.}
\label{Phiext}
\end{figure}
In figure\,\ref{Phiext} we plot the extrapolation of $\Phi(M_P)$ to
$a=0$ for all four schemes listed above. A remarkable feature is that
for $\stg/M_P > 0.21$, the results in the continuum limit are rather
consistent, including the data computed using the KLM norm in
conjunction with the pole mass. Since at fixed $\beta$ the data
computed using the relativistic norm show the wrong qualitative
behaviour (see figure\,\ref{frootm}), this is quite a surprising
result. Thus, at this stage it appears that the details of the
procedure employed to deal with large mass effects and the
normalisation factor $Z_A$ produce a variation of only 3--5\% in the
continuum results for $\Phi(M_P)$ for meson masses
$M_P\,\lesssim\,2.2\,\gev$.  These observations are consistent with
the findings in refs.\,\cite{PCW_prop_93,JLQCD_lat96}. However, the
spread of results at $a=0$ using different procedures increases for
larger masses, as can be seen in figure\,\ref{frootm_cont}. This is an
indication that eventually lattice artefacts make the continuum
extrapolation harder to control if the quark mass is chosen very
large. The slope of $\Phi(M_P)/\stg^{3/2}$ as a function of $a\stg$
increases for larger heavy quark mass, which is the expected behaviour
as the influence of lattice artefacts becomes greater. As
figure\,\ref{Phiext} shows, the results computed using the KLM norm
and the kinetic mass show the smallest slope in the lattice
spacing~$a$ compared to all other procedures employed.
Unlike in the case of the static approximation, the extrapolations of
$\Phi(M_P)/\stg^{3/2}$ for relativistic heavy quarks are fairly
insensitive to the value of $\beta_{\rm min}$. This can be seen from
figure\,\ref{Phiext} where the data for $\Phi(M_P)/\stg^{3/2}$ at
$\beta=5.74$ are not incompatible with the extrapolations performed
for $\beta_{\min}=6.0$. However, below we will still use
$\beta_{\min}=6.0$ for our best estimates as a safeguard against
lattice effects in the final results, especially since the lattice
spacing at $\beta=5.74$ is only $a^{-1}\simeq1.1\,\gev$.

\subsubsection{Interpolation to the physical meson masses}
We are now in a position to perform step~(d) and interpolate
$\Phi(M_P)/\stg^{3/2}$ to the physical meson masses. Setting the scale
by $f_\pi/\stg|_{a=0}=0.285(13)$, we have interpolated our
results for $\Phi(M_P)/\stg^{3/2}$ at $\stg/M_P=0.21,\,0.24$ and~0.26
to $\stg/M_D=0.245(11)$. Our final result for $f_D$ thus is
\be \label{GLresfD}
f_D=191\pm19\,{\rm (stat)}\err{3}{5}{\rm (extr)}\err{0}{20}
{\rm (scale)}\,\mev,
\ee
where the first error is statistical, the second is an estimate of
systematic effects by varying $\beta_{\rm min}$ and the treatment of
large mass effects. The third error is due to choosing $M_\rho$ or
$\SP$ to set the scale. The latter effect is predominantly
due to the known mismatch of $f_\pi/M_\rho$ computed in the quenched
approximation and the experimental result, which is partly ascribed to
quenching. 

In the case that the light quark has been interpolated to the mass of
the strange quark, steps~(a)--(d) of the extrapolation procedure are
entirely analogous, and for $f_{D_s}$ we obtain
\be \label{GLresfDs}
f_{D_s} = 206\pm17\,\,{\rm (stat)}\err{6}{7}{\rm (extr)}\err{0}{21}
{\rm (scale)}\,\mev.
\ee
By interpolation of the data for $\Phi(M_P)/\stg^{3/2}$ between the
points in the static approximation and those at finite mass $M_P$ to
the mass of the $B$~meson we obtain an estimate for $f_B$. Here,
however, there are two more sources of systematic errors which we
include in our final result. The first is the variation on
$f_B\sqrt{M_B}$ in the static approximation by using set\,2 instead
of set\,1 for the central value. The second arises by choosing
different sets of values for $\stg/M_P$ that are included in the
interpolation. Our best estimate is obtained using set\,1 and
$\stg/M_P=0.21,\,0.24$ and $0.26$ as in the case of $f_D$. Our result
for $f_B$ is
\be
f_B=172\pm24\,{\rm (stat)}\err{13}{12}{\rm(fits)} 
\err{0}{15}{\rm (scale)}\,\mev.
\ee
We can now analyse the mass behaviour of $\Phi(M_P)/\stg^{3/2}$ in the
continuum limit and deduce the leading correction in $1/M_P$ to the
scaling law eq.\,(\ref{GLHQ_scaling}). Figure\,\ref{frootm_cont} is
the analogue of figure\,\ref{frootm} at zero lattice spacing. If we
parametrise $\Phi(M_P)$ as
\be \label{GLPhi_param}
  \Phi(M_P)=\Phi(M_P)_\infty\bigg(1+\frac{C}{M_P}+\frac{D}{M_P^2}\bigg)
\ee
we obtain $C=0.9-2.0\,\gev$. Thus, the conclusions reached at finite
values of~$a$ are also valid in the continuum limit. The large
uncertainty in~$C$ is partly due to the variation of
$\Phi(M_P)/\stg^{3/2}$ under changes in $\beta_{\rm min}$, and in
particular to using either set\,1 or set\,2 in the interpolation to
the mass of the $B$ meson (the highest values for $C$ are obtained by
using set\,2 in the static approximation).

\begin{figure}[t]
%
\vspace{4cm}
\hspace{-1.4cm}
\ewxy{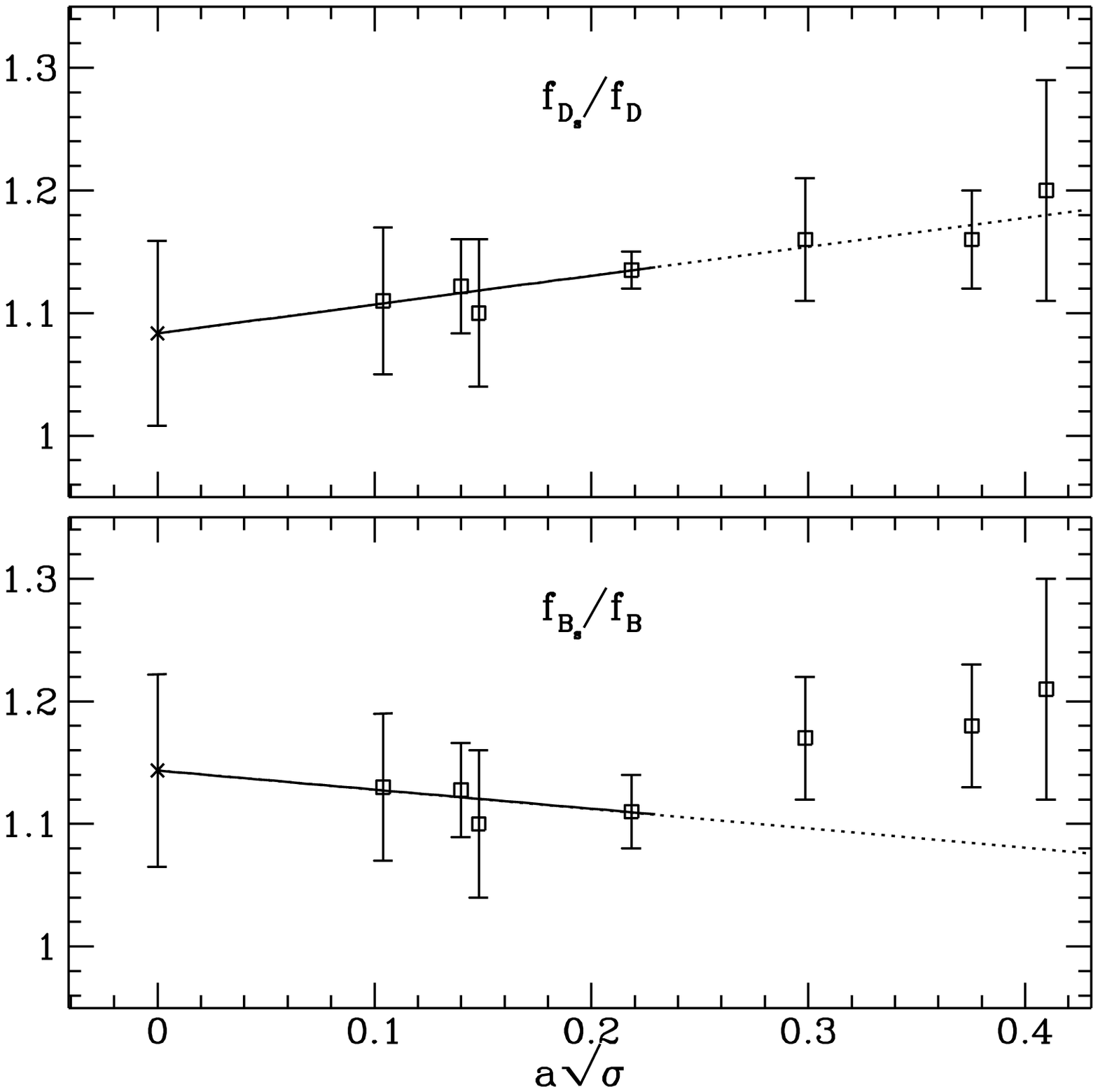}{95mm}
\vspace{-2.6cm}
\fcaption{Extrapolations of ${f_{B_s}}/{f_B}$ and ${f_{D_s}}/{f_D}$ to
the continuum limit as a function of the string tension. The dotted
lines are the continuations of the fits performed for
$\beta\ge6.0$. Data at common values of $a\stg$ have been combined in
a weighted average.}
\label{FfPsoverfP}
\end{figure}

The ratios $f_{D_s}/f_D$ and $f_{B_s}/f_B$ are also of great
interest. Since it can be expected that systematic effects cancel
partially in these ratios, one can attempt a straightforward
extrapolation of the results quoted in
refs.\,\cite{BLS_93,PCW_prop_93,MILC_lat96,CB_priv,BG_95} and
refs.\,\cite{PCW_stat_92,BLS_93,APE_lat94,FNAL_94,Ken_lat93} for
$f_{B_s}/f_B$ in the static approximation. In figure\,\ref{FfPsoverfP}
we have plotted these extrapolations as a function of $a\stg$, and one
can see that the results show a fairly flat behaviour as the continuum
limit is approached. For our final results at $a=0$ we have again
chosen $\beta_{\rm min}=6.0$ and continued the fits to lower values of
$\beta$ in figure\,\ref{FfPsoverfP} for illustrative purposes.  We
obtain
\begin{eqnarray}
& &\frac{f_{D_s}}{f_D} = 1.08(8),\qquad \frac{f_{B_s}}{f_B} = 1.14(8)
\label{GLfPsPd} \\
& &\frac{f_{B_s}}{f_B}\Big|^{\rm stat} = 1.07(5)\err{12}{0},
\end{eqnarray}
where we have used set\,1 for the extrapolation of
${f_{B_s}}/{f_B}$ in the static approximation. If instead one computes
the ratio ${f_{D_s}}/{f_D}$ using our estimates in
eqs.\,(\ref{GLresfD}),\,(\ref{GLresfDs}) one obtains
${f_{D_s}}/{f_D}=1.08(14)\er{1}{0}$ in very good agreement with the above
result, albeit with a larger error.


\subsection{Summary and comparison}
\label{SEKfBsummary}

The analysis of lattice results for heavy-light decay constants shows
that essentially a consistent picture emerges from all three methods
of simulating heavy quarks. The errors are, however, still fairly
large, so that this statement has to monitored as more precise lattice
data become available. Adding all errors in quadrature, our best
estimates for the decay constants $\fd$, $\fds$, $\fb$, $\fbs$ and
$\fbstat$ are listed in tables\,\ref{TfD_comp} and\,\ref{TfB_comp}.

The large $1/M$ corrections to the scaling law for $f_P\sqrt{M_P}$ are
now firmly established. In particular, the behaviour is not changed in
the continuum limit. Furthermore, the heavy quark spin symmetry is
manifest in the infinite mass limit, as $U(M)$, the ratio of matrix
elements of the axial and vector currents indeed approaches
unity. Also for this quantity the corrections in $1/M$ at the $B$ and
$D$ meson masses are sizeable.

The continuum limit requires further investigation. In particular it
is necessary to increase the reliability of the extrapolations by
going to smaller values of the lattice spacing, which correspond to
$\beta$ values larger than~6.3. Furthermore, as an important check
against the influence of lattice artefacts more data using improved
actions and operators need to be computed in the range $\beta\ge6.0$,
in order to allow for a separate continuum extrapolation. So far,
quite a number of results quoted at non-zero lattice spacing are in
agreement with the continuum result derived in this analysis
(c.f. tables\,\ref{TABfD} and\,\ref{TABfB}). In view of the large
errors, however, this should not be taken as evidence that the
continuum limit has essentially been reached already.

\begin{table}[h]
\tcaption{Summary of results for decay constants of charmed mesons.}
\begin{center}
\begin{tabular}{cllll}
\hline
\hline
Source & Author & $\fd\,[\mev]$ & $\fds\,[\mev]$ & $\fds/\fd$ \\
\hline
Lattice & This work & 191\err{19}{28} & 206\err{18}{28} & 1.08(8) \\
        & Martinelli\,\cite{guido_beauty_96},
        Flynn\,\cite{flynn_lat96,flynn_ICHEP_96}  & 205(15)
        & 235(15) & 1.15(5) \\
        & Sommer\,\cite{sommer_review} & 164(20) & & \\[1.0ex]
Sum       & Neubert\,\cite{neub_92} & 170(30) & & \\
Rules     & Dominguez\,\cite{dom_marb_93} & $139-234$ & $174-269$
 & 1.21(6) \\
          & Narison\,\cite{nar_92,nar_94} & 171(16) & 197(20)
 & 1.15(4) \\[1.0ex]
Quark     & Capstick+Godfrey\,\cite{capstick_90} & 240(20) & 290(20)
          & 1.21(13) \\
Models    & Oakes\,\cite{oakes_94} & & & 0.985 \\
          & Hwang+Kim\,\cite{mock1_95} & 253(65) & 265(68) & 1.045 \\
          & Hwang+Kim\,\cite{mock2_95} &  &  & 1.201 \\[1.0ex]
Exp.  & Average\,\cite{PDG_96,Rich_War_96} & $<310$
                & 241(21)(30) & \\
      & L3\,\cite{L3_fDs} &  & 309(58)(33)(38) & \\
\hline
\hline
\end{tabular}
\end{center}
\label{TfD_comp}
\end{table}
\begin{table}[bt]
\tcaption{Summary of results for decay constants of bottom mesons.}
\begin{center}
\begin{tabular}{cllll}
\hline
\hline
Source & Author & $\fb\,[\mev]$ & $\fbstat\,[\mev]$ & $\fbs/\fb$ \\
\hline
Lattice & This work & 172\err{27}{31} & 245\err{50}{47} & 1.14(8) \\
        & Martinelli\,\cite{guido_beauty_96},
        Flynn\,\cite{flynn_lat96,flynn_ICHEP_96}  & 175(25) &
        & 1.15(5) \\
        & Sommer\,\cite{sommer_review} & 180(46) & 276(37)
        & $1.10-1.18$ \\[1.0ex]
Sum       & Neubert\,\cite{neub_92} & 190(50) & $200-300$ & \\
Rules     & Eletskii+Shuryak\,\cite{ele_shur_92} & $150-175$ & $165-200$ & \\
          & Bagan\,\etal\,\cite{BBBD_92} &  & $195-245$ & \\
          & Dominguez\,\cite{dom_marb_93} & $133-183$ & & 1.22(2) \\
          & Narison\,\cite{nar_92,nar_94} & 209(34) & 259(41)
 & 1.16(4) \\[1.0ex]
Quark     & Capstick+Godfrey\,\cite{capstick_90} & 155(15) &  & 1.35(18) \\
Models    & Oakes\,\cite{oakes_94} & & & 0.989 \\
          & Hwang+Kim\,\cite{mock1_95} & 201(51) &  & 1.053 \\
          & Hwang+Kim\,\cite{mock2_95} &  &  & 1.173 \\[1.0ex]
QCD DR    & Boyd\,\etal\,\cite{BoyGrinLeb_94} & $<195$ & & \\
\hline
\hline
\end{tabular}
\end{center}
\label{TfB_comp}
\end{table}

We can now compare continuum lattice results for heavy-light deacy
constants (also using the values quoted in other recent global
analyses of lattice
data\,\cite{flynn_lat96,sommer_review,guido_beauty_96,flynn_ICHEP_96})
to the findings of other theoretical methods and experimental
results. In tables\,\ref{TfD_comp} and\,\ref{TfB_comp} we compare our
best estimates with those from QCD sum rules, potential models and
experimental results where available. The tables highlight the large
uncertainties which are present in all determinations of these
quantities so far. As mentioned before, lattice results suffer from
systematic effects due to quenching, uncertainties in the
extrapolation procedure and the treatment of heavy quarks. Estimates
obtained using QCD sum rules are rather sensistive to the choice of
input parameters in the evaluation, which in the past has sparked a
long controversy among different authors. More recent results from the
sum rule
approach\,\cite{neub_92}${}^{-}$\cite{dom_marb_93,nar_94,nar_95},
however, agree quite well with those obtained on the lattice. It
should be noted that QCD sum rule estimates for $f_{P_s}/f_P$ are
somewhat larger than the lattice values. It is, however, difficult to
estimate the real uncertainties associated with the results obtained
using QCD sum rules. An interesting result is quoted in
ref.\,\cite{BoyGrinLeb_94}, where an upper bound on $\fb$,
i.e. $\fb<195\,\mev$ is derived in a model-independent way, using
rigorous QCD dispersion relations.

Heavy-light decay constants are hard to measure experimentally, since
the leptonic width is small compared to the total width, and also
because the presence of a neutrino in the final state complicates the
analysis. Furthermore, the leptonic decay rate for a $B$~meson is
Cabibbo-suppressed, since it is proportional to $|V_{ub}|^2$, which
puts it beyond the reach of current experiments. For charmed mesons,
only an upper bound can be quoted for $f_D$\,\cite{PDG_96}
(c.f. table\,\ref{TfD_comp}), whereas leptonic decays of $D_s$ are
Cabibbo favoured, since their rate is proportional to
$|V_{cs}|^2$. Thus, $\fds$ has indeed been measured by several
experiments\,\cite{Rich_War_96}${}^{-}$\cite{L3_fDs}.

The observation of large $1/M$ corrections to
eq.\,(\ref{GLHQ_scaling}) in lattice calculations has been confirmed
by QCD sum
rules\,\cite{neub_92}${}^{-}$\cite{BBBD_92,nar_92,nar_93,PBall_94}. Sum
rule estimates for the coefficient~$C$ in eq.\,(\ref{GLPhi_param})
vary in the range 0.7--1.3\,\gev, in good agreement with the lattice
result quoted in subsection\,3.4. So far, only lattice calculations
using NRQCD without the axial current normalisation have produced
substantially larger corrections, such as $C=2.8(5)\,\gev$ quoted in
ref.\,\cite{SGO_decay_96}.

To summarise, we have shown that lattice calculations using unimproved
conventional and static heavy quarks have produced results for
heavy-light decay constants with a total accuracy of 15\% for charmed
and 30\% for bottom mesons. A number of systematic errors necessitate
further studies, as signified by the rather large errors. Preliminary
results from runs with dynamical quarks\,\cite{MILC_lat96,ROME2_96}
suggest that decay constants could increase by up to 10\% in the
unquenched theory. However, unlike some time ago, there is now
agreement between lattice calculations and QCD sum rules, and also the
few experimental results. It is of great importance to increase the
precision of lattice data, especially in view of future dedicated
experiments for $B$ physics.


\section{\bbar\ Mixing}
\label{SEKbbar}

We will now turn the discussion to lattice calculations of the
$B$~parameter $\bb$, which is relevant for \bbar\ mixing.  The lattice
estimates for $\bb$ and the decay constant $f_B$ from the previous
section can then be combined into the quantity $f_B\sqrt{\bb}$, which
is of particular interest, since it constrains the possible values of
the parameters $\rho$ and $\eta$ in the standard Wolfenstein
parametrisation of the CKM matrix. At the end of this section, we will
therefore attempt to quote a common estimate for this quantity using
data from several lattice calculations and study its phenomenological
implications in section\,5.

\subsection{The $\Delta B=2$ four-fermion operator on the lattice}
\label{SEKollatt}

In the continuum the interpolating operator for oscillations between a
$B^0$ and a $\overline{B}^0$ meson is the four-fermion operator $O_L$
defined by
\be
   O_L = \big(\overline{b}\gamma_\mu(1-\gamma_5)q\big)\,
         \big(\overline{b}\gamma_\mu(1-\gamma_5)q\big).
\ee
This operator enters the $\Delta B=2$ effective
Hamiltonian\,\cite{BuJaWe_90}. The amplitude of \bbar\ mixing is
usually expressed in terms of the $B$ parameter, which is the ratio of
the operator matrix element to its value in the so-called vacuum
insertion approximation
\be
\label{GLbbmu_def}
  \bb(\mu) \equiv
  \frac{\langle\overline{B^0}\left|\,O_L(\mu)\,\right|B^0\rangle}
                  {\frac{8}{3}\,|\langle0|A_4|B^0\rangle|^2}
= \frac{\langle\overline{B^0}\left|\,O_L(\mu)\,\right|B^0\rangle}
                  {\frac{8}{3}\,\fb^2\,M_B^2}.
\ee
Here, $\mu$ is a renormalisation scale, and $A_4$ is the temporal
component of the heavy-light axial current
$A_\mu=\overline{b}\gamma_\mu\gamma_5 q$. The dependence of $\bb$ on
the renormalisation scale can be removed by multiplication with a
factor derived from the anomalous dimension of $O_L$. At leading order
(LO) or next-to-leading order (NLO) respectively, a renormalisation
group invariant $B$ parameter can be defined by
\begin{eqnarray}
  \rgbb^{\rm LO}  &=& \alpha_s(\mu)^{-2/\beta_0}\bb(\mu) \label{GLbbLO}
  \\ 
  \rgbb^{\rm NLO} &=& \alpha_s(\mu)^{-2/\beta_0}
	\Big(1+\frac{\alpha_s(\mu)}{4\pi}\,J_{n_f}\Big)
        \bb(\mu), \label{GLbbNLO}
\end{eqnarray}
where $\beta_0=11-2n_f/3$ and $J_{n_f}$ is derived from the one- and
two-loop anomalous dimensions $\gamma_L^{(0)}$ and $\gamma_L^{(1)}$ of
the operator $O_L$. In the $\msbar$ scheme one
obtains\,\cite{BuJaWe_90}
\begin{eqnarray}
  & & J_{n_f} =
  \frac{1}{2\beta_0}\left(\gamma_L^{(0)}\frac{\beta_1}{\beta_0}-\gamma_L^{(1)}\right), \quad \beta_1=102-\frac{38}{3}n_f, \\
  & & \gamma_L^{(0)}=4, \qquad \gamma_L^{(1)}=-7+\frac{4}{9}n_f.
\end{eqnarray}
With these definitions $\rgbb$ is renormalisation group invariant only
up to leading order and next-to-leading order, respectively.

If the operator matrix element is evaluated on the lattice using
Wilson fermions, the contributions of the corresponding operators with
opposite and mixed chirality need also to be taken into account, as
was explained in subsection\,2.3. Since every operator has to be
renormalised separately, the matching between the lattice operators
and their continuum counterparts is potentially an intricate
problem. Hence, in the case of four-fermion operators, the importance
of a non-perturbative determination of the various matching factors is
even greater. Attempts in this direction have been reported in the
context of the $B$ parameter $B_K$\,\cite{mauro_lat95,Z_BK}.

Lattice results for the $B$ parameter $\bb$ have been
published recently using the static
approximation\,\cite{bbar,GimMar_bbar_96,Ken_bbar_96} and also
propagating heavy
quarks\,\cite{JLQCD_lat95,soni_lat95,BBS_lat96}. Earlier attempts are
discussed in refs.\,\cite{abada_92,BDHS88,ELC_stat_91}.

Here we shall describe in some detail the evaluation of $\bb$ in the
static approximation and the subsequent matching of the results in the
effective lattice theory to the $\Delta B=2$ Hamiltonian in the
continuum, which plays the r\^ole of the ``full'' theory, even though
it is an effective theory in itself. The matching is usually performed
in two steps (although there is in principle no need to do so), in
which one first matches the operator in full QCD at scale $\mu=m_b$ to
the continuum effective theory (i.e. HQET). In the second step one
matches the basis of local operators in the continuum effective theory
to those in the effective theory on the lattice. The following diagram
illustrates the two-step process and lists the operators and
renormalisation scales at every stage:
\par\noindent
\unitlength 1.06cm
\begin{picture}(12,3.5)
\put(0.0,1.0){
   \framebox(3.0,2.0){
     \parbox{26mm}{\sloppy
       {\bf Continuum}\\[0.3ex] $\Delta B=2$ effective\\
       Hamiltonian, $O_L(m_b)$
     }
  }
}
\put(4.5,1.0){
   \framebox(2.8,2.0){
     \parbox{26mm}{\sloppy
       {\bf Continuum}\\[0.3ex] HQET\\ scale $\mu$\\[0.4ex]
       $\widetilde{O}_L(\mu),\,\widetilde{O}_S(\mu)$
     }
  }
}
\put(8.8,1.0){
   \framebox(3.0,2.0){
     \parbox{26mm}{\sloppy
       {\bf Lattice}\\[0.3ex] static approx.\\ scale $a^{-1}$\\
       $\widehat{O}_L,\,\widehat{O}_R,\,\widehat{O}_N,\,\widehat{O}_S$
     }
  }
}
\thicklines
\put(3.2,2.0){\vector(1,0){1.3}}
\put(4.5,2.0){\vector(-1,0){1.3}}
\put(3.4,2.3){\mbox{step\,1}}
\put(7.5,2.0){\vector(1,0){1.3}}
\put(8.8,2.0){\vector(-1,0){1.3}}
\put(7.7,2.3){\mbox{step\,2}}
\end{picture}
\par\noindent
Thus, when relating the continuum full theory to the continuum
effective theory the operator $\widetilde{O}_S$ is generated at order
$\alpha_s$, owing to the mass of the heavy quark\,\cite{FHH91}
\be
   O_S = \big(\overline{b}(1-\gamma_5)q\big)\,
         \big(\overline{b}(1-\gamma_5)q\big).
\label{GLdefos}
\ee
Furthermore, due to explicit chiral symmetry breaking, the following
operators mix with $O_L$ in the discretised theory
\begin{eqnarray}
   O_R &=& \big(\overline{b}\gamma_\mu(1+\gamma_5)q\big)\,
           \big(\overline{b}\gamma_\mu(1+\gamma_5)q\big),
\label{GLdefor}\\
   O_N &=&  \big(\overline{b}\gamma_\mu(1-\gamma_5)q\big)\,
            \big(\overline{b}\gamma_\mu(1+\gamma_5)q\big), \nonumber\\
       & & \hspace{-2.5mm}
            +\big(\overline{b}\gamma_\mu(1+\gamma_5)q\big)\,
            \big(\overline{b}\gamma_\mu(1-\gamma_5)q\big), \nonumber\\
       & & \hspace{-2.5mm}
            +2\big(\overline{b}(1-\gamma_5)q\big)\,
            \big(\overline{b}(1+\gamma_5)q\big), \nonumber\\
       & & \hspace{-2.5mm}
            +2\big(\overline{b}(1+\gamma_5)q\big)\,
            \big(\overline{b}(1-\gamma_5)q\big).
\label{GLdefon}
\end{eqnarray}
The relation between the $B$ parameter in the full theory and the
relevant operator matrix elements in the effective lattice theory is
then given by
\be
\bb(m_b)=\frac{ \sum_{X} Z_X^{\rm stat}
  \Big\langle\overline{B^0}\left|\,\widehat{O}_X(a)\,\right|B^0\Big\rangle }
  {\frac{8}{3}\big(\zastat\langle0|A_4|B^0\rangle\big)^2},
\qquad X=L,R,N,S
\label{GLBBmb}
\ee
Here the matching factors $Z_X^{\rm stat}$ and $\zastat$ for the axial
current are the products of the renormalisation factors required for
steps\,1 and\,2, which are chosen such that all dependence on the
scales $\mu$ and $a$ in $\bb(m_b)$ vanishes. So far, only perturbative
estimates exist for these matching factors, and we refer the reader to
appendix\,A for explicit expressions. Here, we only
mention that a complete matching at order $\alpha_s$ can be performed,
since the two-loop anomalous dimensions for step\,1 have been computed
in the $\msbar$ scheme\,\cite{Gim_92}$^{-}$\cite{Buch_96}.

\subsection{Lattice results for $\bb$ and $\bbs$}

Results for the $B$ parameter $\bb$ are compiled in
table\,\ref{Tbbar}. In order to facilitate the comparison of different
calculations we have chosen to quote $\bb(m_b)$ at a reference scale
$m_b=5\,\gev$, although some authors prefer a different value. For
$\mu<5\,\gev$, this conversion has been performed at leading order
according to
\be
  \bb(m_b=5\,\gev) = \bb(\mu)\,\left(\frac{\alpha_s(\mu)}{\alpha_s(m_b)}
			\right)^{-2/\beta_0}.
\ee
We have used the expression for $\alpha_s(\mu)$ at leading order, viz.
\be
   \alpha_s^{\rm LO}(\mu) = \frac{4\pi}{\beta_0\ln(\mu^2/\Lambda^2)},
\ee
which was evaluated for $\Lambda=200\,\mev$ and $n_f=4$ active quark
flavours. Requiring $\alpha_s$ to be continuous at the threshold
$\mu=m_b$, we find $\alpha_s^{\rm LO}(m_b)=0.2342$. The value of
the renormalisation group invariant $B$ parameter $\rgbb^{\rm LO}$ in
table\,\ref{Tbbar} was obtained from eq.\,(\ref{GLbbLO}) using the
value of $\alpha_s^{\rm LO}(m_b)$. Similarly, $\rgbb^{\rm NLO}$ was
obtained from eq.\,(\ref{GLbbNLO}) by inserting the value of
$\alpha_s^{\rm NLO}(m_b)=0.1842$.

\begin{table}[tb]
\tcaption{Results for the $B$ parameter $\protect\bb(m_b)$
obtained in the quenched approximation. $\protect\rgbb^{\rm LO}$ and
$\protect\rgbb^{\rm NLO}$ are the renormalisation group invariant $B$
parameters obtained from $\protect\bb(m_b)$ as explained in the
text. Due to our particular choice of parameters, the results for
$\protect\rgbb$ may differ slightly from those quoted in the original
articles. The static approximation was used in
refs.\,\protect\cite{bbar,GimMar_bbar_96,Ken_bbar_96}. Furthermore,
UKQCD as well as the authors of\,\protect\cite{GimMar_bbar_96} use an
improved action ($\protect\csw=1$) for the light quark sector. 
}
\begin{center}
\begin{tabular}{lcllll}
\hline
\hline
Collab. & $\beta$ & $\bb(\mu)$ & $\bb(m_b)$ & $\rgbb^{\rm LO}$ &
$\rgbb^{\rm NLO}$ \\
\hline
Ken\,\cite{Ken_bbar_96} & 6.0 & 0.98(4) & 0.97(4) & 1.42(6)
 & 1.54(6) \\
 & & $\mu=4.33\,\gev$ & & & \\[0.8ex]
UKQCD\,\cite{bbar} & 6.2 & & 0.69\er{3}{4}\er{2}{1} &
1.02\er{5}{6}\er{3}{2} & 1.10\er{5}{6}\er{3}{2} \\
[0.8ex]
G+M\,\cite{GimMar_bbar_96} & 6.0 & & 0.63(4) & 0.92(6) & 1.00(6) \\
                               & & & 0.73(4) & 1.07(6) & 1.16(6) \\[0.8ex]
JLQCD\cite{JLQCD_lat95} & 6.3 & & 0.840(60) & 1.23(9) & 1.34(10) \\
                        & 6.1 & & 0.895(47) & 1.31(7) & 1.42(8)  \\[0.8ex]
B+S\,\cite{soni_lat95} & ``$\infty$'' & 0.96(6)(4) & 0.89(6)(4) &
1.30(9)(6) & 1.42(10)(6) \\
                       & & $\mu=2\,\gev$ & & & \\[0.8ex]
ELC\,\cite{abada_92} & 6.4 & 0.86(5) & 0.84(5) & 1.24(7) & 1.34(8) \\
                     &  & $\mu=3.7\,\gev$ & & & \\[0.8ex]
BDHS\,\cite{BDHS88} & 6.1 & 1.01(15) & 0.93(14) & 1.36(20) &
1.48(22) \\
                    & & $\mu=2\,\gev$ & & & \\

\hline
\hline
\end{tabular}
\end{center}
\label{Tbbar}
\end{table}

By comparing the results for $\bb(m_b)$ in the table one observes a
cluster of values in the range $0.80-0.90$, as well as a few results
clearly above and below, despite the fact that similar lattice
spacings have been used. As was first noted in ref.\,\cite{bbar},
ambiguities in the perturbative matching procedure are largely
responsible for this. As the matching factors $\zastat$, $Z_X^{\rm
stat}$, $X=L,R,N,S$ in eq.\,(\ref{GLBBmb}) are themselves products of
matching factors known to $O(\alpha_s)$, the ratios $Z_X^{\rm
stat}/(\zastat)^2$ contain some contributions of order
$\alpha_s^2$. By including or excluding these next-to-next-to-leading
order terms, the following three possibilities can be applied:
\begin{itemize}
\item[(${\rm M}_1$)] include $O(\alpha_s^2)$ contributions by using the
ratios $Z_X^{\rm stat}/(\zastat)^2$.
\item[(${\rm M}_2$)] exclude $O(\alpha_s^2)$ contributions in the
individual $Z$-factors by expanding $Z_X^{\rm stat}$ and $\zastat$
separately to order $\alpha_s$. 
\item[(${\rm M}_3$)] exclude $O(\alpha_s^2)$ contributions by
expanding the ratios $Z_X^{\rm stat}/(\zastat)^2$ to
order\,$\alpha_s$. 
\end{itemize}
In ref.\,\cite{Ken_bbar_96} it was suggested that the low values for
$\bb(m_b)$ reported by UKQCD were due to using procedure $({\rm
M}_2)$, and that therefore the large discrepancy between
refs.\,\cite{Ken_bbar_96} and\,\,\cite{bbar} could largely be
explained by the amplification of ambiguities by employing a
particular matching procedure, rather than the different values of
$\csw$ or $\beta$ in the two simulations. In view of this issue, we
wish to test the consistency of results from different groups, by
listing in table\,\ref{Tbbar_comp} the values for $\bb(m_b)$ from
refs.\,\cite{bbar,GimMar_bbar_96,Ken_bbar_96}, which were obtained by
applying methods $({\rm M}_1)$, $({\rm M}_2)$ and $({\rm M}_3)$ for
uniform choices of input parameters. Indeed it turns out that method
$({\rm M}_2)$ systematically produces lower values of
$\bb(m_b)$. However, for each common procedure the discrepancy between
the results for $\csw=0$ (ref.\,\cite{Ken_bbar_96}) and $\csw=1$
(refs.\,\cite{bbar,GimMar_bbar_96}) cannot fully be resolved, so that
other effects must also contribute to the observed differences.

One concludes that, although different prescriptions should be
equivalent in principle, there are discrepancies of up to 25\% in the
final values of $\bb(m_b)$ for $\csw=1$ and even larger variations for
unimproved Wilson fermions. As each of the methods $({\rm M}_1) -
({\rm M}_3)$ is perfectly justified, these differences may be viewed
as a systematic error on $\bb$ as a result of perturbative
matching. As far as lattice results for $\bb$ in the static
approximation are concerned, these ambiguities do not arise
exclusively from the discretisation of the theory, since the matching
step between the continuum full and continuum effective theories
contains a large $O(\alpha_s)$ contribution (see also
appendix\,A). This problem can only be adressed in a
two-loop calculation for this matching step.

\begin{table}[tb]
\tcaption{Results for $\protect\bb(m_b)$ in the static approximation
for matching procedures $({\rm M}_1)-({\rm M}_3)$, using results
for the individual matrix elements from different groups. The
couplings $\alpha_s(\mu)$, $\alpha_s(m_b)$ were evaluated using
$m_b=5\,\gev$, $n_f=4$, $\Lambda^{(4)}=200\,\mev$. Only statistical
errors are shown. Tadpole improvement has not been applied, except
that a mean field improved coupling $\alpha_V(q^*)$ at $aq^*=2.18$
has been used.}
\begin{center}
\begin{tabular}{cllll}
\hline
\hline
method & $\alpha_s(m_b)$ & UKQCD\,\cite{bbar} &
G+M\,\cite{GimMar_bbar_96} & Ken\,\cite{Ken_bbar_96} \\
\hline
$({\rm M}_1)$ & 0.2342 & 0.81(4) & 0.82(3) & 0.87(3) \\
              & 0.1842 & 0.80(5) & 0.81(3) & 0.85(3) \\[0.8ex]
$({\rm M}_2)$ & 0.2342 & 0.60(4) & 0.59(3) & 0.42(3) \\
              & 0.1842 & 0.64(5) & 0.64(3) & 0.51(3) \\[0.8ex]
$({\rm M}_3)$ & 0.2342 & 0.87(5) & 0.83(2) & 0.96(3) \\
              & 0.1842 & 0.88(5) & 0.85(3) & 0.98(3) \\

\hline
\hline
\end{tabular}
\end{center}
\label{Tbbar_comp}
\end{table}

Different procedures should eventually yield the same answer in the
continuum limit, provided that lattice artefacts are not so large as
to spoil the extrapolation $a\to0$. Clearly, the currently available
data for $\bb$ do not allow for such an extrapolation as in the case
of decay constants discussed in the previous section. However,
preliminary results in the continuum limit for $\bb$ and some
SU(3)-flavour breaking ratios involving the $B$ parameter have been
reported\,\cite{soni_lat95,BBS_lat96}. Also, by comparing the results
from refs.\,\cite{bbar,GimMar_bbar_96}, which were both obtained for
$\csw=1$, one observes only a weak dependence on the lattice
spacing~$a$.

In the absence of further data which might clarify the issues of
matching as well as the influence of lattice artefacts and the size of
$1/M$ corrections to the results in the static approximation, we
resort to quoting a central result for $\bb$ with an error which
encompasses the observed variations. Our best estimate is
\be
   \bb(m_b) = 0.85\err{13}{22},
\ee
which corresponds to 
\be
   \rgbb^{\rm LO}  = 1.24\err{18}{32},\qquad 
   \rgbb^{\rm NLO} = 1.35\err{21}{35},
\ee
so that our final estimate for the renormalisation group invariant $B$
parameter can be summarised as
\be
    \rgbb = 1.3\er{2}{3}.
\ee
These numbers can be compared with the results from QCD sum rules. The
authors of ref.\,\cite{NarPiv_94} obtain $\bb(5\,\gev)=0.99(15)$,
which, in view of the large errors, is in agreement with our estimate.
An earlier attempt to calculate the matrix element of $O_L$ in the sum
rule approach resulted in a very low value\,\cite{pich_88}.

We can combine our value for $\rgbb$ with the result for the
decay constant $f_B$ in the previous section. Using
$\fb=172\err{22}{31}\,\mev$ from table\,\ref{TfB_comp}, we find
\be
   \fb\sqrt{\rgbb} = 195\err{30}{40}\,\mev.
\ee
This is one of the central results of this review and can be taken as
the present estimate for this quantity treating as many individual
results as possible on an equal footing.

Now we turn the discussion to SU(3)-flavour breaking ratios involving
the $B$ parameter and the decay constant. As becomes clear from
table\,\ref{Tbbsbd} the $B$ parameter shows very little dependence on
the mass of the light quark, quite in contrast to the decay
constant. The results for $\bbs/\bbd$ can therefore be summarised by
the value
\be
    \bbs/\bbd = 1.00(2).
\label{GLbbsbd}
\ee
The result can be combined with our estimate of $\fbs/\fbd=1.14(8)$ in
table\,\ref{TfB_comp} and eq.\,(\ref{GLfPsPd}) to compute the ratio
\be
   \xisd^2 = \frac{\fbs^2\bbs}{\fbd^2\bbd} = 1.30(18).
\label{GLxisd_hw}
\ee
More results for $\xisd^2$, which have not been used in the derivation
of this result, are listed in table\,\ref{Tbbsbd}. It is worth noting
that apart from the estimates for $\xisd^2$ in the static
approximation (ref.\,\cite{bbar,GimMar_bbar_96}), there is also a
preliminary result\,\cite{BBS_lat96} which treats heavy quarks in the
conventional approach. Moreover, in that study $\xisd^2$ was computed
directly from the ratio of operator matrix elements, viz.
\be
   r_{sd} \equiv
\frac{\langle\overline{B_s^0}\big|\,\widehat{O}_L\,\big|B_s^0\rangle}
{\langle\overline{B_d^0}\big|\,\widehat{O}_L\,\big|B_d^0\rangle}
 = \xisd^2\frac{M_{B_s}}{M_{B_d}}
\ee
over a range of lattice spacings.  The results in table\,\ref{Tbbsbd}
and their statistical errors can be combined in a weighted average to
yield
\be
  \xisd^2= 1.38(15),
\label{GLxisd_lat}
\ee
where we have added in quadrature a systematic error estimated from
the spread of central values in the table. 

\begin{table}[tb]
\tcaption{Results for the SU(3)-flavour breaking ratio 
$\protect\bbs/\protect\bbd$ and the ratio $\protect\xisd^2$ defined
in eq.\,(\protect\ref{GLxisd_hw}). Note that
$\protect\rgbbs/\protect\rgbbd=\protect\bbs(m_b)/\protect\bbd(m_b)$,
and thus this ratio is independent of the details of the conversion to
the renormalisation group invariant $B$ parameter.}
\begin{center}
\begin{tabular}{lclll}
\hline
\hline
Collab. & $\beta$ & $\bbs/\bbd$ & $\xisd^2$ & Comments \\
\hline
Ken\,\cite{Ken_bbar_96} & 6.0 & 0.99\er{1}{1}(1) & & static, $\csw=0$ \\[0.8ex]
G+M\,\cite{GimMar_bbar_96} & 6.0 & 1.01(1) & 1.38(7) & static, $\csw=1$ \\[0.8ex]
UKQCD\,\cite{bbar} & 6.2 & 1.02(2) & 1.34\er{9}{8}\er{5}{3} 
& static, $\csw=1$ \\[0.8ex]
JLQCD\,\cite{JLQCD_lat95} & 6.3 & 1.05(3) &  & conv., $\csw=0$ \\
                        & 6.1 & 0.99(3) &  & conv., $\csw=0$ \\[0.8ex]
BBS\,\cite{BBS_lat96} & ``$\infty$'' &  & 1.49(13)(31) & conv., $\csw=0$ \\[0.8ex]
this work &  &  & 1.30(18)  & eqs.\,(\ref{GLfPsPd},\,\ref{GLbbsbd}) \\
\hline
\hline
\end{tabular}
\end{center}
\label{Tbbsbd}
\end{table}

\section{Analysis of CP violation}
\label{SEKCPviol}

In this section we shall combine lattice results with experimental
data and other theoretical estimates to place constraints on the most
poorly known elements of the CKM matrix and on the unitarity triangle.

\subsection{Parametrisations of the CKM matrix and the unitarity
triangle}

We start the discussion by recalling the basic definitions and
relations involving the CKM matrix. As is well known, the CKM matrix
$\Vckm$ relates the gauge eigenstates appearing in the SM Lagrangian
to mass eigenstates. For flavour-changing charged current transitions
this has the consequence that, in addition to the dominant transitions
between up- and down-type quarks, $u\leftrightarrow d$,
$c\leftrightarrow s$ and $t\leftrightarrow b$, there are further
transitions of lesser strength. The corresponding Lagrangian
describing flavour-changing charged current interactions in the
hadronic sector has the form
\be
{\cal L}_{\rm CC}^{\rm had}=-\frac{g}{\sqrt{8}}
   (\bar{u},\bar{c},\bar{t})\gamma^\mu(1-\gamma_5)\,\Vckm\,
   \left(\begin{array}{c}
	d\\ s \\ b \\ 
   \end{array}\right)\,W_\mu^\dagger + {\rm h.c.},
\ee
where the CKM matrix
\be
\Vckm=
\left(\begin{array}{ccc}
V_{ud} & V_{us} & V_{ub} \\
V_{cd} & V_{cs} & V_{cb} \\
V_{td} & V_{ts} & V_{tb} \\
\end{array}\right)
\ee
is a unitary matrix in flavour space. Since the dominant transitions
are proportional to the diagonal matrix elements $V_{ud}$, $V_{cs}$
and $V_{tb}$, one expects a hierarchical structure of $\Vckm$. An
approximate parametrisation of the CKM matrix which takes account of this
structure is due to Wolfenstein\,\cite{Wolf_83}. By expanding $\Vckm$
in powers of the Cabibbo angle $|V_{us}|=\lambda\simeq0.22$ to order
$\lambda^3$ one obtains
\be
\Vckm\simeq\left(\begin{array}{ccc}
1-\frac{\lambda^2}{2} & \lambda & A\lambda^3(\rho-i\eta) \\
-\lambda & 1-\frac{\lambda^2}{2} & A\lambda^2 \\
A\lambda^3(1-\rho-i\eta) & -A\lambda^2 & 1 \\
\end{array}\right),
\ee
with the remaining parameters $A$, $\rho$ and $\eta$ of order one.

In the case of three generations of quarks and leptons, $\Vckm$ can be
parametrised in terms of three angles and one observable complex
phase, of which the latter is required to describe CP~violation. As
has been pointed out in ref.\,\cite{BuLauOs_94}, a consistent
treatment of CP-violating effects necessitates the inclusion of
higher terms in the Wolfenstein parametrisation, viz.
\be
\Vckm\simeq\left(\begin{array}{ccc}
1-\frac{\lambda^2}{2} & \lambda & A\lambda^3(\rho-i\eta) \\
-\lambda-iA^2\lambda^5\eta & 1-\frac{\lambda^2}{2} & A\lambda^2 \\
A\lambda^3(1-\overline{\rho}-i\overline{\eta}) 
& -A\lambda^2-iA\lambda^4\eta & 1 \\
\end{array}\right),
\ee
where the rescaled parameters $\overline{\rho}$ and $\overline{\eta}$
are expanded as
\be
\overline{\rho}=\rho\Big(1-\frac{\lambda^2}{2}+O(\lambda^4)\Big),\qquad
\overline{\eta}=\eta\Big(1-\frac{\lambda^2}{2}+O(\lambda^4)\Big).
\ee

The relative size of CKM matrix elements is easily recognised in the
Wolfenstein parametrisation: the diagonal elements are of order one,
and $|V_{us}|$, $|V_{cd}|$ are both of order 20\,\%. The relative size
of $|V_{cb}|$ and $|V_{ts}|$ are 4\,\%, whereas $|V_{ub}|$ and
$|V_{td}|$ are of order 1\,\%.

The Wolfenstein parameters $\lambda$ and $A$ are rather well
determined experimentally\,\cite{PDG_96,Gibbons_ICHEP96,Czar_96,alilon_96}
\begin{eqnarray}
\lambda & = & |V_{us}| = 0.2205\pm0.0018 \\
A & = & \left|\frac{V_{cb}}{V_{us}^2}\right| = 0.81\pm0.058,
\end{eqnarray}
whereas the elements $|V_{ub}|$ and $|V_{td}|$ have uncertainties of
around 30\,\%, so that the parameters $\rho$ and $\eta$ are rather
poorly known.

The unitarity of $\Vckm$ imposes the following conditions on its
elements
\be
    V_{ij}\,V_{ik}^* = 0, \qquad j\not=k.
\ee
There are six such conditions, each of which can be represented
graphically by a triangle. The triangle relation for the most poorly
known CKM matrix elements reads
\be
   V_{ud}V_{ub}^* + V_{cd}V_{cb}^* + V_{td}V_{tb}^* = 0,
\label{GLtriangle}
\ee
and its representation is shown in figure\,\ref{Ftriangle}.

\begin{figure}[tb]
\begin{center}
\unitlength 1.06cm
\begin{picture}(12,3.5)
\thicklines
\put(0.7,1.0){\line(3, 4){1.5}}
\put(2.2,3.0){\line(3,-2){3.0}}
\put(0.7,1.0){\line(1, 0){4.5}}
\put(2.2,0.5){\makebox(1.0,0.5)[b]{$V_{cd}V_{cb}^*$}}
\put(0.3,2.0){\makebox(1.0,0.5)[l]{$V_{ud}V_{ub}^*$}}
\put(3.9,2.0){\makebox(1.0,0.5)[l]{$V_{td}V_{tb}^*$}}
\put(0.8,1.0){\makebox(0.5,0.5)[r]{$\gamma$}}
\put(1.85,2.4){\makebox(0.5,0.5)[r]{$\alpha$}}
\put(4.0,1.0){\makebox(0.5,0.5)[r]{$\beta$}}
\put(6.7,1.0){\line(3, 4){1.5}}
\put(8.2,3.0){\line(3,-2){3.0}}
\put(6.7,1.0){\line(1, 0){4.5}}
\put(6.2,0.5){\makebox(1.0,0.5)[b]{(0,0)}}
\put(10.7,0.5){\makebox(1.0,0.5)[b]{(1,0)}}
\put(7.7,3.2){\makebox(1.0,0.5)[b]{$(\overline{\rho},\overline{\eta})$}}
\put(6.5,2.0){\makebox(1.0,0.5)[l]{$\overline{\rho}+i\overline{\eta}$}}
\put(9.9,2.0){\makebox(1.0,0.5)[l]{$1-\overline{\rho}-i\overline{\eta}$}}
\put(6.8,1.0){\makebox(0.5,0.5)[r]{$\gamma$}}
\put(7.85,2.4){\makebox(0.5,0.5)[r]{$\alpha$}}
\put(10.0,1.0){\makebox(0.5,0.5)[r]{$\beta$}}
\end{picture}
\end{center}
\fcaption{The unitarity triangle corresponding to
eq.\,(\protect\ref{GLtriangle}) and its rescaled version obtained in
the $(\overline{\rho},\overline{\eta})$-plane by diving all sides by
$V_{cd}V_{cb}^*$.} 
\label{Ftriangle}
\end{figure}

In order to constrain the values of $\rho$ and $\eta$ one needs
information on $|V_{td}|$, which can be extracted from $\bbar$
mixing. The expression for the mass difference $\Delta M_d$ between
the $B^0_d$ and $\overline{B^0_d}$ states reads
\be
   \Delta M_d = \frac{G_F^2\,M_W^2}{6\pi^2} \eta_{B_d}S(m_t/M_W)
   \fbd^2\rgbbd\,|V_{td}V_{tb}^*|^2,
\label{GLxd}
\ee
where $G_F$ is the Fermi constant, and $M_W$ is the mass of the
$W$-boson. The quantity $\eta_{B_d}=0.55\pm0.01$\,\cite{BuJaWe_90}
parametrises QCD short-distance corrections, and $S(m_t/M_W)$ is a
slowly varying function of $m_t/M_W$\,\cite{Inami_Lim,Buras_81}. In
the following we shall evaluate the above expression using our lattice
estimate for $\fbd\sqrt{\rgbbd}$.

\subsection{Constraints on the Wolfenstein parameters $\rho$ and
$\eta$}

Using lattice data in conjunction with experimental data and other
theoretical estimates, we now want to analyse $\bbard$ and $\bbars$
mixing in order to place constraints on the parameters $\rho$ and
$\eta$ discussed in the last subsection.

As is obvious from eq.\,(\ref{GLxd}), the quantity $\fbrootb{d}$ plays
a central r\^ole in this study. Further input parameters are the
running mass of the top quark, $\overline{m}_t(m_t)$, the
short-distance corrections $\eta_{B_d}$ and the experimentally
measured value of $\Delta M_d$. For $\bbars$ mixing, the quantity
$\xi_{sd}^2$ is also required. A compilation of input parameters for
this study is listed in table\,\ref{Tinput}. 

\begin{table}[b]
\tcaption{Input parameters for the study of $\rho$ and $\eta$.}
\begin{center}
\begin{tabular}{cll}
\hline
\hline
Quantity & Value & Source \\
\hline
$\fbrootb{d}$   &  $195\err{30}{40}\,\mev$  & Lattice, this work  \\
$\xi_{sd}^2$   &  $1.38\pm0.15$            & Lattice, this work  \\
$M_{B_d}$      &  $5279\,\mev$             & Exp.\,\cite{PDG_96} \\
$M_{B_s}$      &  $5369\,\mev$             & Exp.\,\cite{PDG_96} \\
$\Delta M_d$   &  $0.464\pm0.018\,{\rm ps}^{-1}$ &
Exp.\,\cite{Gibbons_ICHEP96}\\ 
$\overline{m}_t(m_t)$ &  $165\pm9\,\gev$   &
Exp.\,\cite{Tipton_ICHEP96,alilon_96} \\
\hline
\hline
\end{tabular}
\end{center}
\label{Tinput}
\end{table}

Solving for $|V_{td}|$ in eq.\,(\ref{GLxd}) and inserting the relevant
input parameters as well as $\eta_B=0.55\pm,0.01$, one finds
\be
   |V_{td}|=\big(8.94\,{}^{+1.43}_{-1.88}\big)\times10^{-3},
\label{GLvtd}
\ee
and thus the relative uncertainty in $|V_{td}|$ according to this
estimate amounts to $15 - 20$\,\%. The results for $|V_{td}|$ can be
translated into a constraint on $\overline{\rho}$ and
$\overline{\eta}$ via the unitarity triangle
\be
\sqrt{(1-\overline{\rho})^2+\overline{\eta}^2} =
\frac{1}{\lambda}\left|\frac{V_{td}}{V_{cb}}\right| =
1.03\,{}^{+0.18}_{-0.23}, 
\ee
where we have used the present determination of
$|V_{cb}|$\,\cite{Gibbons_ICHEP96,Czar_96} 
\be
   |V_{cb}| = 0.0393\pm,0.0028.
\ee
Further constraints on $\rho$ and $\eta$ can be obtained from indirect
CP violation in the \mbox{$K^0$--$\overline{K^0}${}} system through a
theoretical analysis of the parameter
$\epsilon_K$\,\cite{HerrNier_95,HerrNier_96,Herr_ICHEP96}. This
requires as additional input an estimate for the kaon $B$ parameter
$\rgbk$. Several theoretical methods, including lattice
simulations,\cite{soni_lat95,BarBuGe_88}$^{-}$\cite{sharpe_lat96}
quote values for $\rgbk$ which are compatible with the range
\be
   \rgbk=0.75\pm0.10.
\ee
There has been considerable progress in calculating $\rgbk$ on the
lattice\,\cite{sharpe_lat96}. However, a number of systematic effects
such as the continuum limit, the chiral behaviour of $\rgbk$ computed
using Wilson fermions, and the dependence on the number of flavour
require further
studies\,\cite{JLQCD_BK1_lat96}$^{-}$\cite{KilPekVen_lat96}. A more
detailed discussion of lattice results for $\rgbk$ is, however, beyond
the scope of this review.

Herrlich and Nierste\,\cite{HerrNier_96} have obtained constraints on
$\overline{\rho}$ and $\overline{\eta}$ from a theoretical analysis of
$\epsilon_K$ using the complete $\Delta S=2$ effective hamiltonian at
next-to-leading order. Using very similar estimates for $\fbrootb{d}$
and $\xi_{sd}^2$ (i.e. $\fbrootb{d}=200\pm40\,\mev$,
$\xi_{sd}^2=1.32\pm0.12$) compared to this review, they find
\be
   -0.20 \leq \overline{\rho} \leq 0.22,\qquad
    0.25 \leq \overline{\eta} \leq 0.43,
\label{GLrhoeta}
\ee
in good agreement with a similar study by Ali and
London\,\cite{alilon_95,alilon_96}. 

Now we turn our attention to $\bbars$ mixing. The ratio of mass
differences for the $\bbard$ and $\bbars$ systems is
\be
   \frac{\Delta M_s}{\Delta M_d} = 
   \frac{\eta_{B_s}}{\eta_{B_d}}\frac{\fbs^2\rgbbs}{\fbd^2\rgbbd}
   \frac{M_{B_s}}{M_{B_d}}\frac{|V_{ts}|^2}{|V_{td}|^2}.
\ee
Since $\eta_{B_s}=\eta_{B_d}$\,\cite{Buras_BEAUTY95} this becomes
\begin{eqnarray}
   \frac{\Delta M_s}{\Delta M_d} & = & \xi_{sd}^2 
   \frac{M_{B_s}}{M_{B_d}}\frac{|V_{ts}|^2}{|V_{td}|^2} \nonumber\\
   & = & (1.40\pm0.15) \frac{|V_{ts}|^2}{|V_{td}|^2},
\end{eqnarray}
where we have used the parameters listed in table\,\ref{Tinput}. This
result can be used together with the measured value of $\Delta M_d$ to
predict $\Delta M_s$, provided that the allowed range of
$|V_{ts}|^2/|V_{td}|^2$ is determined. This has been performed by Ali
and London\,\cite{alilon_96}, who have analysed $|V_{td}|/|V_{ts}|$ as
a function of $\fbrootb{d}$, assuming that $\rgbk=0.75\pm0.10$. Using
our estimate of $\fbrootb{d}=195\err{30}{40}\,\mev$ one reads off the
allowed range of $|V_{td}|/|V_{ts}|$ from figure\,4 in
ref.\,\cite{alilon_96}
\be
   \frac{|V_{td}|}{|V_{ts}|}=0.228\pm0.040\,{}^{+0.050}_{-0.035},
\ee
where the first error is due to the uncertainty in $\rgbk$, and the
second reflects the error on $\fbrootb{d}$. This yields
\be
  \Delta M_s = 12.5\pm0.48\pm1.3\pm4.4\,{}^{+5.5}_{-3.8}\,{\rm ps}^{-1}.
\ee
Here the first error is due to the experimental error on $\Delta M_d$,
the second due to the error in $\xi_{sd}^2$, and the third and fourth
to the uncertainties in the matrix elements $\rgbk$ and $\fbrootb{d}$,
respectively. Adding the errors in quadrature one gets
\be
   \Delta M_s = 12.5\,{}^{+7.2}_{-6.0}\,{\rm ps}^{-1}.
\ee
Using the updated results for the $B_s$ lifetime,
$\tau_{B_s}=1.61\pm0.10\,\rm ps$\,\cite{PDG_96}, this corresponds to a
value for the $\bbars$ mixing parameter $x_s$ of
\be
   x_s = 20.1^{+11.6}_{-~9.7}.
\ee
Our estimates for $\Delta M_s$ and $x_s$ can be compared to the
experimental lower bound of
\be
   \Delta M_s > 7.8\,{\rm ps}^{-1}, \quad x_s>12.6 \qquad {\rm
   (95\,\%~C.L.)}
\ee
quoted by the ALEPH
Collaboration\,\cite{Gibbons_ICHEP96,Zeitnitz_96,ALEPH_ICHEP96}. This
indicates that, although the experimental lower bound is still smaller
than the central value of the SM prediction, experiments start to
exclude parts of the allowed range of $\Delta M_s$. Therefore,
experimental data for $\Delta M_d$ and $\Delta M_s$ can further
restrict the allowed range of $(\overline{\rho},\,\overline{\eta})$, a
fact that has already been exploited in deriving the constraints in
eq.\,(\ref{GLrhoeta})\,\cite{Herr_ICHEP96}.

The constraints on CKM parameters have sharpened significantly in
recent years, not least thanks to more accurate and consistent
determinations of weak matrix elements using lattice techniques.

\section{Concluding Remarks}

In this review we have combined a variety of lattice results for
heavy-light decay constants and the $B$ parameter $\bb$ into a common
estimate for $\fb\sqrt{\bb}$. A central part of this study was the
analysis of the continuum limit of decay constants and the assessment
of systematic errors. Our result for $\fb\sqrt{\bb}$ has been combined
with other theoretical and experimental input to constrain some of the
CKM matrix elements. Although the overall errors on lattice results in
the heavy quark sector are still rather large, the progress in lattice
calculations over the past years has helped a lot in order to sharpen
the bounds on CKM matrix elements and the unitarity triangle.

The interplay between different theoretical tools plays an important
r\^ole in this analysis, as it serves to extract a consistent picture
in situations where each method has its own limitations. An example
for this is the study of $1/M$ corrections to the scaling law for
decay constants. Lattice simulations are currently limited by a number
of systematic effects, such as quenching, which cannot fully be
addressed at present due to lack of resources. However, in order to
increase the precision of lattice calculations, not only larger
computing power but also new theoretical ideas such as
non-perturbative improvement, are of great importance.

Here we have concentrated on leptonic decays of heavy quarks. This is
by no means the only area where lattice calculations have made an
impact on the determination of CKM matrix elements. Indeed, another
poorly known element, $V_{ub}$, can be determined by considering
semi-leptonic $B$ decays, such as $B\rightarrow\pi\ell\nu_\ell$ or
$B\rightarrow\rho\ell\nu_\ell$. Lattice simulations offer
model-independent estimates for the relevant form factors, and a
variety of calculations has already been
performed\,\cite{flynn_lat96}. Likewise, lattice results for rare
decays such as $B\rightarrow K^*\gamma$ have been reported in the past
few years. All of these calculations serve ultimately to test the
consistency of the Standard Model, and it is evident that the lattice
is a versatile tool in these analyses.

With the advent of dedicated experimental facilities for $B$~physics,
such as HERA-B, BaBar and LHC-B, together with an improved theoretical
understanding of heavy quark physics, the scene is set for an exciting
new period in particle physics.

\nonumsection{Acknowledgements}
I would like to thank my colleagues in the UKQCD Collaboration, and in
particular Chris Sachrajda for the fruitful collaboration on some of
the topics reviewed in this work. Furthermore, I would like to thank
Dina Alexandrou, Chris Allton, Claude Bernard, Jonathan Flynn, Shoji
Hashimoto, Fred Jegerlehner, Rainer Sommer and Mike Teper for
interesting discussions. Special thanks go to Chris Allton, Claude
Bernard and Shoji Hashimoto for communicating some of their results
prior to publication, and to Vicente Gim\'enez for useful exchanges
about matching factors for the $B$ parameter. The support of the
Particle Physics and Astronomy Research Council (PPARC) through the
award of an Advanced Fellowship is gratefully acknowledged.


\appendix{Matching Factors for the Static Approximation}
\label{APPZfactors}

In this appendix we list the expressions for the perturbative matching
factors for the heavy-light axial current and the $\Delta B=2$
four-fermion operator in the static approximation. As described in
subsection\,4.1, the matching is usually performed in a
two-step process, where one first matches the full theory in the
continuum to the effective continuum theory, which is then matched to
the effective theory on the lattice.

Throughout this appendix we use the following convention for the one-
and two-loop coefficients of the perturbative $\beta$ function
\be
    \beta_0=11-\frac{2}{3}n_f,\qquad \beta_1=102-\frac{38}{3}n_f,
\ee
where $n_f$ is the number of active quark flavours.

\subappendix{The axial current}

A full $O(\alpha_s)$ matching between the axial current $A_\mu$ in
full QCD and its counterpart $\widetilde{A}_\mu$ in the continuum
effective theory can be performed, since the two-loop anomalous
dimension of $\widetilde{A}_\mu$ has been computed\,\cite{Gim_92}. For
typical hadronic scales $\mu<m_b$, and using the na\"\i{}ve
dimensional regularisation scheme one
obtains\,\cite{eichten_hill_90_1,Gim_92}
\be
A_\mu = \left(\frac{\amb}{\amu}\right)^{d_{1,A}}
  \left\{1+\frac{\amu-\amb}{4\pi}\,J_A+\frac{\amb}{4\pi}\,C_2\right\}\,
\widetilde{A}_\mu,
\ee
where
\be
    C_2=-8/3
\ee
and 
\begin{eqnarray}
  & & d_{1,A}=\frac{\gamma_A^{(0)}}{2\beta_0},\qquad
      J_A=\frac{\gamma_A^{(0)}}{2\beta_0}\left(\frac{\beta_1}{\beta_0}
               -\frac{\gamma_A^{(1)}}{\gamma_A^{(0)}}\right), \\
  & & \gamma_A^{(0)}=-4,\qquad
      \gamma_A^{(1)}=-\frac{254}{9}-\frac{56}{27}\pi^2
      +\frac{20}{9}\,n_f.
\end{eqnarray}
The general expression for the matching of the current
$\widetilde{A}_\mu$ to the current $\widehat{A}_\mu$ in the effective
theory on the lattice is
\be
\widetilde{A}_\mu=\left(1+\frac{\alpha_s^{\rm latt}(\mu)}{4\pi}\big[
2\ln(\mu^2a^2)+D_A\big]\right)\,\widehat{A}_\mu.
\label{GLefflatt}
\ee
Numerical values of $D_A$ depend on the choice of $\csw$ and are
tabulated in table\,\ref{Tmatcoeff}.

Eq.\,(\ref{GLefflatt}) is usually evaluated using a tadpole improved
or boosted value for the lattice coupling $\alpha_s^{\rm latt}$, for
instance $\widetilde{\alpha}_s=\alpha_s^{\rm latt}/u_0^4$, where $u_0$
is the average link, $u_0=\langle\frac{1}{3}\,{\rm
Re\,Tr}\,P\rangle^{1/4}$, or alternatively
$u_0=1/(8\kcrit)$. Combining measured values for $u_0$ with its
perturbation expansion, viz.
\be
   u_0=1+u_0^{(1)}\alpha_s+O(\alpha_s^2)
\ee
one can write down the tadpole improved version of eq.\,(\ref{GLefflatt})
\be
\widetilde{A}_\mu=
\sqrt{u_0}\,\left(1+\frac{\widetilde{\alpha}_s(\mu)}{4\pi}\big[
2\ln(\mu^2a^2)+D_A-\textstyle\frac{1}{2}u_0^{(1)}\big]\right)\,
\widehat{A}_\mu.
\ee
The square root of $u_0$ has to be taken since the static-light axial
current contains only one relativistic quark field. The matching
factor between $\widetilde{A}_\mu$ and $\widehat{A}_\mu$ could in
principle also be determined non-perturbatively using the method in
ref.\,\cite{MPSTV_95}. Combining the factors for the two matching
steps one finds
\begin{eqnarray}
  \zastat &=&
\left(\frac{\amb}{\amu}\right)^{d_{1,A}}
  \left\{1+\frac{\amu-\amb}{4\pi}\,J_A-\frac{8}{3}\frac{\amb}{4\pi}\right\}
  \nonumber\\ 
  & & \times \sqrt{u_0}\left(1+\frac{\widetilde{\alpha}_s(\mu)}{4\pi}\big[
2\ln(\mu^2a^2)+D_A-\textstyle\frac{1}{2}u_0^{(1)}\big]\right).
\end{eqnarray}

\subappendix{The four-fermion operator}

The perturbative matching for the four-fermion operator is more
complicated due to operator mixing. Apart from the operator $O_L$, the
basis of local operaotrs in the continuum effective theory consists
also of the operator $O_S$ defined in eq.\,(\ref{GLdefos}). The full
$O(\alpha_s)$ matching relation between $O_L$ and the operaotrs
$\widetilde{O}_L$ and $\widetilde{O}_S$ in the continuum effective
theory is given by
\begin{eqnarray}
O_L(m_b) & = & 
\left(\frac{\amb}{\amu}\right)^{d_{1}}
  \left\{1+\frac{\amu-\amb}{4\pi}\,J+\frac{\amb}{4\pi}\,B_1\right\}\,
  \widetilde{O}_L(\mu) \nonumber\\
& & +\left\{\left(\frac{\amb}{\amu}\right)^{d_{2}}
           -\left(\frac{\amb}{\amu}\right)^{d_{1}}\right\}
    \frac{\gamma_{21}^{(0)}}{\gamma_{22}^{(0)}-\gamma_{11}^{(0)}}
     \frac{\amb}{4\pi}\,B_2\,  \widetilde{O}_L(\mu) \nonumber\\
& & +\left(\frac{\amb}{\amu}\right)^{d_{2}}\frac{\amb}{4\pi}\,B_2\,
     \widetilde{O}_S(\mu),
\end{eqnarray}
where in na\"\i{}ve dimensional regularisation\,\cite{FHH91}
\be
  B_1=-14,\qquad B_2=-8
\ee
and\,\cite{FHH91,Gim_93}${}^{-}$\cite{Buch_96}
\begin{eqnarray}
 & &   d_i=\frac{\gamma_{ii}^{(0)}}{2\beta_0},\qquad 
  J=\frac{\gamma_{11}^{(0)}}{2\beta_0}\left(\frac{\beta_1}{\beta_0}
               -\frac{\gamma_{11}^{(1)}}{\gamma_{11}^{(0)}}\right), \\
 & &  \gamma^{(0)}\equiv\left(\begin{array}{cc}
                              \gamma_{11}^{(0)} & \gamma_{12}^{(0)} \\
                              \gamma_{21}^{(0)} & \gamma_{22}^{(0)}
                              \end{array}\right)
                       =\left(\begin{array}{rr}
                              -8 & 0 \\ 4/3 & -8/3
                              \end{array}\right),\\[0.5 ex]
 & &  \gamma_{11}^{(1)}=-\frac{808}{9}-\frac{52}{27}\pi^2+\frac{64}{9}n_f.
\end{eqnarray}
When matching $\widetilde{O}_L(\mu)$ to its lattice counterpart
$\widehat{O}_L(\mu)$, one has to consider the lattice operators
$\widehat{O}_R(\mu)$ and $\widehat{O}_N(\mu)$
(eqs.\,(\ref{GLdefor},\ref{GLdefon})), such that
\begin{eqnarray}
\widetilde{O}_L(\mu) &=& \left(1+\frac{\alpha_s^{\rm latt}(\mu)}{4\pi}
  \big[4\ln(\mu^2a^2)+D_L\big]\right)\,\widehat{O}_L(a) \nonumber\\
   & & +\frac{\alpha_s^{\rm latt}(\mu)}{4\pi}D_R\,\widehat{O}_R(a)
       +\frac{\alpha_s^{\rm latt}(\mu)}{4\pi}D_N\,\widehat{O}_N(a).
\end{eqnarray}
Since $\widetilde{O}_S$ is generated at order $\alpha_s$ one finds
\be
\widetilde{O}_S(\mu) = \widehat{O}_S(a) + \hbox{higher orders}.
\ee
Values for the coefficients $D_L,\,D_R$ and $D_N$ have been determined
for $\csw=0$ and~1\,\cite{FHH91,BorPit92,BPFG_93} and are listed in
table\,\ref{Tmatcoeff}. We have chosen the convention that the
static-light meson propagator is proportional to $e^{-{\cal E}t}$, such
that the reduced value of the quark self-energy has been
used\,\cite{eichten_hill_90_2} in the evaluation of $D_A$ and $D_L$.

\begin{table}[tb]
\tcaption{Values for the coefficients $D_A,\,D_L,\,D_R$ and $D_N$ for
unimproved and improved Wilson fermions according to
refs.\,\protect\cite{BorPit92,FHH91,BPFG_93}. In the case of the four-fermion
operator for $\csw=1$, the conventions of ref.\,\protect\cite{BPFG_93}
are used.}
\begin{center}
\begin{tabular}{cr@{.}lr@{.}lr@{.}lr@{.}l}
\hline
\hline
$\csw$ & \multicolumn{2}{c}{$D_A$} & \multicolumn{2}{c}{$D_L$}
       & \multicolumn{2}{c}{$D_R$} & \multicolumn{2}{c}{$D_N$}  \\
\hline
0 & $-$27&17 & $-$38&91 & $-$1&61 & $-$14&40  \\
1 & $-$20&20 & $-$22&50 & $-$5&40 & $-$14&00  \\
\hline
\hline
\end{tabular}
\end{center}
\label{Tmatcoeff}
\end{table}

Combining the two matching steps and using tadpole improvement, we
find the following expressions for the renormalisation factors
$Z_X^{\rm stat},\,X=L,\,R,\,N,\,S$ 
\begin{eqnarray}
Z_L^{\rm stat} &=& \left(\frac{\amb}{\amu}\right)^{d_1}
  \left\{1+\frac{\amu-\amb}{4\pi}\,J-14\frac{\amb}{4\pi}\right\}
  \nonumber\\ 
  & & \times u_0\left(1+\frac{\widetilde{\alpha}_s(\mu)}{4\pi}\big[
4\ln(\mu^2a^2)+D_L-u_0^{(1)}\big]\right) \nonumber\\
& &  +\left\{\left(\frac{\amb}{\amu}\right)^{d_2}
            -\left(\frac{\amb}{\amu}\right)^{d_1}\right\} 
     (-2)\frac{\amb}{4\pi}  \\
Z_R^{\rm stat} &=& \left(\frac{\amb}{\amu}\right)^{d_1}
  \left\{1+\frac{\amu-\amb}{4\pi}\,J-14\frac{\amb}{4\pi}\right\}
  \nonumber\\
  & & \times \frac{\widetilde{\alpha}_s(\mu)}{4\pi}
             \big[4\ln(\mu^2a^2)+D_R\big] \\
Z_N^{\rm stat} &=& \left(\frac{\amb}{\amu}\right)^{d_1}
  \left\{1+\frac{\amu-\amb}{4\pi}\,J-14\frac{\amb}{4\pi}\right\}
  \nonumber\\
  & & \times \frac{\widetilde{\alpha}_s(\mu)}{4\pi}
             \big[4\ln(\mu^2a^2)+D_N\big] \\
Z_S^{\rm stat} &=& \left(\frac{\amb}{\amu}\right)^{d_2}
                   (-8)\frac{\amb}{4\pi}.
\end{eqnarray}

One can now use the above expressions directly in the calculation of
$\bb(m_b)$ or, alternatively, perform a systematic expansion of
$\zastat$ and $Z_X^{\rm stat},\,X=L,\,R,\,N,\,S$ to order
$\alpha_s$. Since the $B$ parameter is a ratio of matrix elements,
another possibility is an expansion of the ratios $Z_X^{\rm
stat}/(\zastat)^2,\,X=L,\,R,\,N,\,S$ to order $\alpha_s$.

\appendix{Continuum Limit in the Light Quark Sector}
\label{APPscales}

Here we describe the continuum extrapolation of quantities
in the light quark sector, which are commonly used to set the scale
in lattice gauge theories. We will focus on the rho meson mass
$M_\rho$, the pion decay constant $f_\pi$ and the 1P--1S splitting in
charmonium, $\SP$.

For these three quantities the question arises whether or not they can
be computed with high enough precision and reliably be extrapolated to
the continuum limit:
\begin{itemize}
\item precise lattice estimates for $f_\pi$ require knowledge of the
current normalisation constant $Z_A$, which for most of the available
lattice data is only known in perturbation theory.
\item $M_\rho$ is usually obtained with small errors, but
studies with high statistics\,\cite{JLQCD_rho_lat95} show oscillations
in the relevant correlation functions, which, despite the fact that
they can be modelled, makes it more difficult to extract a precise
mass value. Furthermore, as has been discussed in
ref.\,\cite{sommer_review}, the $a$~dependence of $M_\rho$ computed
with unimproved Wilson fermions is very strong, such that the
continuum extrapolation may not be too reliable.
\item the splitting $\SP$ shows only a weak $a$~dependence. However,
the estimates published so far have still relatively large statistical
errors\,\cite{FNAL_alpha_92}. 
\end{itemize}
This discussion underlines the importance of obtaining low energy
quantities with higher precision, and this may now be possible using
the recently proposed non-perturbatively improved Wilson
action\,\cite{alphaIII}, which is expected to yield lattice data that
are completely free of lattice artefacts at order~$a$.  Since also the
current normalisation constants have been determined
non-perturbatively\,\cite{alphaIV}, one can expect highly accurate
continuum results for $f_\pi$ as well. For the time being, however, we
must restrict our attention to the unimproved Wilson case, as lattice
data for improved actions are not yet available over a large enough
range in $\beta$ to allow for a continuum extrapolation.

The extrapolation of hadronic quantities in the light quark sector
proceeds by forming dimensionless ratios, say $f_\pi/\stg$, using
lattice data for the string tension, and extrapolating it to the
continuum limit as a function of $a\stg$. In table\,\ref{Tstring} we
list the lattice data for the string tension that have been used
throughout this review. In cases when several estimates are available
at one $\beta$ value they have been combined in a weighted
average. Estimates of $a\stg$ at ``intermediate'' values of $\beta$
(e.g. $\beta=6.17$) were then obtained by linear interpolation of
$\ln(a\stg)$ as a function of $\beta$.

\begin{table}[tb]
\tcaption{Lattice results for the string tension $a\protect\stg$.}
\begin{center}
\begin{tabular}{cr@{.}ll}
\hline
\hline
$\beta$ & \multicolumn{2}{c}{$a\stg$} & Author  \\
\hline
6.8 & 0&0730(12) & Bali\,\&\,Schilling\,\cite{bali_sch_93} \\
6.5 & 0&1068(10) & UKQCD\,\cite{UKQCD_pot65} \\[0.8ex]
6.4 & 0&1215(12) & Bali\,\&\,Schilling\,\cite{bali_sch_93} \\
    & 0&1218(5)(23) & UKQCD\,\cite{HW_lat94} \\
    & 0&1216(11) & combined \\[0.8ex]
6.2 & 0&1619(19) & Bali\,\&\,Schilling\,\cite{bali_sch_93} \\
    & 0&1609(28) & UKQCD\,\cite{light_hadrons_93} \\
    & 0&1608(7)(16) & UKQCD\,\cite{HW_lat94} \\
    & 0&1612(12) & combined \\[0.8ex]
6.0 & 0&2265(55) & Bali\,\&\,Schilling\,\cite{bali_sch_93} \\
    & 0&2182(21) & Michael\,\&\,Perantonis\,\cite{cm_sjp_90} \\
    & 0&2154(10)(40) & UKQCD\,\cite{HW_lat94} \\
    & 0&2185(18) & combined \\[0.8ex]
5.9 & 0&2702(37) & MTc\,\cite{MTc_91} \\
5.8 & 0&3302(30) & MTc\,\cite{MTc_91} \\
5.7 & 0&4099(24) & MTc\,\cite{MTc_91} \\
\hline
\hline
\end{tabular}
\end{center}
\label{Tstring}
\end{table}

\begin{table}[tb]
\tcaption{Lattice results for the hadronic scale $r_0/a$.}
\begin{center}
\begin{tabular}{cr@{.}ll}
\hline
\hline
$\beta$ & \multicolumn{2}{c}{$r_0/a$} & Author  \\
\hline
6.5 & 11&23(21) & UKQCD\,\cite{UKQCD_pot65,alpha_su3} \\
6.4 &  9&90(54) & Bali\,\&\,Schilling\,\cite{bali_sch_93,alpha_su3} \\
    &  9&75(17) & UKQCD\,\cite{HW_lat94} \\[0.8ex]
6.2 &  7&38(25) & Bali\,\&\,Schilling\,\cite{bali_sch_93,alpha_su3} \\
    &  7&29(17) & UKQCD\,\cite{HW_lat94} \\
    &  7&27(3)  & SESAM\,\cite{SESAM_pot_96} \\[0.8ex]
6.0 &  5&44(26) & Bali\,\&\,Schilling\,\cite{bali_sch_93,alpha_su3} \\
    &  5&47(11) & UKQCD\,\cite{HW_lat94} \\
    &  5&35\er{2}{3}  & SESAM\,\cite{SESAM_pot_96} \\
\hline
\hline
\end{tabular}
\end{center}
\label{Tr0}
\end{table}

\begin{figure}[tb]
\vspace{4.0cm}
\hspace{-1.4cm}
\ewxy{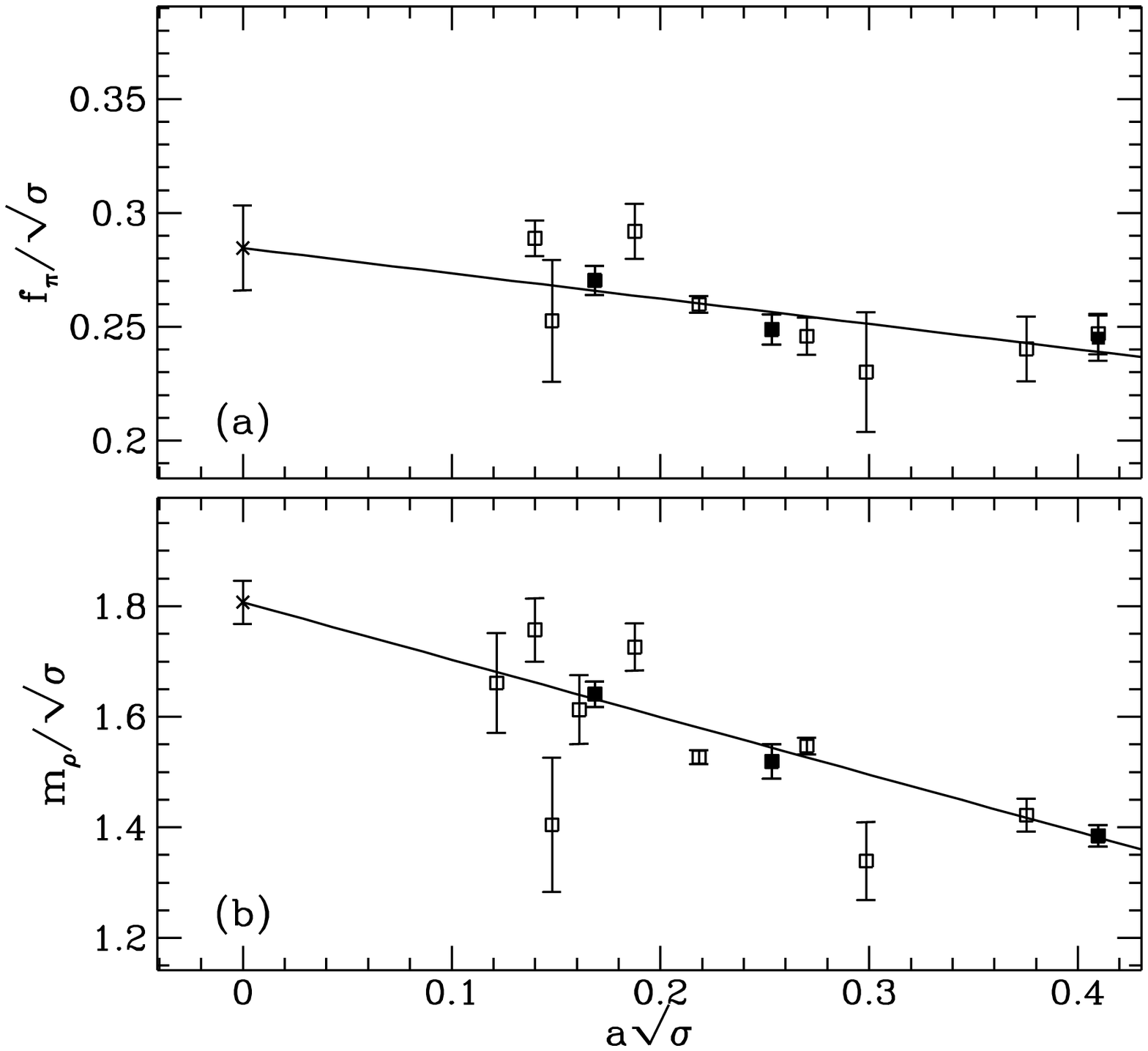}{95mm} 
\vspace{-2.6cm}
\fcaption{
Extrapolation of: (a) $f_\pi/\protect\stg$ from
ref.\,\protect\cite{GF11_fpi}, (b) $M_\rho/\protect\stg$ from
ref.\,\protect\cite{GF11_spec} to the continuum limit (full squares).
Open squares denote the data for $f_\pi$ from
refs.\,\protect\cite{BLS_93,PCW_prop_93,BG_95,FNAL_94,APE_stat60_93,APE_cab_91,QCDPAX_95,APE_spec_96}
and for $M_\rho$ from
refs.\,\protect\cite{PCW_stat_92,FNAL_94,APE_stat60_93,APE_cab_91}${}^{-}$\protect\cite{APE_spec_96}.
Data at the same value of $\beta$ have been combined.}
\label{Ffpi_mrho_cont} 
\end{figure}

The conceptual advantages of the hadronic scale $r_0$ described in
ref.\,\cite{sommer_r0_93} make it an ideal quantity to perform the
extrapolation to the continuum limit. A compilation of lattice results
for $r_0/a$ can be found in table\,\ref{Tr0}. Since $r_0/a$ is so far
only available for $\beta\ge6.0$, the Wilson data at lower $\beta$
values (in both the light and heavy quark sectors) cannot be used in
the extrapolations to $a=0$. Thus, we have decided to use the string
tension, keeping in mind that eventually $r_0$ should be used when
more precision results for hadronic quantities become available.

In order to obtain $f_\pi/\stg$ and $M_\rho/\stg$ in the continuum
limit, we shall use the results quoted by the GF11
collaboration\,\cite{GF11_fpi,GF11_spec}. Their simulations were
intended as a comprehensive study of the continuum and infinite volume
limits of the light hadronic sector in the unimproved Wilson
theory. There are, in fact, many more lattice results for $f_\pi$ and
$M_\rho$\,\cite{PCW_stat_92,BLS_93,PCW_prop_93,abada_92,BG_95,FNAL_94,APE_stat60_93,APE_cab_91}${}^{-}$\cite{APE_spec_96},
also for improved
actions\,\cite{APE_stat60_93,QCDSF_96}${}^{-}$\cite{tadpole}, and
mostly with small statistical errors. However, a reliable continuum
extrapolation of all available data is difficult to perform
without a consistent treatment of systematic errors. This is the main
reason why we have chosen to use only GF11's results in the present
study. The accuracy and reliability of the extrapolated values can be
checked by comparing to other simulations.

In figure\,\ref{Ffpi_mrho_cont}\,(a) we show the continuum extrapolation of
$f_\pi/\stg$. For the axial current normalisation factor $Z_A$ we used
the tadpole improved perturbative expression evaluated using the
boosted coupling $g_P^2=g_0^2/u_0^4$. It is seen that most other
simulations are in fact consistent with GF11's data. In principle the
error on the extrapolated result could be reduced by including more
data points. However, this would imply a very selective use of the
available data, since overall compatibility of all results is hard to
achieve. 

At zero lattice spacing we obtain
\be
   \frac{f_\pi}{\stg}\Big|_{a=0} = 0.285(13).
\ee
We have checked the stability of this result by using different
prescriptions to evaluate $Z_A$ and find that the resulting variation
is much smaller than the quoted error.

Figure\,\ref{Ffpi_mrho_cont}\,(b) shows the corresponding extrapolation of
$M_\rho/\stg$. The result at $a=0$ is
\be
   \frac{M_\rho}{\stg}\Big|_{a=0} = 1.807(39).
\ee
The figure illustrates the strong $a$~dependence of this ratio in the
unimproved Wilson theory (note that $M_\rho/\stg$ is plotted on a much
smaller scale than $f_\pi/\stg$). It is, however, reassuring to notice
that more recent simulations using a non-perturbative value of
$\csw$\,\cite{QCDSF_96,UKQCD_prep} obtain values for $M_\rho/\stg$
which are consistent with the extrapolated Wilson result, whilst
showing very little residual $a$~dependence.

In order to compute $\SP/\stg$ in the continuum limit we have
extrapolated the results quoted in ref.\,\cite{FNAL_alpha_92} and
obtain 
\be
   \frac{\SP}{\stg}\Big|_{a=0} = 1.01(12).
\ee
It should be emphasised that the non-perturbatively improved
action\,\cite{alphaIII} could be used to obatin much more precise
estimates of the above quantities. One will thus be able to set the
scale much more reliably in future simulations.

\nonumsection{References}

\end{document}